\documentclass[sigconf, nonacm]{acmart}

\usepackage[T1]{fontenc}
\usepackage[utf8]{inputenc}
\usepackage{booktabs}
\usepackage{multirow}
\usepackage{balance}
\usepackage{xspace}
\usepackage{lscape}
\usepackage{colortbl}
\usepackage{hhline}
\usepackage{adjustbox}
\usepackage{colorframed}

\usepackage{tcolorbox}
\usepackage{verbatim}
\usepackage{graphicx}
\usepackage{balance}  
\usepackage{caption}
\usepackage{subcaption}
\usepackage{setspace}
\usepackage{float}
\usepackage{tikz}
\usepackage{amsmath}
\usepackage{breqn}
\usepackage{algorithm}
\usepackage{multicol}
\usepackage[noend]{algpseudocode}
\usepackage{listings}
\definecolor{key-color}{rgb}{0.8, 0.47, 0.196}

\usepackage{framed}
\usepackage{graphicx}
\usepackage{color}
\usepackage{balance}  
\usepackage{tabularx}
\usepackage{caption}
\usepackage{amsthm}
\usepackage{booktabs}
\usepackage{multirow}
\usepackage{enumerate}

\definecolor{shadecolor}{RGB}{235,235,235}

\newenvironment{squishedlist}
{
  \begin{list}{$\bullet$}
  {
    \setlength{\itemsep}{0.2pt}
    \setlength{\parsep}{1pt}
    \setlength{\topsep}{0.3em}
    \setlength{\partopsep}{0.2pt}
    \setlength{\leftmargin}{1.4em}
    \setlength{\labelwidth}{1em}
    \setlength{\labelsep}{0.5em}
  }
}
{
  \end{list}
}

\definecolor{codegreen}{rgb}{0,0.6,0}
\definecolor{codegray}{rgb}{0.5,0.5,0.5}
\definecolor{codepurple}{rgb}{0.58,0,0.82}
\definecolor{backcolour}{rgb}{0.95,0.95,0.92}

\lstdefinestyle{mystyleCpp}{
language=C++,
backgroundcolor=\color{backcolour},   
commentstyle=\color{codegreen},
keywordstyle=\color{magenta},
numberstyle=\tiny\color{codegray},
stringstyle=\color{codepurple},
basicstyle=\ttfamily\footnotesize,
breakatwhitespace=false,         
breaklines=true,
captionpos=b,                    
keepspaces=true,                 
numbers=left,                    
numbersep=5pt,                  
showspaces=false,                
showstringspaces=false,
showtabs=false,                  
tabsize=2,
literate={-}{{-}}1,    
morekeywords={MATCH, RETURN, WHERE}
 }

\lstdefinestyle{mystyleSQL}{
language=SQL,
backgroundcolor=\color{backcolour},   
commentstyle=\color{codegreen},
keywordstyle=\color{magenta},
numberstyle=\tiny\color{codegray},
stringstyle=\color{codepurple},
basicstyle=\ttfamily\footnotesize,
breakatwhitespace=false,         
breaklines=true,
captionpos=b,                    
keepspaces=true,                 
numbers=left,                    
numbersep=5pt,                  
showspaces=false,                
showstringspaces=false,
showtabs=false,                  
tabsize=2,
morekeywords={MATCH, RETURN, WHERE}
 }

\usepackage{enumitem}
\usepackage{hyphenat}

\usepackage{hyperref}
\expandafter\def\expandafter\UrlBreaks\expandafter{\UrlBreaks\do\/\do\-\do\.} 
\definecolor[named]{Purple}{cmyk}{0.55,1,0,0.15}
\definecolor[named]{DarkBlue}{cmyk}{1,0.58,0,0.21}
\hypersetup{bookmarksnumbered,unicode,naturalnames,colorlinks,breaklinks,
	linkcolor=Purple,
	citecolor=Purple,
	urlcolor=DarkBlue,
	filecolor=DarkBlue}

\graphicspath{ {images/} }

\newcommand{\revision}[1]{{\color{black} {#1}}}

\newtheorem{example}{Example}

\newcommand{\Kuzu}{Kuzu}

\tikzstyle{io} = [trapezium, trapezium left angle=70, trapezium right angle=110, minimum width=3cm, minimum height=0.7cm, inner sep=1pt, text badly centered, text width=1.2cm, draw=black, fill=blue!10]
\tikzstyle{decision} = [diamond, minimum width=1cm, minimum height=1cm, text width=1.5cm, text badly centered, inner sep=1pt, draw=black, fill=green!10]
\tikzstyle{process} = [rectangle, minimum width=3cm, minimum height=1cm, text width=2.5cm, text centered, draw=black, fill=red!10]
\tikzstyle{arrow}=[draw, -latex]

\newcommand{\OneTOneS}{\texttt{1T1S}}
\newcommand{\nTOneS}{\texttt{nT1S}}
\newcommand{\nTkS}{\texttt{nTkS}}
\newcommand{\nTkMS}{\texttt{nTkMS}}

\newcommand{\DuckPGQnTkS}{\texttt{Duck-nTkS}}

\newcommand\vldbdoi{10.14778/3749646.3749706}
\newcommand\vldbpages{4465 - 4477}
\newcommand\vldbvolume{18}
\newcommand\vldbissue{11}
\newcommand\vldbyear{2025}
\newcommand{\HUGE}{\fontsize{22pt}{26pt}\selectfont}
\newcommand\vldbauthors{\authors}
\newcommand\vldbtitle{\shorttitle} 

\newcommand\vldbavailabilityurl{https://github.com/anuchak/kuzu}
\newcommand\vldbpagestyle{empty} 



\begin{document}
\title{Robust Recursive Query Parallelism in Graph Database Management Systems}

\newif\iflong\longtrue     
\longtrue

\newif\ifcv
\cvfalse

\author{Anurag Chakraborty}
\affiliation{%
  \institution{University of Waterloo}
  \streetaddress{200 University Ave W}
  \city{Waterloo}
  \state{Ontario}
  \country{Canada}
}
\email{a8chakra@uwaterloo.ca}

\author{Semih Salihoğlu}
\orcid{0000-0002-1825-0097}
\affiliation{%
  \institution{University of Waterloo}
  \streetaddress{200 University Ave W}
  \city{Waterloo}
  \state{Ontario}
  \country{Canada}
}
\email{semih.salihoglu@uwaterloo.ca}


\ifcv
  \renewcommand{\thesection}{CL\arabic{section}}
    \thispagestyle{\vldbpagestyle}   
  \twocolumn[%
    \vspace*{1em}                   
    \begin{center}
      {\HUGE\bfseries Cover Letter\par}
      \vspace{1em}                  
    \end{center}
  ]


We thank all the reviewers and the meta-reviewer for their constructive feedback. In our revi\textit{}sed paper,
we highlight the changes we have made to address 
the review comments  in blue font. In the cover letter, we have a section for each reviewer's main comments.
In our responses, we give pointers to the necessary sections to identify the main text changes that address reviewers’ comments. 

Addressing the comments required many edits as well as 2 additional pages of material to be added to our main text. To make space for these changes, we removed previous Figure 3, which
was showing the morsel-driven evaluation of the second pipeline of a hash join 
query evaluation. We merged Listings 1 and 3 into a single Listing. These were very similar pseudocodes describing the serial execution of the IFE subroutine. We also removed
Figures 10-12, which were presenting the data 
in Tables 3 and 4 as line charts. We present
these figures in a new longer version of the paper
we prepared~\cite{chakraborty:recjoins-tr}. 
We further replotted several other figures to reduce space and
shortened many paragraphs throughout the text. We do not highlight those paragraphs. 
We addressed the presentation issues reviewers have
brought up, e.g., the ones highlighted in the metareview and do not respond to them explicitly in the cover letter. Finally, in our response CLX, refers to Section X of the cover letter.

\section{Meta-Reviewer}

\subsection{Clarify and highlight takeaways [R3O1]}
\label{meta:r1}
\begin{snugshade*}
\noindent It is not clear what the major takeaways from the paper are. The only recommendation is ".. the \nTkS\ policy as a robust policy to parallelize recursive queries." 
The evaluation results in Tables 3 and 4 and Figure 10, 11 and 12 otherwise do not provide any non-trivial insights. It would be useful to highlight interesting takeaways 
throughout the paper, and have a section on the implications of these insights for researchers and practitioners.
\end{snugshade*}

In our revised paper, we highlighted our 
main takeaways in italics at the ends of: (i) Section \ref{subsection:many-sources-workloads}, which
is where we finish our major experiments comparing the space
of scheduling policies we consider against other baselines; (ii) Section \ref{subsection:choice-of-k}, where we study the effects of $k$; and (iii) Section \ref{subsection:ms-bfs-morsel-optimization}, where we study the effects of the multi-source morsels. 
Our first takeaway, as the reviewer observes, is that we recommend the \nTkS\ policy as the robust scheduling policy to use in systems
(highlighted in Section \ref{subsection:many-sources-workloads}). Second, after our new
drill-down experiments, we show that the 
optimal choice of $k$ depends on the average density of the input
graphs, which systems can consider when using the \nTkS\ policy (Section \ref{subsection:choice-of-k}).
Third, we provide a much more nuanced advice on the use of the 
MS-BFS optimization than prior work. In prior work this optimization
was tested only on queries with enough source nodes to saturate
each the 64 ``lanes'' of each multi-source morsel and only on 
queries that returned path lengths. We show that if the system
cannot saturate enough lanes, e.g., if the query contains less than 8 source nodes, the costs of MS-BFS overweigh its benefits, and systems are better off not using 
this optimization. Further, if queries return actual paths,
the memory costs of running many multi-source morsels can be prohibitively large, forcing systems to run out of memory (highlighted in Section \ref{subsection:ms-bfs-morsel-optimization}).
Another high-level takeaway from our 
paper is that the two different parallelism approaches adopted in 
GDBMS and parallel graph analytics systems and the MS-BFS 
optimization can all be integrated into DBMSs as different morsel 
dispatching policies. We find this the appropriate framework 
for our readers to interpret these techniques. We also highlighted this
takeaway in the conclusions section of our revised paper.
Finally, we believe our implementation details and designs will be valuable to readers, as they serve as a blueprint for how to implement these policies in other systems.

\subsection{More details, drill-downs, and explanation of results (R4W1/W2) }
\label{meta:r2}
\begin{snugshade*}
\noindent R4W1: While I really like the scope of the experiments, I find the presentation of experimental results quite confusing. In particular, drilldowns into numbers are missing and explanations for the observed trends are quite vague. The discussion of differences of measured absolute latencies between systems is also not included. \\
R4W2: Table 1 shows runtime for each frontier level as number of threads increases which clearly indicates for that single use case that system is under utilized. However, CPU utilization is not measured nor discussed in any other experiment and speedups compared to a single thread are only reported.
\end{snugshade*} 
As the reviewer notes, we already had a drill-down experiment explaining the limited parallelism behavior of \nTOneS\ policy in Table \ref{tab:frontier-runtime}. Our revised manuscript enhances our drill-down analyses in several ways. 
First, we measured the CPU utilization in our main experiments from Sections~\ref{subsection:single-source-workloads} to~\ref{subsection:many-sources-workloads}, which are reported in Tables \ref{tab:shortest-path-length-workloads} and \ref{tab:shortest-path-workloads}.
As expected, the CPU utilization numbers correlate with and explain the
runtime behaviors of \OneTOneS, \nTOneS, and \nTkS. 
In particular they validate the limited CPU utilization of
\nTOneS\ and Ligra on some of the datasets and very low utilization
of \OneTOneS\ and Neo4j on single and 8-source workloads. 
Second, we have provided 3 additional drill-down experiments
in Section \ref{subsection:choice-of-k} to 
explain the performance degradation of the \nTkS\ policy on the Spotify graph when increasing $k$ (see our response in Section~\ref{meta:r4} below). Third, we provided details about the memory requirements of the auxiliary data structures in Sections \ref{subsec:msntks} and \ref{subsection:ms-bfs-morsel-optimization} to explain
the experiments in which the \nTkMS\ policy runs out of memory.  We believe with the addition of these detailed analyses and experiments, we provide more insights into the system behaviors we highlight in our experiments.

\subsection{Show additional results on hardware with more cores (if available) (R2W2/D2.1)}
\label{meta:r3}
\begin{snugshade*}
\noindent The authors use one hardware platform with a specific number of virtual cores (32). As the paper’s focus is on the scaling behavior of morsel dispatching strategies with the number of cores, it would have been very interesting to see at least one more hardware platform with a different number of virtual cores (maybe a scale-up machine with 100+ virtual cores).
\end{snugshade*}

Unfortunately, our own research group does not have access to a machine with larger number of cores. This is the largest machine we have in our cluster, whose machines we have used in our previous recent VLDB experiments and analysis publications~\cite{graindb} and ~\cite{cardinalitygraphs} as well. We were able to gain limited access to a machine with 24 physical and 48 virtual cores to run several of our experiments (Intel Xeon Platinum 8160 CPU @2.10GHz processor, 32 KB of L1 data cache, 32KB L1 instruction cache, 1 MB L2 cache and 33 MB L3 cache). This machine does not have a meaningfully larger number of cores than the one we used in our paper, e.g., 100+ as the reviewer suggested. 
However, to verify that our main results recommending
the \nTkS\ policy is the same on this machine,
we repeated some of our main experiments from Section~\ref{sec:evaluation}. We are providing these experiments in the cover letter. Table ~\ref{tab:many-src-pathlen-48core-workload} repeats our experiment from Table 3c in the original 
paper. As can be seen, the behavior of \nTkS\ and \nTOneS\ and \OneTOneS\ policies follow a similar pattern as in Table 3c. Similarly, Figure ~\ref{fig:value-of-k-more-cores} repeats our experiment from Figure \ref{fig:choice-of-k-single-source} on Spotify in the original paper, 
and confirms that the overall trend in this experiment is also the same as in the original paper.

\begin{figure}[ht]
    \centering
    \includegraphics[width=\linewidth]{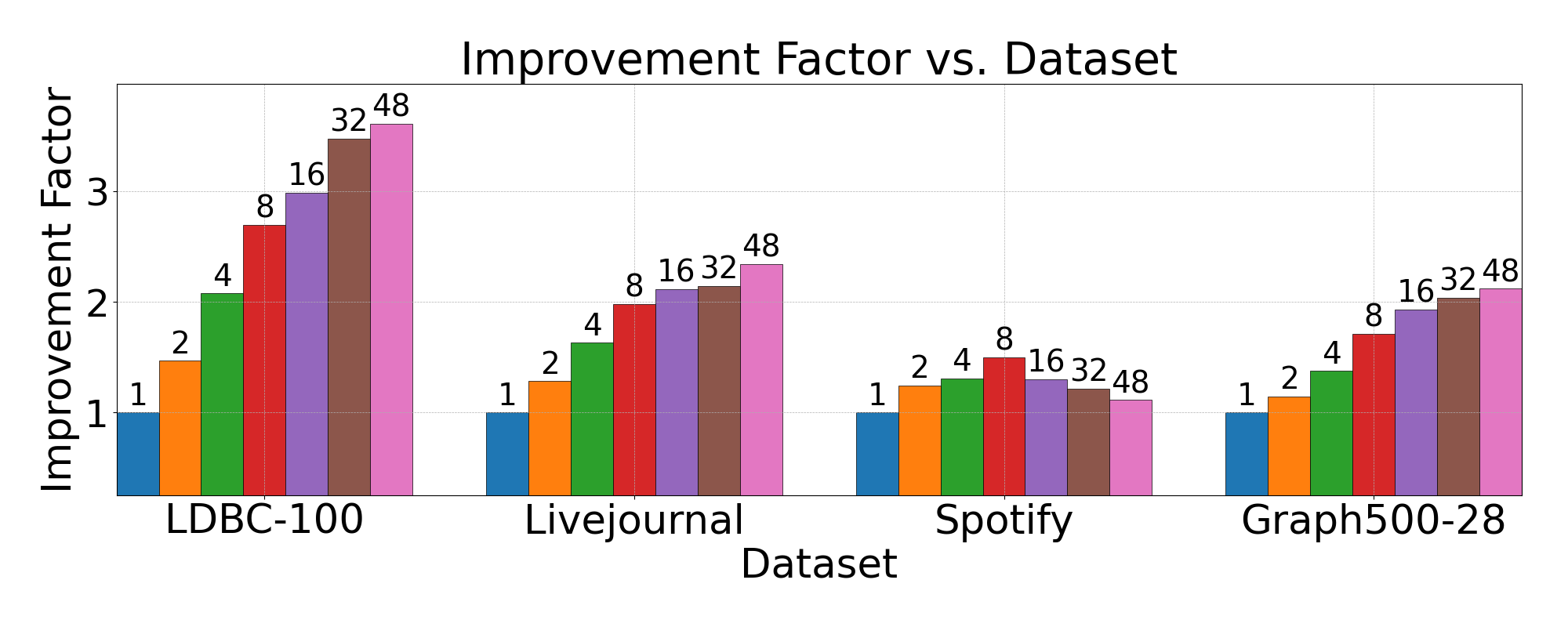}
    \vspace{-15pt}
    \caption{Varying $k$ (48-core machine)}
    \label{fig:value-of-k-more-cores}
\end{figure}

\begin{table}[ht]
\centering
\small
\setlength{\tabcolsep}{2pt}
\begin{tabular}{|c|c|c|c|c|c|}
\hline
                         & \begin{tabular}[c]{@{}c@{}}Threads\\ ($\rightarrow$)\end{tabular} & 1       & 8             & 32             & 48             \\ \hline
\multirow{6}{*}{LDBC}    & nTkS                                                                & 15736   & 2048 (7.7x)   & 573 (28.0x)    & 469 (33.5x)    \\ \cline{2-6} 
                         & nT1S                                                                & 15490   & 3099 (5.0x)   & 1658 (9.4x)    & 1475 (10.5x)   \\ \cline{2-6} 
                         & 1T1S                                                                & 15094   & 2026 (7.4x)   & 696 (21.0x)    & 548 (27.5x)    \\ \cline{2-6} 
                         & Ligra                                                               & 3163    & 585 (5.4x)    & 355 (8.9x)     & 282 (11.2x)    \\ \cline{2-6} 
                         & Neo4j                                                               & 7081    & 957 (7.4x)    & 348 (20.3x)    & 289 (24.5x)    \\ \cline{2-6} 
                         & D-nTkS                                                              & 4129    & 1119 (3.7x)    & 1115 (3.7x)    & 1089 (3.8x)    \\ \hline
\multirow{6}{*}{LJ}      & nTkS                                                                & 113133  & 13967 (8.1x)  & 4334 (26.1x)   & 3459 (32.7x)   \\ \cline{2-6} 
                         & nT1S                                                                & 113237  & 19866 (5.7x)  & 9515 (11.9x)   & 8514 (13.3x)   \\ \cline{2-6} 
                         & 1T1S                                                                & 109828  & 13729 (8.0x)  & 5132 (21.4x)   & 4256 (25.8x)   \\ \cline{2-6} 
                         & Ligra                                                               & 59456   & 9290 (6.4x)   & 4608 (12.9x)   & 4100 (14.5x)   \\ \cline{2-6} 
                         & Neo4j                                                               & 112830  & 14466 (7.8x)  & 5698 (19.8x)   & 4927 (22.9x)   \\ \cline{2-6} 
                         & D-nTkS                                                              & 27923   & 3534 (7.9x)   & 1131 (24.7x)   & 886 (31.5x)    \\ \hline
\multirow{6}{*}{Spotify} & nTkS                                                                & 820176  & 106751 (7.7x) & 38005 (21.5x)  & 35978 (22.8x)  \\ \cline{2-6} 
                         & nT1S                                                                & 810616  & 114085 (7.1x) & 38750 (20.1x)  & 38156 (21.2x)  \\ \cline{2-6} 
                         & 1T1S                                                                & 811096  & 109211 (7.4x) & 44528 (18.2x)  & 39565 (20.5x)  \\ \cline{2-6} 
                         & Ligra                                                               & 1065394 & 183688 (5.8x) & 54080 (19.7x)  & 44577 (23.9x)  \\ \cline{2-6} 
                         & Neo4j                                                               & 2381885 & 345200 (6.9x) & 133813 (17.8x) & 104012 (22.9x) \\ \cline{2-6} 
                         & D-nTkS                                                              & 818891  & 104986 (7.8x) & 34552 (23.7x)  & 31740 (25.8x)  \\ \hline
\multirow{6}{*}{G-28}    & nTkS                                                                & 3959436 & 557667 (7.1x) & 179160 (22.1x) & 159013 (24.9x) \\ \cline{2-6} 
                         & nT1S                                                                & 3929836 & 770556 (5.1x) & 293271 (13.4x) & 248724 (15.8x) \\ \cline{2-6} 
                         & 1T1S                                                                & 3929406 & 613970 (6.4x) & 188009 (20.9x) & 172342 (22.8x) \\ \cline{2-6} 
                         & Ligra                                                               & 3040976 & 515420 (5.9x) & 201389 (15.1x) & 172783 (17.6x) \\ \cline{2-6} 
                         & Neo4j                                                               & 4243249 & 614963 (6.9x) & 245275 (17.3x) & 204002 (20.8x) \\ \cline{2-6} 
                         & D-nTkS                                                              & 2034338 & 271245 (7.5x) & 86567 (23.5x)  & 76192 (26.7x)  \\ \hline
\end{tabular}
\caption{64-Source Workload (48 cores)}
\label{tab:many-src-pathlen-48core-workload}
\end{table}

\subsection{Explain counter-intuitive results for Spotify dataset (R2W2/D2.3, R4W3)}
\label{meta:r4}
\begin{snugshade*}
\noindent Experimental results deviating from the general trend are not explained. For instance, in Figure 13 (a), the query runtime does not improve for k > 4. This fact is merely mentioned in Section 5.5, but not explained.
\end{snugshade*}

We have addressed this comment in Section \ref{subsection:choice-of-k}, where we provide several drill-down experiments to explain why Spotify behaves differently. Our first drill-down experiment correlates the runtime improvement factors we observe across different datasets with CPU cache hits, which we measure
using the last-level cache load throughput (LLC-Load metric in Linux perf outputs) across our experiments in this section. Here we make two observations. First, in Spotify, LLC-Load throughput increases
until $k=4$ and then decreases. In contrast, in every other dataset LLC-Load
throughput only increases with increasing $k$.
Second, the LLC throughput is significantly larger in Spotify (between 38-50M loads/second) compared to other datasets (at most 24M loads/second). Therefore even when $k$ is small, Spotify is achieving very good cache locality compared to other datasets. This analysis shows that the CPU cache locality perfectly correlates with the runtime numbers we see.

We then explained this cache behavior through a structural property of the Spotify graph. Specifically, Spotify has a much higher average density (535) than other datasets (at most 44). This leads to better cache locality because on average nodes have more common neighbors in Spotify. Therefore, if threads are working on the same source morsel, then during frontier extensions they access same locations in the auxiliary data structures we use, specifically the visited array structure. We performed additional experiments to explicitly measure the amount of common accesses to the same visited array locations, which we present in the longer version of our paper~\cite{chakraborty:recjoins-tr} due to space constraints. These experiments verify that in Spotify, there are many more common accesses to the same visited array than in other datasets. The conclusions from these drill-down analyses is that increasing $k$ has two effects. On the one hand, it has the positive effect of increasing total parallelism, since each thread works on separate source morsels. On the other hand, on graphs with high average degrees, it can also decrease the cache locality because threads start working on different auxiliary data structures. Therefore, on graphs with high degree, the loss of cache locality can offset the parallelism benefits of increasing $k$, and even lead to performance degradation.

To further validate this phenomenon, we performed a controlled experiment on a set of random graphs, where we progressively increased the average degree of a graph with 5 million nodes, from 25 to 500, and showed that starting from an average degree of 100, we see a Spotify-like behavior. We believe these detailed explanations clarify the deviating behavior of Spotify.

\subsection{Report certain parameters and explain their impact (R4W4)}
\label{meta:r5}
\begin{snugshade*}
\noindent Sizes of auxiliary data structures and availability of resources are mentioned in multiple places in the evaluation section as a limiting factor, but the values are never reported so it is hard for the reader to fully understand their impact.
\end{snugshade*}

We have extended Section \ref{subsec:data-structures} to specify the size of the auxiliary data structures used in our implementation. Additionally, we also specify the size of these data structures in Section \ref{subsection:ms-bfs-morsel-optimization} that cause the queries to run out of memory.

\subsection{Report on stability of measurements (R2W2/D2.4)}
\label{meta:r7}
\begin{snugshade*}
\noindent (D2.4) Section 5.1: “Any reported runtimes are averages of 3 repeated executions.” Three repetitions doesn’t sound much. Did you verify how stable the numbers are? It would be great if you could report how much skewed the runtime results are.
\end{snugshade*}

We have specified the average deviation of our execution numbers in Section \ref{sec:exp-setup} and the median deviation across the datasets. 
In addition to this we clarify in  Section \ref{sec:evaluation} that all queries on the system are run on a warmed up instance. We first run the queries to warm up the database buffer, discard the first runtime and report the average of three more runs.


\section{Reviewer 1}

\subsection{D2 and part of D3: Extended evaluation}
\label{r2:c1}
\begin{snugshade*}
\noindent Must extends its evaluation significantly. Implementing the policies in one other system is a good step towards that direction but clearly not sufficient. I'd recommend evaluating two or more additional existing systems that implement one of the policies and adding a larger system and a metric (CPU utilization) to the evaluation.
\end{snugshade*}
\begin{snugshade*}
\noindent At the same time, some comparison is possible across systems. However, implementing the policies *also* in another system would still strengthen the claimed generality of these insights.
\end{snugshade*}
\noindent 
To address this comment, we have added DuckPGQ as a new baseline system and integrated
the \nTkS\ policy in it. We call this 
baseline \DuckPGQnTkS. We present
\DuckPGQnTkS\ numbers to our main experiments
in Sections~\ref{subsection:single-source-workloads}-\ref{subsection:many-sources-workloads}. 
DuckPGQ creates an in-memory
CSR indices to store adjacency lists
completely from scratch for each query. 
We exclude the CSR index construction pipeline
from our reported numbers, which dominates
end-to-end runtime. However,
if we focus only the IFE computation
part of DuckPGQ, \DuckPGQnTkS\ scales very similarly to \Kuzu-\nTkS\ and is even
faster. This is because
\DuckPGQnTkS\ 
accesses neighbors directly in memory to scan neighbors while Kuzu's adjacency lists 
are disk-based and are accessed through
Kuzu's buffer manager. In short, this baseline,
verifies the effectiveness of the \nTkS\
policy in a separate system.
We now also report the CPU utilization metrics.
Please see our metareview response in \ref{meta:r2}.

\subsection{D3: Affiliation of the authors and Kuzu}
\begin{snugshade*}
\noindent It must be noted that the main reason for the authors to pick K\`uzu is the fact that they are the initiator of that project and co-founder of the start-up that now continues its development. That is, in principle, fine and in some way strengthens the paper because the implementation is done by people who intimately know the target system and there are real chances for the prototype to be used by the system eventually. However, the paper should mention the affiliation of the authors and cite prior work on Kuzu as work by the main authors, both of which are currently not the case.
\end{snugshade*}
\noindent The reviewer is right that Kuzu started as a research project in the second author's research group at University of Waterloo. The second author is also the co-founder of the company that is now commercializing Kuzu. We added a footnote about this in the introductory section of the paper. However, the research behind this paper is carried out in the second author's role as the PhD advisor of the first author of the paper, so we find it more appropriate to list only his university affiliation in the author block.

\section{Reviewer 2}
\subsection{D2.1: Experiments on more cores}
\begin{snugshade*}
\noindent ... it would have been very interesting to see at least one more hardware platform with a different number of virtual cores (maybe a scale-up machine with 100+ virtual cores).
\end{snugshade*}
For this point, please see our response in \ref{meta:r3}.
\subsection{D2.2: NUMA effects}
\begin{snugshade*}
\noindent The used hardware platform has two NUMA nodes, so one could expect NUMA effects to play a role. However, NUMA effects are not investigated in the evaluation.
\end{snugshade*}  
We did not investigate NUMA effects in our implementation as our implementation does not contain optimizations to be NUMA-aware. We leave the implementation of NUMA-aware parallelization policies to future work.

\subsection{D2.3 and D3: Different Spotify behavior}
\begin{snugshade*}
\noindent Experimental results deviating from the general trend are not explained. For instance, in Figure 13 (a), the query runtime does not improve for k > 4. This fact is merely mentioned in Section 5.5, but not explained.
\end{snugshade*}
\noindent For this point, please see our response in \ref{meta:r4}. 

\subsection{D2.4: Runtime averaging details}
\begin{snugshade*}
\noindent Section 5.1: “Any reported runtimes are averages of 3 repeated executions.” Three repetitions doesn’t sound much. Did you verify how stable the numbers are? It would be great if you could report how much skewed the runtime results are.
\end{snugshade*}
\noindent
For this point, please see our response in \ref{meta:r7}.

\subsection{D2.5: Selection of source nodes}
\begin{snugshade*}
\noindent Section 5.1: How did you select the source nodes? 
\end{snugshade*}
We selected the sources nodes randomly and ensured that from each selected source node, we can perform three levels of IFE computations. In our revised paper, we explain this detail in Section~\ref{sec:exp-setup}.

\subsection{D3: Insights}

\begin{snugshade*}
\noindent Except for the behavior on the Spotify dataset in Figure 13 (a) (which is not explained in the text and not in-line with the story), all experimental results are pretty much as one could expect from the description of the morsel dispatching strategies. For a regular research paper, this would be good. For an EA\&B paper, though, this means readers don’t get a lot of fundamentally new insights.
\end{snugshade*}

For this point, please see our response in \ref{meta:r1}.

\subsection{D5 and D6: Related Work}
\label{r2:d26}
\begin{snugshade*}
\noindent As far as I understand, you focus on the single-node (but multi-core) graph processing. I was wondering how recursive queries are parallelized in distributed graph processing systems (multiple nodes). Is there any related work you could draw inspiration from?
\end{snugshade*}
\begin{snugshade*}
\noindent Looking at Listing 1, it seems like the core of the problem you investigate is how to best parallelize the work of two nested for-loops, whereby the iterations of each for-loop can be processed independently and in parallel. The goal is to fully utilize the hardware resources (threads/vcores). I understand that your focus is specifically on query processing in graph databases. However, I could imagine that the underlying problem is more general and might have been investigated before in adjacent fields.
\end{snugshade*}
In our original paper, we had cited the Pregel distributed system, which adopt vertex-centric parallelism. Pregel is similar to Ligra and adopts the frontier-based parallelism that is captured in our \nTOneS\ policy. So, our work is directly related to these prior works that study distributed systems. 
In our revised work, we added a new paragraph in related work that briefly connects our work to the work on communication-avoiding parallel algorithms from the HPC literature. In this literature, IFE-base graph algorithms are modeled as sparse matrix multiplication algorithms, where a matrix represents the neighbors of nodes. This literature aims to design different ways to partition and replicate parts of matrices across a distributed set of compute nodes to minimize communication across machines. Instead, we assume a DBMS that has a fixed layout of the graph where parallel threads can access entire adjacency lists (this corresponds to accessing each row of a matrix). We note also that graph DBMSs differ from the setting assumed in this literature in other ways, e.g., graph databases compute and output paths and assume an operator-based tuple-at-a-time or vector-at-a-time processors. So our work is closer to the DBMS literature we cover. Finally, we cited 
the Boehm et al. paper the reviewer mentioned
in the first paragraph of related work. Specifically,
we mention task- and data-level parallelism as the
two broad approaches to parallelism in data systems.
We position morsel-driven parallelism as a form
of data parallelism and mention the Boehm et al. work as an example of a system that adopts a hybrid approach.

\section{Reviewer 3}
\subsection{O1: Key takeaways}
\begin{snugshade*}
\noindent It is not clear what the major takeaways from the paper are. The only recommendation (Section 7 Conclusions) is ".. the nTkS policy as a robust policy to parallelize recursive queries." The evaluation results in Tables 3 and 4 and Figure 10, 11 and 12 otherwise do not provide any non-trivial insights. It would be useful to highlight interesting takeaways throughout the paper, and have a section on the implications of these insights for researchers and practitioners.
\end{snugshade*}
\noindent 
For this point, please see our reponse in \ref{meta:r1}.

\subsection{O2: Choice of OSS systems to modify}
\begin{snugshade*}
\noindent The authors have modified some systems (Ligra) to adhere to the proposed scheduling policies, but chose not to modify other open-source systems (such as DuckPGQ). Having results for a widely used and popular system like DuckPGQ can strengthen the paper's contributions.
\end{snugshade*}
\noindent 
For this point, please see our response in \ref{r2:c1}.

\subsection{O3: Improvements and clarifications in evaluation}
\begin{snugshade*}
\noindent Some clarifications are required in the 
evaluation: \\
Beyond the vertex and edge sizes, what are the 
characteristics of the considered benchmark datasets. As 
mentioned before, Table 1 was interesting to me as it unveiled 
some of the internal properties of the graph such as the number 
of IFE nodes at each level. More specific insights like this 
about each of the datasets, and mapping how the characteristics 
influence performance on the considered queries would be useful.
\end{snugshade*}
\noindent 
One characteristic that has an effect on the overall scalability of the \nTOneS\ and \nTkS\ policies is the average degrees of the nodes, which can affect
the CPU cache locality of the \nTkS\ policy and the optimal
choice of $k$. We have emphasized this structural property
of graphs in our revised manuscript. Please see our response
in \ref{meta:r4} for details of this property.

\section{Reviewer 4}
\subsection{W1-W2: Drill-down experiments and CPU utilization}
\begin{snugshade*}
\noindent While I really like the scope of the experiments, I find the presentation of experimental results quite confusing. In particular, drilldowns into numbers are missing and explanations for the observed trends are quite vague. The discussion of differences of measured absolute latencies between systems is also not included.
\end{snugshade*}
\noindent 
\begin{snugshade*}
\noindent (W2) Table 1 shows runtime for each frontier level as number of threads increases which clearly indicates for that single use case that the system is under utilized. However, CPU utilization is not measured nor discussed in any other experiment and speedups compared to a single thread are only reported.
\end{snugshade*}
\noindent 
For these points, please see our response in \ref{meta:r2}.

\subsection{W3: Different Spotify behavior}
\begin{snugshade*}
\noindent (W3) Discussion in section 5.5 regarding tuning parameter k focuses on observing performance differences that are similar for all datasets except Spotify which is not clearly explained. Is structure of this dataset different compared to others in some way, e.g. are sizes of frontier levels different? 
\end{snugshade*}
\noindent 
For this point, please see our response in \ref{meta:r4}.

\subsection{W4: More experimental details}
\begin{snugshade*}
\noindent (W4) Sizes of auxiliary data structures and availability of resources are mentioned in multiple places in the evaluation section as a limiting factor, but the values are never reported so it is hard for the reader to fully understand their impact.
\end{snugshade*}
\noindent 
For this point, please see our response in \ref{meta:r5}.

\clearpage
\balance

    \clearpage
  \setcounter{page}{1}
  \setcounter{section}{0}
  \setcounter{subsection}{0}
  \setcounter{subsubsection}{0}
  \setcounter{figure}{0}
  \setcounter{table}{0}
  \renewcommand{\thesection}{\arabic{section}}
\fi

\begin{abstract}
Efficient multi-core parallel processing of recursive join queries is critical for 
achieving good performance in graph database management systems (GDBMSs). 
Prior work adopts two broad approaches. First is the state of the 
art morsel-driven parallelism, whose vanilla application in GDBMSs parallelizes computations at the source node level. Second is to parallelize each iteration of the
computation at the {\em frontier} level.
We show
that these approaches can be seen as part of
a design space of morsel dispatching policies based on picking
different granularities of morsels. We then empirically study 
the question of which policies parallelize better in practice
under a variety of datasets and query workloads that contain
one to many source nodes.
We show that these two policies can be combined 
in a hybrid policy that issues morsels both at the source node and frontier levels. We then show that the multi-source breadth-first search optimization from prior work can also be modeled as 
a morsel dispatching policy that packs multiple source nodes into {\em multi-source morsels}.
We implement these policies inside a single system, 
the \Kuzu\ GDBMS, and evaluate them both within \Kuzu\ and across other systems. We show that 
the hybrid policy captures the behavior
of both source morsel-only and frontier morsel-only policies in cases when these approaches
parallelize well, and outperform them on queries when they are limited,
and propose it as a robust approach to parallelizing recursive queries. We further show that assigning multi-sources is beneficial, as it reduces the amount of scans, but only when there is enough sources in the query.

\end{abstract}
\maketitle

\pagestyle{\vldbpagestyle}
\begingroup\small\noindent\raggedright\textbf{PVLDB Reference Format:}\\
\vldbauthors. \vldbtitle. PVLDB, \vldbvolume(\vldbissue): \vldbpages, \vldbyear.\\
\href{https://doi.org/\vldbdoi}{doi:\vldbdoi}
\endgroup
\begingroup
\renewcommand\thefootnote{}\footnote{\noindent
This work is licensed under the Creative Commons BY-NC-ND 4.0 International License. Visit \url{https://creativecommons.org/licenses/by-nc-nd/4.0/} to view a copy of this license. For any use beyond those covered by this license, obtain permission by emailing \href{mailto:info@vldb.org}{info@vldb.org}. Copyright is held by the owner/author(s). Publication rights licensed to the VLDB Endowment. \\
\raggedright Proceedings of the VLDB Endowment, Vol. \vldbvolume, No. \vldbissue\ %
ISSN 2150-8097. \\
\href{https://doi.org/\vldbdoi}{doi:\vldbdoi} \\
}\addtocounter{footnote}{-1}\endgroup

\vspace{-10pt}
\ifdefempty{\vldbavailabilityurl}{}{
\vspace{.3cm}
\begingroup\small\noindent\raggedright\textbf{PVLDB Artifact Availability:}\\
The source code, data, and/or other artifacts have been made available at 
\url{https://github.com/anuchak/kuzu}.
\endgroup
}

\section{Introduction}
\label{sec:introduction}

Modern graph database management systems (GDBMS) follow the property graph data model, 
which supports modeling application records in the form of nodes and edges.
 An important feature of GDBMSs is that their query languages have the notion of paths as a first-class citizen 
 and special clauses to ask several common recursive queries. For example,
the Cypher query language~\cite{opencypher} has the arrow-based syntax to specify paths
or the Kleene star (``*'') syntax followed by 
the ``* SHORTEST'' keyword to compute shortest paths between nodes.
We refer to queries that require such recursive evaluation as \textit{recursive queries} and the specialized query language clauses in GDBMSs as \textit{recursive clauses}.

\begin{figure}[t!]
\centering
  \includegraphics[keepaspectratio, width=8 cm]{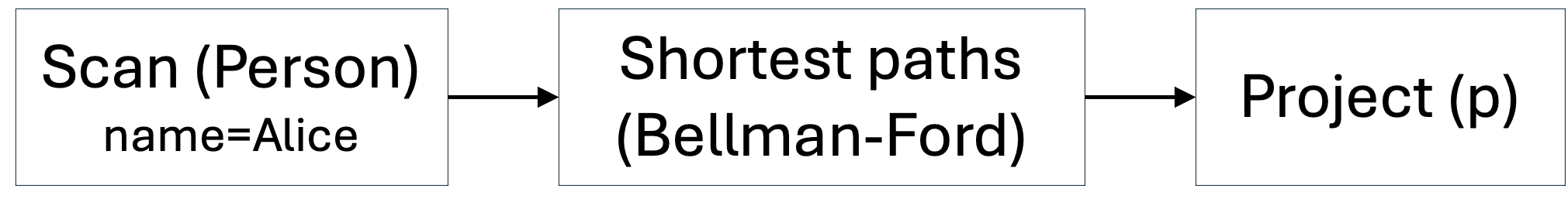}
  \caption{Example query plan with a recursive operator.}
  \label{fig:ex-gdbms-plan}
\end{figure}

We study the problem of how to parallelize recursive queries in GDBMSs. To explain the core problem that motivates this paper,
consider the following Cypher query that finds the shortest paths that consist of \texttt{Knows} edges  
from each Person node with name Alice to other Person nodes:

\begin{lstlisting}[language=SQL,style=mystyleSQL]
MATCH p = (a:Person)-[r:Knows* SHORTEST]->(b:Person)
WHERE a.name = Alice RETURN p
\end{lstlisting}

\noindent An example standard plan, drawn left to right, corresponding to this query is shown in Figure~\ref{fig:ex-gdbms-plan}. 
The plan contains a specialized  
shortest path operator,
which scans source nodes and runs the Bellman-Ford shortest path algorithm from the sources. Often, the core computation inside the recursive algorithm is a breadth-first search like computation which can be expressed as an {\em iterative frontier extensions} (IFE) subroutine~\cite{ammar:ife}. Briefly, in the IFE subroutine, neighbors of a {\em frontier} (i.e., a set) of ``active'' nodes' are explored to form a new frontier of active nodes, until a convergence criterion is met, such as when the next frontier after an iteration is empty.


A common approach to parallelizing queries in DBMSs is \textit{morsel-driven parallelism}~\cite{leis:morsel-driven}. 
It breaks a query plan into one or more subplans (``tasks''/``pipelines''), each of which
starts with a table scan operator, which scans tuples from a base or intermediate table (henceforth ``leaf table'').
The system executes tasks in some order and parallelizes each task $T$ by assigning small fragments of inputs, called {\em morsels}, from the leaf table to worker threads, which work in parallel on $T$ until 
the leaf table is consumed.  
This approach is adopted across many RDBMSs and GDBMSs, such as Hyper~\cite{5767867}, Umbra~\cite{Neumann2020UmbraAD},  
DuckDB~\cite{Raasveldt2019DuckDBAE}, Neo4j~\cite{neo4j} and Kuzu~\cite{kuzu:cidr}. In vanilla morsel-driven parallelism, if there
are not enough morsels available for all threads, e.g., when
there is a single Person node with name Alice, the system can end
up assigning morsels to a few threads while keeping other
threads idle. However, a recursive query from even a single
source node can be expensive and amenable to parallelization, as
real-world graph databases tend to be heavily connected. 
This paper studies the question of: {\em How should a GDBMS that adopts morsel-driven parallelism parallelize recursive computations?}

IFE is a common subroutine used not only to perform
recursive path computations but also other graph algorithms that are executed in parallel or distributed graph analytics systems, such as Ligra~\cite{Ligra} and Pregel~\cite{Pregel}. These systems adopt an alternative approach to parallelize the IFE subroutine.  
Specifically, these systems parallelize each frontier of the IFE subroutine by assigning subsets of active nodes to different threads in each iteration. Therefore, these systems parallelize the work of each iteration of a recursive computation from a source node.

In this paper, we first describe the design space of parallelization approaches that a system can adopt based on the granularity of morsels the system can pick. We call these approaches {\em morsel dispatching policies} and show that this space captures and
generalizes the commonly adopted approaches from above.
Vanilla morsel-driven approach of GDBMSs issues {\em source morsels} to threads. We
refer to this policy as {\texttt 1T1S}, for {\bf 1}-{\bf T}hread-to-{\bf 1}-{\bf S}ource scheduling.
In contrast, parallel graph analytics systems issue morsels from the frontiers of a single source node.
We refer to these as
{\em frontier morsels} and to this policy
as {\texttt nT1S}, for {\bf n}-{\bf T}hreads-to-{\bf 1}-{\bf S}ource scheduling. We then identify a hybrid policy 
that issues both source morsels and frontier
morsels to threads. We refer to this hybrid
approach as \texttt{nTkS}, for {\bf n}-{\bf T}hreads-to-{\bf k}-{\bf S}ource nodes policy. 

Next, we empirically analyze
the pros and cons of different policies under a variety
of datasets and query workloads that contain from one to hundreds of sources.
We show: (i) the \texttt{1T1S} policy parallelizes well
when there are many sources but degrades on queries with few sources, (ii) the \texttt{nT1S} policy achieves limited parallelism with few sources (this limited 
parallelism persists with many sources). The hybrid \texttt{nTkS} approach captures the desired parallelism behavior of 
\texttt{1T1S} and \texttt{nT1S} on queries where they parallelize well, and outperforms them on queries where they demonstrate limited
parallelism. As such, we recommend it as a robust morsel dispatching policy for GDBMSs adopting morsel-driven parallelism.

Finally, we revisit the multi-source breadth-first search optimization from prior work~\cite{msbfs}, which performs concurrent breadth first
searches from a batch of source nodes (implemented in DuckPGQ ~\cite{DuckPGQ}). We show that this optimization
can also be modeled as a morsel dispatching policy, in which
multiple source nodes are packed into {\em multi-source morsels}.
We describe a hybrid policy called \texttt{nTkMS},
for {\bf n}-{\bf T}hreads-to-{\bf k}-{\bf M}ulti-{\bf S}ource nodes policy.
Instead of dispatching source morsels that contain single source nodes, \texttt{nTkMS} 
dispatches work both as {\em multi-source morsels}, containing up to 64 source nodes, and frontier morsels.
We show empirically that the \texttt{nTkMS} policy outperforms the \texttt{nTkS} policy, as it can reduce the amount of scans performed from the database, 
however only when there are enough source nodes in the query to saturate the 64-size groups.

We have implemented all of these policies in \Kuzu~\cite{kuzu:cidr}, which is a columnar GDBMS that adopts morsel-driven parallelism.
\footnote{
\revision{\Kuzu\ started as a research prototype in our research group and is now actively being developed in a spinoff company co-founded by the second author of this paper.}}
We compare our own implementations in \Kuzu\ with Neo4j and Ligra systems, as well as our implementation
of the \nTkS\ policy in the DuckPGQ system.
This allows us to demonstrate the behavior of these policies both
in a controlled manner in a single system as well as 
on different system implementations that are at
different performance levels. 
Aside from the morsel dispatching suggestions we make in the paper, the details of our implementation in \Kuzu\ can be of independent interest to readers and serve as a blueprint for how these policies can be implemented in other GDBMSs. 




\section{Background}
\label{sec:background}
We first cover morsel-driven parallelism in more detail with an example. 
Then, we cover the {\em iterative frontier extensions (IFE)}~\cite{Ligra, Ammar_2022}
algorithmic subroutine. Lastly, we describe
the basic query plan structure that 
we assume that a GDBMS generates to evaluate
recursive clauses in this paper.
We note that throughout the paper, the terms ``source'' and ``destination''  do {\em not} indicate
any direction (forward or backward) in the recursive computations.
``Source'' indicates the nodes from which a recursive computation finds paths 
to a set of ``destination'' nodes in some direction. 

\subsection{Morsel-Driven Parallelism}

Consider the following SQL query that consists of a join
between an \texttt{Employee} and \texttt{Department} records.

\begin{lstlisting}[language=SQL,style=mystyleSQL]
SELECT e.name, b.name WHERE e.age > 55 
FROM Employee e, Department d WHERE e.dID = d.ID;
\end{lstlisting}

\begin{figure}[ht]
\includegraphics[keepaspectratio,width=8.5cm]{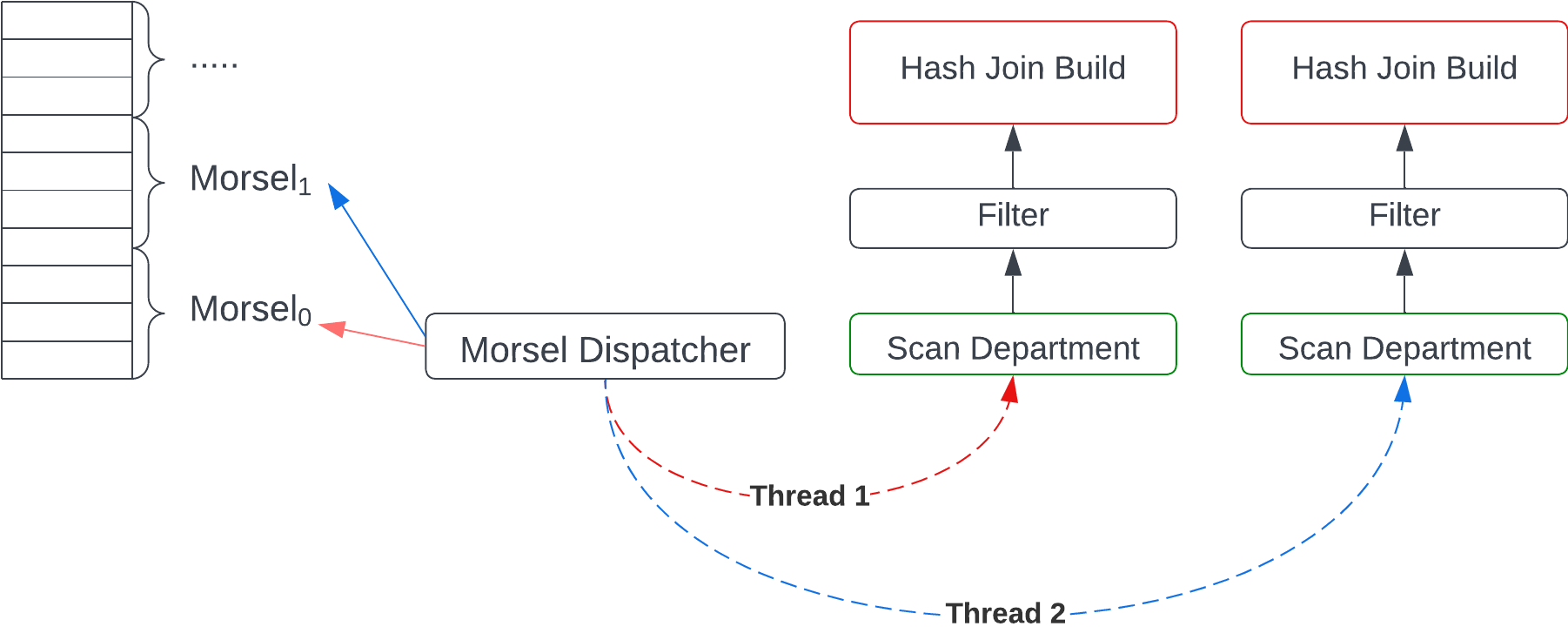}
\centering
\caption{Morsel-driven execution of hash join build task.}
\label{fig:morsel1}
\end{figure}

\noindent In morsel-driven parallelism~\cite{leis:morsel-driven}, a DBMS
breaks the query plans into subplans (a.k.a. tasks/pipelines) that are executed in some order. 
Consider a simple hash join-based plan for this query that builds a hash table on the \texttt{Department} table, which
is probed by each  \texttt{Employee} tuple. A standard approach is to 
break this plan into two tasks: (i) Task$_1$ is the subplan that builds the hash table (Figure \ref{fig:morsel1});
and (ii) Task$_2$ is the subplan that probes the hash table from Task$_1$.
Each task is a linear chain of operators that starts
with a leaf operator that scans a {\em leaf} table, which are distributed to multiple threads for parallel execution of the task.

Each thread $W_i$ creates a copy of the task and 
scans morsels of tuples, e.g., 100K from the leaf table. These tuples are processed by the rest of the operators in the task 
until $W_i$ needs to grab another morsel of tuples. This parallel computation continues until all of the leaf table's tuples are exhausted. The logic of assignment of morsels to threads is 
implemented by a piece of code termed \textit{morsel dispatcher} \cite{leis:morsel-driven}. 

Figure \ref{fig:morsel1} shows a possible execution of Task$_1$ under two threads. 
The morsel dispatcher assigns \texttt{Morsel$_0$} to \texttt{Thread$_1$} and
\texttt{Morsel$_1$} to  \texttt{Thread$_2$}. 
After grabbing their morsels, threads execute the rest of the operators independently except at the last operators
of subplans, which form a {\em pipeline break}. This is
where synchronization may be needed, e.g., 
to build a global hash table out of local thread-level hash tables. 
Once Task$_1$ is finished, the system starts executing Task$_2$.


\begin{lstlisting}[float=tb, caption={Serial IFE subroutine. Outer-most for loop starts an IFE subroutine from a set of source nodes.},abovecaptionskip=0pt,belowcaptionskip=-10pt,label={lst:serial-ife},language=C++,style=mystyleCpp,escapechar=^]
for (src : srcNodes): ^\label{line:first-for-loop}^
   nextFrontier.setActive(src); ^\label{line:ife-line-start}^
   while (!curFrontier.isEmpty()): ^\label{line:ife-subroutine}^
      swapCurNextFrontiers()
      for (node : graph->nodes()): ^\label{line:second-for-loop}^
         if (curFrontier->isActive(node)):
            for (nbr : node.scanFwd()): ^\label{line:third-for-loop}^
               if (edgeCompute(node, nbr)): 
                  nextFrontier->setActive(nbr)
      curFrontier.reset() ^\label{line:ife-subroutine-end}^
   outputResults() ^\label{line:ife-line-end}^
\end{lstlisting}

\subsection{Iterative Frontier Extensions (IFE)}
\label{subsec:ife}

IFE is a BFS-like high-level algorithmic subroutine
based on message passing between nodes and their neighbors. 
IFE is at the core of many recursive path finding operators implemented in GDBMSs, such as Neo4j \cite{gds-neo4j-github}, \Kuzu \cite{kuzu:cidr}, Memgraph \cite{memgraph-github}, or DuckPGQ \cite{DuckPGQ}. As such, in this paper, we focus on parallelizing
recursive operators that execute IFE subroutines.
In an IFE subroutine, the computation starts from an initial {\em frontier}, which is a set of nodes from which the 
recursive computation is triggered.
Then, in iterations, the neighbors of each vertex in the current frontier are explored to 
construct the next frontier of active nodes, until a convergence criterion is met. Depending on the particular 
recursive clause, a node may be visited multiple times or just once while performing the recursive 
computation.
Listing \ref{lst:serial-ife} shows the pseudocode of the IFE sbroutine. Let us ignore the outer-most for loop for now. The core subroutine is between lines~\ref{line:ife-subroutine} and~\ref{line:ife-subroutine-end}.
The pseudocode is written using the \texttt{edgeCompute()} interface of systems like Pregel or 
Ligra. This is the interface we use in our implementation as well. 
For each `active' vertex $u$, IFE executes \texttt{edgeCompute()} on each $e = (u, v)$
edge of $u$ and returns true if $v$ should be put in the next frontier.
Different algorithms implement different \texttt{edgeCompute()} functions using
different auxiliary data structures to store algorithm-specific per-vertex values. Below, we give an example 
for computing unweighted shortest path (``shortest paths'' for short) lengths from a single source.
Other recursive path finding algorithms, such as finding variable-length paths or path lengths, are expressed in a similar manner. 


\begin{example}
\label{ex:bellman-ford}
{\em 
Listing~\ref{alg:unweighted-sp} shows a pseudocode \texttt{edgeCompute()} for computing the shortest path lengths from a single source $s$ to the rest of the nodes in a graph.
The algorithm keeps the lengths of the shortest paths from $s$ to each vertex in a \texttt{len} array.
This array is initialized to -$\infty$ except $s$, which is set to 0. 
Then at each iteration, we update any node $v$ that is visited for the first time from a currently active node $u$ as follows. We first
set \texttt{len[v]} to \texttt{len[u]} + 1 and then
we put $v$ into the next frontier, by returning true. 
}
\end{example}

\subsection{Overall Query Plan Structure}
\begin{figure}[t!]
\includegraphics[keepaspectratio,width=7.0cm]{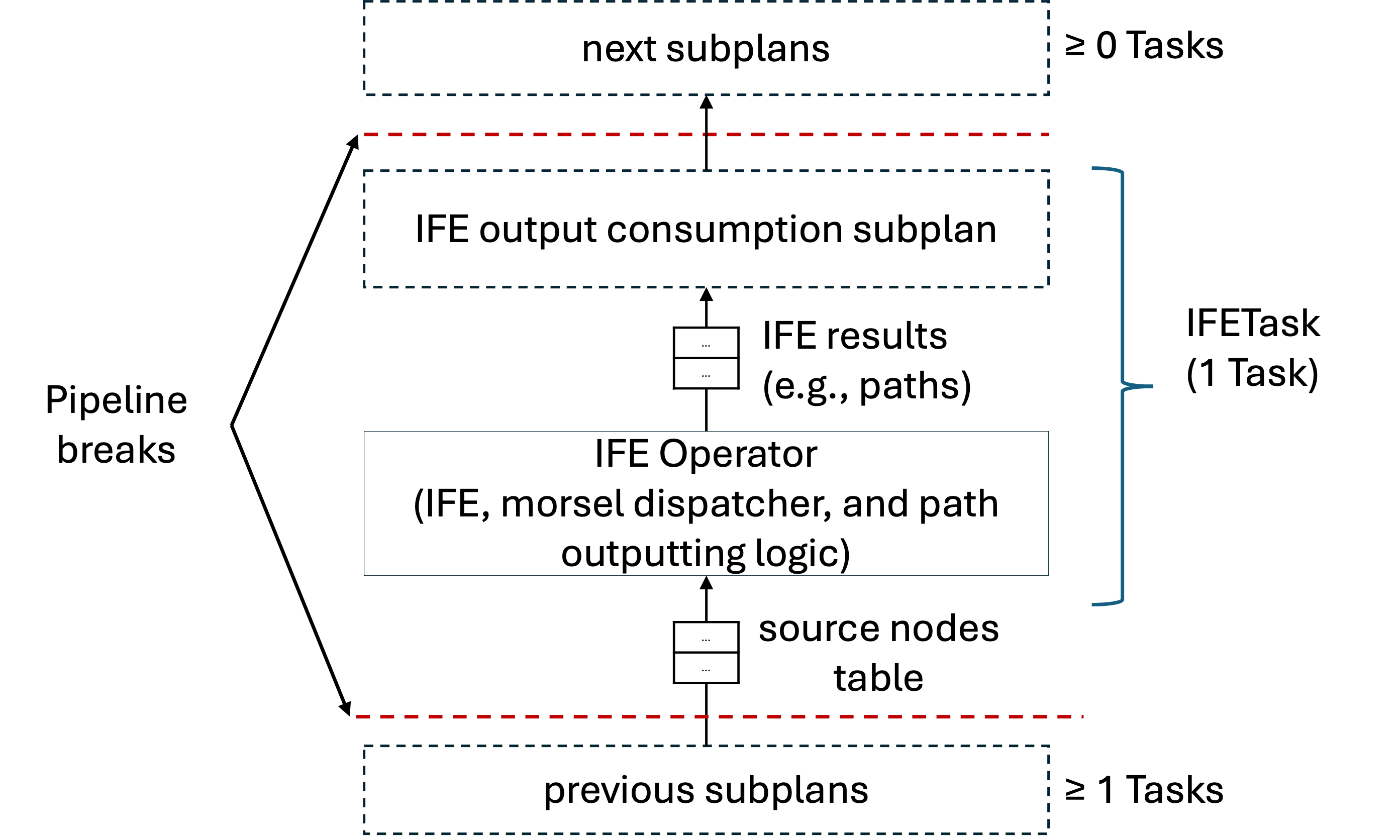}
\centering
\caption{Recursive Clause Query Plan}
\label{fig:plan-structure}
\end{figure}

Throughout this paper, we assume that a GDBMS compiles
each recursive clause into an operator that performs IFE subroutine executions.
We assume that the subplan/task in which the IFE operator runs looks as in
Figure~\ref{fig:plan-structure}.
Let us refer to the IFE operator's task as the IFETask. 
We assume that a set of prior subplans 
execute prior to the IFETask (``previous subplans'' in the figure). The last of these subplans
computes the source nodes from which a recursive path
computation should be performed and passes these as
a {\em source nodes table} to the IFETask. 
The IFETask starts with the IFE operator,
which implements the logic of morsel dispatching,
e.g., scanning source nodes from the {\em source nodes table}, performing the frontier extensions, as well as pipelining
the output paths or path lengths of the IFE subroutines to
the rest of the operators in the IFETask (``IFE output consumption subplan in the figure''). If IFETask
is not the last task in the query plan, other suplans 
may execute after the IFETask (``next subplans'' in the figure) .

\begin{lstlisting}[float=th, caption={\texttt{edgeCompute()} for  shortest path lengths.},abovecaptionskip=0pt, belowcaptionskip=-10pt,label={alg:unweighted-sp},language=C++,style=mystyleCpp,escapechar=^]
class ShortestPathLengths {
int len[numNodes]; // initialized to UINT64_MAX
void init(src) { 
  len[src] = 0
  curFrontier.add(src)
}
bool edgeCompute (e=(u, v)) {
  if (len[v] == UINT64_MAX):
    len[v] = len[u] + 1
    return true
  else: 
    return false
}
\end{lstlisting}

\section{Design Space of Scheduling Policies}
\label{subsec:scheduling-policies}
We next describe a design space of natural parallelization approaches 
for executing recursive path finding algorithms based on IFE, which covers and extends the two popular approaches in prior literature. 
Recall Listing~\ref{lst:serial-ife}, which showed the pseudocode of the serial implementation of an IFE-based algorithm. The outer-most for loop loops through each source node $s$ in the source nodes table on line~\ref{line:first-for-loop}.
Then for each $s$, the algorithm runs a separate IFE subroutine from $s$ (inside a while loop). 
The design space we describe is based on which parts of this serial IFE-based algorithm is split into morsels for dispatching to worker threads.
In Listing~\ref{lst:serial-ife}, there are 2 for loops that contain scans that can be parallelized.
These are on lines~\ref{line:first-for-loop} and~\ref{line:second-for-loop}. 
The policies we describe are based on which or both of these loops they parallelize.

\begin{figure}[t!]
\includegraphics[keepaspectratio,height=3.6cm,width=7.75cm]{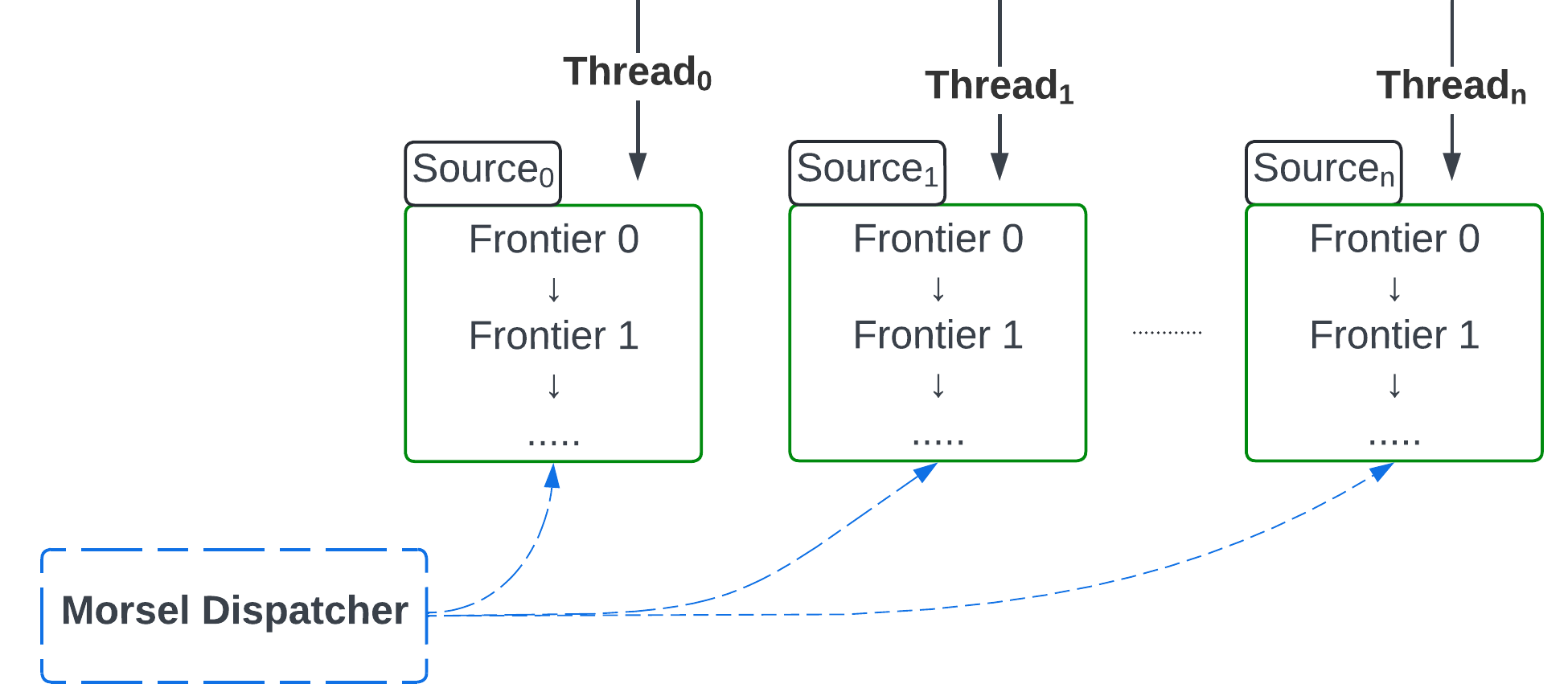}
\centering
\caption{\OneTOneS\ scheduling policy.}
\label{fig:1T1S}
\end{figure}

\subsection{1T1S Policy}
The first natural approach dispatches each IFE subroutine execution from a source as a unit of work  (Line \ref{line:ife-line-start}-\ref{line:ife-line-end} in
Listing~\ref{lst:serial-ife}). 
Specifically, the first for loop on line~\ref{line:first-for-loop} scans each source node $s$ 
and starts a new IFE subroutine from $s$. 
We call these IFE subroutines as {\em source morsels}.
In vanilla morsel-driven parallelism, e.g., adopted in Neo4j~\cite{neo4j}, DuckPGQ~\cite{DuckPGQ} and 
existing approach in \Kuzu, the scan in this first for loop
is done by scans from a source table, and the recursive computation is parallelized
at this scan. 
This is the approach we call \OneTOneS\, for {\bf 1}-{\bf T}hread-to-{\bf 1}-{\bf S}ource node policy. 
Figure~\ref{fig:1T1S} shows the high-level execution of this policy 
when there are $n$ sources to run IFE subroutines from.

The advantage of \OneTOneS\ is that when there are many sources from which an IFE subroutine
should be executed, \OneTOneS\ can easily keep threads busy. Further, 
under the \OneTOneS\ policy, the IFE subroutine implementations can use 
fast data structures that do not contain any synchronization primitives, such as hardware or software locks.
This is because \OneTOneS\ guarantees that only one thread works on any IFE computation.
At the same time, this approach will not be able to utilize multiple threads efficiently 
on queries that contain fewer sources than there are threads, e.g., only one source. 
Worse, if the system
uses large morsel sizes, e.g., 100K as in the original morsel-driven parallelism
paper~\cite{leis:morsel-driven} or 131K as in the DuckPGQ system, this approach can fail to parallelize even when
there are many sources in the query.

\subsection{nT1S Policy}
\begin{figure}[t]
\includegraphics[keepaspectratio,height=3.6cm,width=7.75cm]{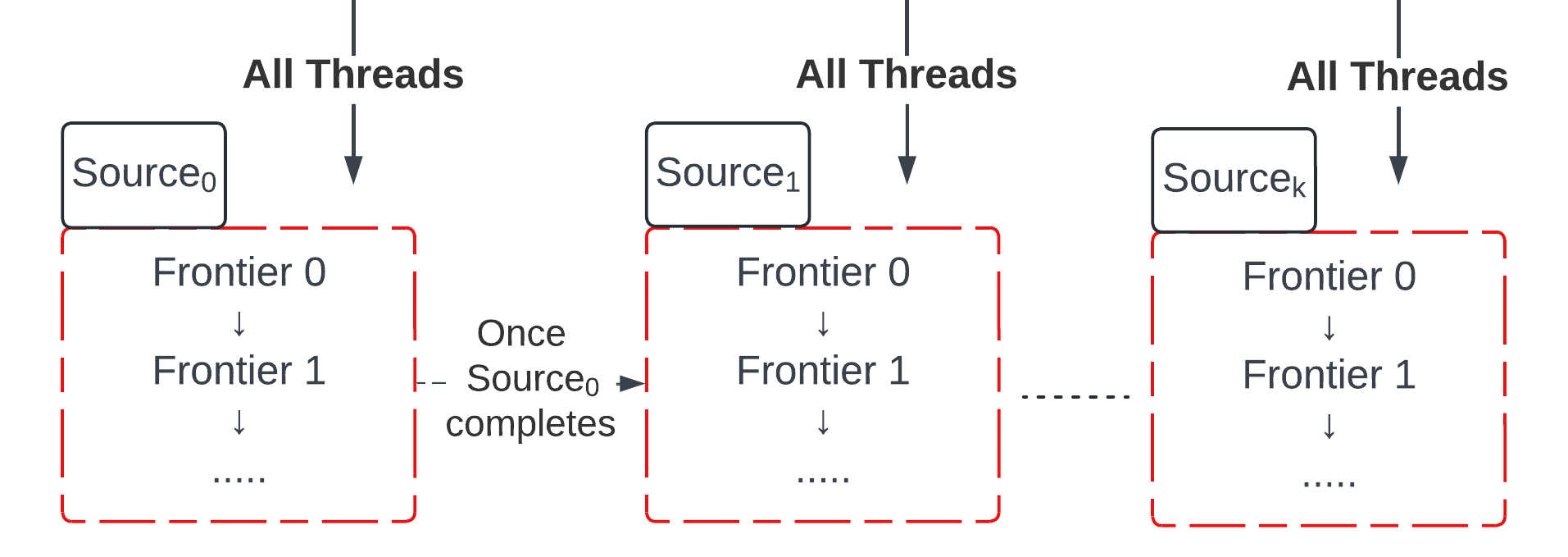}
\centering
\caption{\nTOneS\ scheduling policy.}
\label{fig:nT1S}
\end{figure}

The second natural approach dispatches work only from the inner for loop
on line~\ref{line:second-for-loop} in Listing~\ref{lst:serial-ife}.
This policy takes each source $s$ one by one and splits the scan of the current frontier of 
each iteration of the IFE subroutine from $s$ into morsels.
We refer to these as {\em frontier morsels}.
Therefore, each frontier morsel is a set of active nodes
and a single thread is responsible for executing the \texttt{edgeCompute()} function on each neighbor 
of each node in the frontier morsel.
This is the approach implemented in parallel graph analytics systems,
such as Ligra~\cite{Ligra} and Pregel~\cite{Pregel}. 
We call this approach \nTOneS, for
{\bf n}-{\bf T}hreads-to-{\bf 1}-{\bf S}ource node policy. Figure~\ref{fig:nT1S} shows the high-level execution of this policy.

The advantage of \nTOneS\ is that unlike \OneTOneS, \nTOneS\ can share work
when there are fewer sources in the query than there are threads. 
However, this approach is limited by how much
a single IFE subroutine from one source can parallelize.
Due to Amdahl's Law for parallelism \cite{10.1145/1465482.1465560},
the scalability of any program will be limited to
its parallelizable parts. 
In recursive path finding algorithms, 
the frontiers tend to start sparse, grow large, and shrink again and become sparse. The sparse
frontiers limit how much parallelism can be achieved cumulatively.

Table~\ref{tab:frontier-runtime} is an example experiment demonstrating this behavior
in our implementation of the \nTOneS\ policy in \Kuzu. The experiment runs a
shortest path query from a single source to all destinations
on the LDBC100 graph, returning the lengths of paths. LDBC100 graph contains 448K nodes and 19.9M edges. The table presents
how many nodes exist in each frontier and how much speedup there is on each frontier as we scale
the number of worker threads on the system from 1 to 32.
As shown, when a frontier is 
dense, as in level 4, we can obtain good scalability of 11.9x, yet on other frontiers, which cumulatively
add up to 36\% of the computation, there is at most 4.1x speedup and often much less. This leads to an overall speedup of 4.8x. This phenomenon
limits the scalability of using \nTOneS\ policy alone.

\begin{table}[t!]
\centering
\footnotesize
\begin{tabular}{|p{2cm}|*{6}{c|}p{0.7cm}|}
\hline
\multicolumn{1}{|l|}{} 
  & \multicolumn{6}{c|}{\textbf{Threads~($\rightarrow$)}} 
  & \multicolumn{1}{c|}{\textbf{Speedup}} \\ 
\hline
\textbf{IFE Level ($\downarrow$)} 
  & \textbf{1} & \textbf{2} & \textbf{4} & \textbf{8} & \textbf{16} & \textbf{32} 
  & \\ 
\hline
L0 (src node)             &    1    & 1  & 2  & 1  & 1  & 1  & \textbf{1.0x} \\
\hline
L1 (17 nodes)             &    3    & 2  & 2  & 1  & 1  & 2  & \textbf{1.5x} \\
\hline
L2 (2053 nodes)           &    6    & 4  & 4  & 2  & 2  & 3  & \textbf{2.0x} \\
\hline
L3 (64326 nodes)          &   65    & 38 & 20 & 14 & 11 &  7  & \textbf{9.3x} \\
\hline
{L4 (276175 nodes)}       &  190    &109 & 60 & 38 & 20 & 16  & \textbf{11.9x} \\
\hline
L5 (56731 nodes)          &   31    & 21 & 14 &  6 &  4 &  6  & \textbf{5.2x} \\
\hline
L6 (6044 nodes)           &    5    &  4 &  2 &  2 &  2 &  3  & \textbf{1.7x} \\
\hline
L7 (1465 nodes)           &    3    &  2 &  2 &  2 &  1 &  2  & \textbf{1.5x} \\
\hline
L8 (458 nodes)            &    2    &  2 &  1 &  2 &  1 &  2  & \textbf{1.0x} \\
\hline
L9 (93 nodes)             &    1    &  1 &  1 &  1 &  1 &  1  & \textbf{1.0x} \\
\hline
L10 (27 nodes)            &    1    &  1 &  1 &  1 &  1 &  1  & \textbf{1.0x} \\
\hline
L11 (7 nodes)             &    1    &  1 &  1 &  1 &  1 &  1  & \textbf{1.0x} \\
\hline
\textbf{Total Runtime}    &  331    &198 &138 & 95 & 72 & 68  & \textbf{4.8x} \\
\hline
\end{tabular}
\caption{Scalability of each frontier level (in \textbf{ms}).}
\label{tab:frontier-runtime}
\end{table}

\subsection{nTkS Policy}
\label{subsec:nTkS}
\begin{figure}[t!]
\includegraphics[height=3.6cm,width=7.75cm]{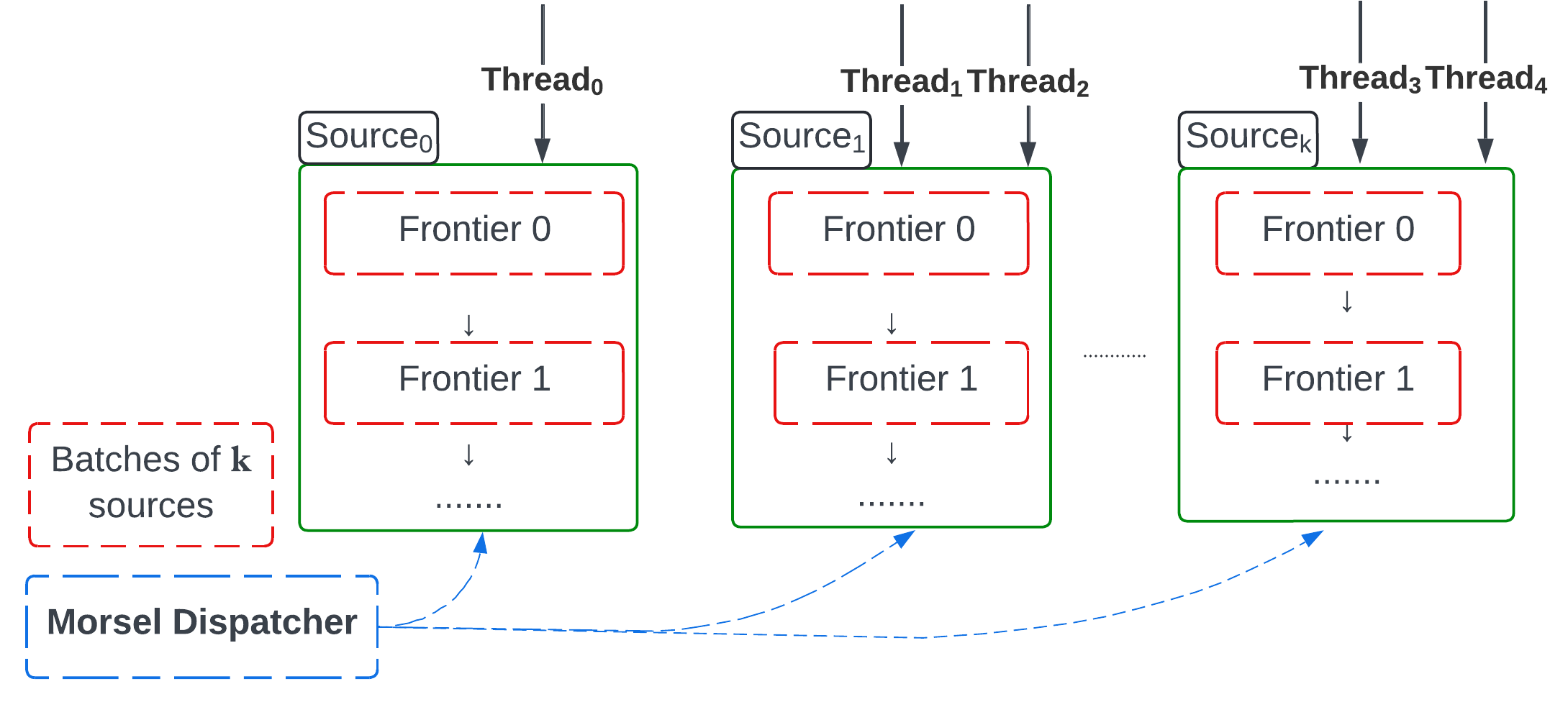}
\centering
\caption{\nTkS\ scheduling policy.}
\label{fig:nTkS}
\end{figure}

We next identify a third hybrid policy that combines parallelization of both loops.
We call this approach \nTkS\ policy, for {\bf n}-{\bf T}hreads-to-{\bf k}-{\bf S}ource nodes.
\nTkS\ maintains the execution of multiple $k$ concurrent IFE subroutines, i.e., source morsels. 
However, it is not the source morsels that are given as morsels to threads. Instead, each frontier of each source morsel is split into morsels but different threads can be dispatched frontier morsels from different source morsels.
Figure~\ref{fig:nTkS} shows the high-level execution of this policy.

The advantage of the \nTkS\ policy is that when there are few source nodes in the query, \nTkS\ aims to mimic the behavior of \nTOneS. When there are many source nodes, \nTkS\ can
outperform \nTOneS\ as when frontiers of a source morsel gets sparse, it can keep idling threads active on other source morsels with denser frontiers. 
When there are many source nodes, \nTkS\ behaves similar to \OneTOneS. However, \nTkS\ can outperform \OneTOneS\ 
because whenever there are fewer source morsels than number of threads, 
while \OneTOneS\ starts keeping threads idle, \nTkS\ starts behaving
more like \nTOneS\ where multiple threads work on the same source morsel.
This situation can arise in two ways: (i) either the original query does not have many source nodes; or (ii) the query contains many source nodes initially but as the computation progresses some source morsels finish, and there are fewer source morsels left to work on than the number of threads.

\subsection{Multi-Source Morsels}
\label{subsec:msntks}
We next review the multi-source BFS (MS-BFS) optimization from reference~\cite{msbfs}, which is also  
used by the DuckPGQ system~\cite{DuckPGQ}. MS-BFS
organizes up to 64 IFE subroutines, IFE$_1$, ..., IFE$_{64}$,  as a unit and runs them concurrently. 
For example, if a shortest path query contains 64 or more source nodes,
MS-BFS runs the shortest path computation from 64 sources at a time.
That is, first, the first frontiers of all 64 IFE subroutines
are extended concurrently, then the second frontiers, then the third frontiers, so on and so forth. For each vertex $u$ in the graph, we store 64 bits to represent $u$'s active state in each of the 64 concurrent IFEs, i.e., the $i$'th bit represents whether
$u$ is active in IFE$_i$. These bits are referred to as ``lanes'' in reference~\cite{msbfs}.
If $u$ is active in iteration $j$
for $t$ IFE subroutines, say IFE$_{\ell_1}$,...,IFE$_{\ell_t}$, then instead of scanning the neighbors of $u$ $t$ times, we can scan them once and run $t$ \texttt{edgeCompute()} on each of those neighbors. 
MS-BFS also reduces the amount of writes performed to set nodes active in frontiers. Specifically, one can 
set a neighbor $v$ of $u$ active in up to $t$ IFE subroutines with a small number of bitwise operations.

We can model MS-BFS also as a morsel dispatching policy that packs multiple source nodes   on line~\ref{line:first-for-loop} of Listing~\ref{line:ife-subroutine} into {\em multi-source morsels}.
Using this optimization,
we can have variants of the previous policies.
For example, if the query
contains 128 source nodes,  the \texttt{nTkMS} policy,
for {\bf n}-{\bf T}hreads-to-{\bf k}-{\bf M}ulti-{\bf S}ource nodes, can launch two multi-source morsels
and have multiple threads grab frontier morsels from each multi-source morsel.
We will empirically evaluate the behavior of \nTkMS\
in Section~\ref{sec:evaluation}.

\section{Implementation Details}
\label{subsec:implementation-details}
We next describe the details of our implementation of these 
policies in \Kuzu. Our implementation can be found here~\cite{anurag-github}. Briefly, \Kuzu\ is a columnar, disk-based GDBMS. It has a vectorized~\cite{dan-abadi-column-db, kuzu:cidr}, i.e., batch-at-a-time, query processor that adopts morsel-driven
parallelism. The query processor is pull-based, where parent operators
pull data from their children tuples via \texttt{getNextTuples()} function calls.
\Kuzu\ stores the adjacency lists of nodes in disk-based compressed sparse-row (CSR) structures, and access to adjacency lists
happens through the system's buffer manager.

 \begin{lstlisting}[caption={Pseudocode of the IFE operator.},abovecaptionskip=2pt,belowcaptionskip=-10pt,label={lst:ife-op},language=C++,style=mystyleCpp,escapechar=^][t!]
struct SourceMorsel {
  Frontier currentFrontier, nextFrontier;
  int curIter; // current iteration of the IFE subroutine
  IFEPhase phase; // one of enum {FRONTIER_EXTENSION, OUTPUT}
  AuxState auxState; // other algorithm-specific data, e.g., Parents
}
class IFE : PhysicalOperator {
MorselDispatcher md;
EdgeCompute ec; // recursive clause-specific edgeCompute function
Graph graph; // provides iterator interface to scan nbrs
DestinationNodeMask targetDsts; // possible destination nodes
bool getNextTuples() {
 while (true) {
   SrcMorsel sm = md.grabSrcMorselIfNecessary(&sm); ^\label{line:sm-grab}^
   if (sm == null): return false; // exit ^\label{line:exit}^
   if (sm.phase == OUTPUT):
     OutputMorsel om = md.grabOutputMorsel(&sm); ^\label{line:out-morsel-begin}^
     if (om == null): continue;
     outputPaths(om, sm, dsts); // pipeline outputs to parent op ^\label{line:out-morsel-end}^
     return true;
   else if (sm.phase == EXTEND_FRONTIER):
     FrontierMorsel fm = md.grabFrontierMorsel(&sm); ^\label{line:frontier-morsel-begin}^
     if (fm == null): continue;
     extendFrontier(fm, sm, ec, graph); ^\label{line:frontier-morsel-end}^
     // if frontier finished, update phase or start new frontier
     sm.checkIfFrontierFinished(); ^\label{line:check-if-last-finishing-thread}^
  }}}
void extendFrontier(FrontierMorsel fm, SourceMorsel sm,
                    EdgeCompute ec, Graph graph) {
  for (auto u : fm):
    if (sm.cFrontier->isActive(u)):
      for (auto v : graph->scanFwd(u)):
        if (ec.compute(u, v, sm)):
          sm.nFrontier->setActive(v);
\end{lstlisting}

\begin{figure}[t!]
\includegraphics[height=4.35cm,width=8.25cm]{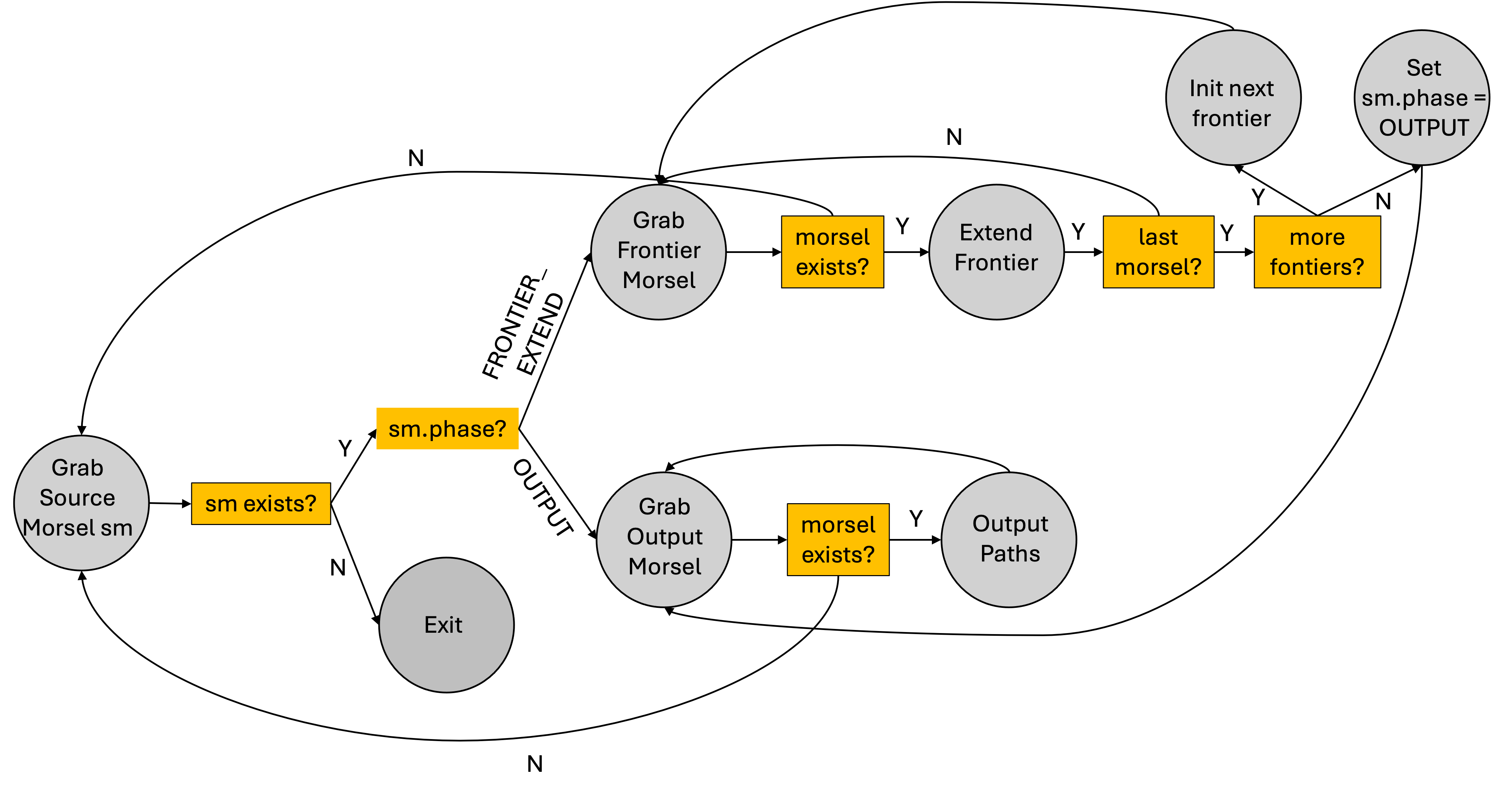}
\centering
\caption{Control flow of IFE operator and morsel dispatcher.}
\label{fig:ife-control-flow}
\end{figure}

\subsection{IFE Operator}
Listing~\ref{lst:ife-op} shows the pseudocode implementation
of the generic IFE operator, which we implemented as a 
standard physical operator in \Kuzu. The operator is constructed with several fields:
\begin{squishedlist}
    \item {\bf \texttt{MorselDispatcher}} implements
the different policies we outlined in Section~\ref{subsec:scheduling-policies} (discussed more momentarily).
   
    \item {\bf \texttt{Graph}} is an interface that provides functions, such as \texttt{scanFwd}, that provide an iterator interface to scan neighbors of vertices in the database. Internally, it translates these calls
    to calls that read database records through the system's buffer manager.
    
    \item {\bf \texttt{EdgeCompute}} implements the \texttt{edgeCompute()} function that implements a specific instance of an IFE-based recursive algorithm. We implemented different \texttt{edgeCompute()} functions for the different recursive clauses in Cypher and whether the query requires computing outputs actual paths or only the path lengths, which require operating on separate auxiliary data structures.\footnote{We found the \texttt{Graph} and \texttt{edgeCompute()} interfaces very helpful in implementing different IFE-based recursive algorithms and recommend it to system developers. These higher-level interfaces are better fits for implementing recursive path finding algorithms than the standard tuple or vector-based query processor interfaces of DBMSs to scan and operate on database records.} For reference, Listing~\ref{lst:single-sp-paths} shows an example \texttt{edgeCompute()} computing the shortest paths clause that returns the actual paths (instead of path lengths).

    \item {\bf \texttt{DestinationNodeMask}} \texttt{targetDsts} is a ``node mask'' that keeps track of the specific nodes in the graph, for which an output path or path length needs to be returned. The mask is an array of boolean values of size the total nodes in the graph.
\end{squishedlist}

 \begin{lstlisting}[caption={\texttt{edgeCompute} function that computes shortest paths.},abovecaptionskip=2pt,belowcaptionskip=-10pt,label={lst:single-sp-paths},language=C++,style=mystyleCpp,escapechar=^][t!]
class ShortestPaths : EdgeCompute {
void compute(nodeID u, nodeID v, SourceMorsel sm) {
   if (!sm.auxState.visited[v])
      sm.auxState.parents[v].addParentEdge(u, sm.curIter);
      sm.auxState.visited[v] = true; // atomic memory_relaxed op
}}
\end{lstlisting}

\vspace{1cm}
\noindent Figure~\ref{fig:ife-control-flow} summarizes the high-level control flow of the IFE operator assuming 
a morsel dispatcher that gives source morsels instead of multi-source morsels. 
The thread $W$ running the operator, inside a while loop, first grabs a source morsel \texttt{sm} from the
morsel dispatcher \texttt{md} by calling \texttt{grabSrcMorselIfNecessary()} (line~\ref{line:sm-grab}). It is inside this function that different morsel dispatching policies are implemented.

The \texttt{SourceMorsel} structure, shown on top of Listing~\ref{lst:ife-op}, represents
the state of an IFE subroutine execution from a single source $s$. It contains the data structures that are needed by the IFE subroutine, which include the current and next frontiers, current iteration of the IFE subroutine, and the auxiliary data structures to store the intermediate paths that are computed (\texttt{auxState} field). It further contains a \texttt{phase} field, which can take one of two values:
\begin{squishedlist}
\item FRONTIER_EXTENSION indicates that the IFE subroutine has not yet finished, i.e., the frontiers have not yet converged.
\item OUTPUT indicates that the IFE subroutine has finished, i.e., all paths from $s$ have been computed and now these paths need to be pipelined to the parent operator.
\end{squishedlist}

After $W$ calls \texttt{grabSrcMorselIfNecessary()}, if there are no more source morsels (line~\ref{line:exit}), the computation has finished and the operator exits. 
Otherwise, if \texttt{sm}'s phase is
FRONTIER\_EXTENSION, $W$ grabs a frontier morsel from
\texttt{sm} and runs \texttt{edgeCompute()} on each active node
in this frontier morsel. Obtaining frontier morsels is a  simple
operation that is independent of morsel dispatching policy, and returns back a range of integer node IDs. After $W$ finishes
its frontier morsel it calls \texttt{sm.checkIfFrontierFinished()} (line~\ref{line:check-if-last-finishing-thread}), which 
checks if all active nodes in the current frontier are processed and
if $W$ is the last thread to finish its frontier morsel. If so, then
$W$ either moves the computation to the next frontier or if there
are no active nodes in the next frontier, then sets \texttt{sm}'s phase to OUTPUT. 


If \texttt{sm}'s phase is
OUTPUT, then the thread grabs an {\em output morsel}. 
The output morsel is a range of node IDs that represent destination nodes. For each valid destination node $d$, 
if the computation found paths from $s$ to $d$, then $W$
outputs each path or path length from $s$ to $d$. 



\subsection{Data Structure Implementations}
\label{subsec:data-structures}

\noindent {\bf  Frontier:} We use a dense frontier implementation to store 
active nodes in the current and next frontier. Dense frontiers are arrays that store one boolean value per vertex in the graph.
When frontiers are swapped at the end of each iteration, we have
a variant of the {\em sparse frontier optimization} from Ligra~\cite{Ligra}. Specifically, if the number of nodes in the next frontier 
is less than 1/8th of all nodes, we construct an additional sparse version of the
frontier in a single threaded manner and write  the ID of every active node to an array. 

\begin{figure}[t!]
\includegraphics[height=2.5cm,width=8cm]{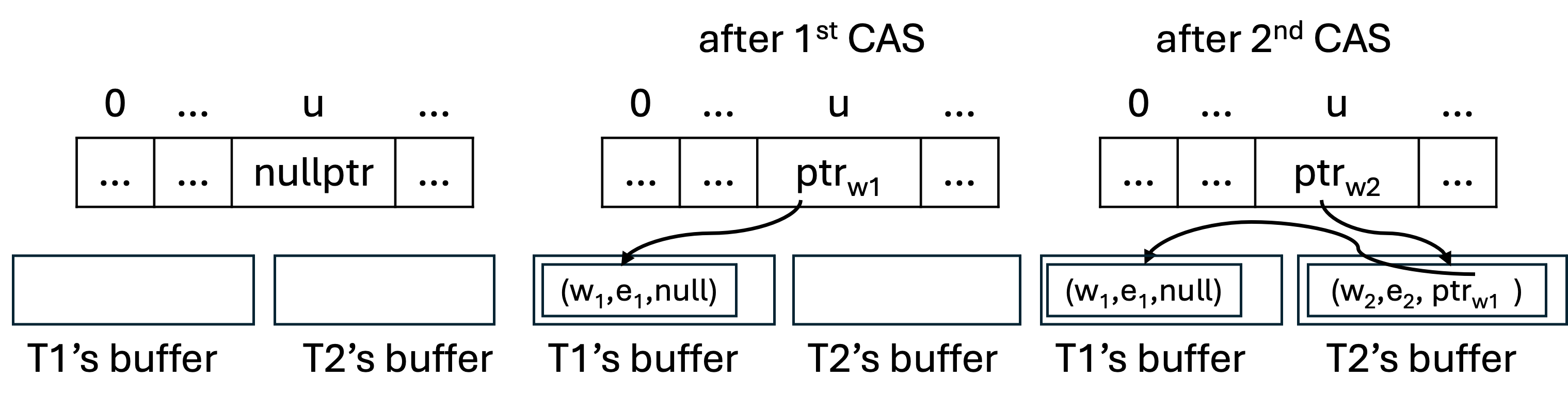}
\centering
\caption{Parents data structure to keep multiple paths.}
\label{fig:parents}
\end{figure}

\noindent {\bf Parents:} When computing paths, we keep paths compactly by keeping track of the ``parents'', i.e., last edges that were used
to visit each node. Figure~\ref{fig:parents} gives an overview of the data structure we use.
Our data structure consists of: (i) a dense pre-allocated array that keeps an 8-byte pointer for each vertex (initialized to null pointers) and that is shared across all threads;
and (ii) a set of memory buffers that are owned by and written to by separate threads. 
When thread $Ti$ needs to write a parent $v$ for node $u$, it writes to its memory buffer a tuple with the information
about $v$, specifically $v$'s ID and the edge ID of the ($v$, $u$) edge, and an 8-byte pointer to the next parent edge on the path. Therefore, for each edge of each path we compute, we store an additional 24 bytes. Then $Ti$ updates $u$'s pointer at the dense array to point to 
this tuple using compare-and-swap operation (CAS). This computation happens inside \texttt{addParentEdge()} function on the \texttt{Parents} data structure,
as shown in the example \texttt{edgeCompute()} function in Listing~\ref{lst:single-sp-paths}.
Figure~\ref{fig:parents} shows an example when two threads $T1$ and $T2$ attempt to add two parent edges to $u$, respectively from $w_1$ and $w_2$.

\noindent {\bf Shortest Paths-specific structures:} Shortest paths computations has the property that each node $u$
can be active only once in one frontier. To ensure nodes are not
put into the frontiers multiple times, we maintain a global {\em visited array}, which is a dense 
structure that keeps a boolean value per vertex. Further, if a shortest paths computation computes only path lengths instead of using the \texttt{Parent} structure, we use dense structures to store only the lengths of paths, which we store as 1 byte in our implementation, which is enough to store the path lengths in the datasets in our experiments.

\noindent {\bf Data structures for multi-source shortest paths implementations:} Here, we followed the overall implementation in reference~\cite{msbfs} and use 
64 bits (for 64 lanes) to represent the active state of each node in frontiers. 
We used two frontiers for current and next and a third \texttt{visited} data structure
that also stores 8 bytes (64 bits) per node that indicate for which of the up to 64 IFE subroutines, IFE$_1$, ..., IFE$_\ell$, a node is visited or in the frontier of. 
Therefore, these auxiliary data structures require 3*8=24 bytes per node per multi-source morsel.
Operations on frontiers and visited arrays can be done with efficient bitwise operations to update a single node's value concurrently for multiple IFE sub-routines.
We also require auxiliary data structures, such as path lengths and parents 
data structures for each source in a multi-source morsel. 
In total, for each multi-source morsel
if there are 64 sources in it, the upfront memory requirements
are as follows. If we are returning only path lengths,
we allocate 24 + 1*64 = 88 bytes per node in the graph.
We do not require further memory during the computation.
If we are returning paths, we allocate upfront 24 + 8*64= 536 bytes per node and 
allocate more data during the computation for the thread-level memory buffers
to store additional edges in the paths computed.

Finally, we note that suppose a node $u$ is active in iteration $i$ for at least one IFE subroutine. We can tell this by inspecting that its frontier 
value $X$ is not equal to 0. 
Further, to update the correct auxiliary data structures, we need to compute which bits of $X$ are exactly 1. 
For this, we use the builtin C++ function \cite{gcc:builtin} that gives the index of the first 1 
bit in $X$ and we use it iteratively to read the index of every 1 bit.

\subsection{Scheduling Policy Implementations}
Different scheduling policies are implemented as different
morsel dispatcher logics for assigning source morsels to threads.
This is represented by the \texttt{grabSrcMorselIfNecessary(SourceMorsel sm)} function of the
\texttt{MorselDispatcher} class in Listing~\ref{lst:ife-op}. Our implementations 
of \OneTOneS\ is straightforward. We discuss only
\nTkS\ since  \nTOneS\ is a special case of \nTkS. 
\texttt{nTkS} is configured with 
a $k\ge 1$ value and launches up to \texttt{k} source morsels as follows.
Whenever a thread $W$ asks for source morsel to work on, as long as there
are fewer than $k$ source morsels that are already launched, and 
there are more sources in the query, $W$ is dispatched a new source morsel \texttt{sm}. Then, $W$ keeps working on \texttt{sm} as long as it can
grab frontier morsels from it (or if \texttt{sm} is in OUTPUT phase, then as long as $W$ can output paths from \texttt{sm}). That is, our implementation of \texttt{nTkS} is ``sticky''. $W$
can be dispatched another source morsel
if it cannot find a frontier morsel from \texttt{sm}, which 
happens at the end of each iteration $i$ of the IFE subroutine.

We also implemented the \texttt{nTkMS} policy. In this policy, threads get a \texttt{MultiSourceMorsel} struct that represents up to 64 concurrent IFE subroutines and the necessary auxiliary data structures. This policy is also configured with a $k \ge 1$ value and follow the same logic for dispatching new multi source morsels as the  \texttt{nTkS} policy. 

\section{Evaluation}
\label{sec:evaluation}

We next evaluate our different morsel dispatching policies
under a variety of queries that contain different numbers
of sources. We further study what is a good value of $k$ for both
 \nTkS\ and \texttt{nTkMS} policies and the performance of the \texttt{nTkMS} policy.
 All of our datasets and queries can be found in our code repo~\cite{anurag-github}.
 

\subsection{Experimental Setup}
\label{sec:exp-setup}
\noindent {\bf Baselines and \Kuzu\ configurations}: 
We compare 4 different policies that we implemented in \Kuzu\ with three additional baselines:

\begin{squishedlist}
\item \Kuzu-\OneTOneS\ , \Kuzu-\nTOneS\ , \Kuzu-\nTkS\ ,
and \Kuzu-\nTkMS, represent the policies we 
implemented in \Kuzu. Throughout our evaluations,
we set $k$ to 32 in \Kuzu-\nTkS\ configuration, except in Section~\ref{subsection:choice-of-k}.
For experiments using \Kuzu-\nTkMS, we specify the $k$ values explicitly.

\item Neo4j: We use Neo4j's v5.16, which implements
a morsel dispatching policy that is akin to the \OneTOneS\ policy. The difference is instead of source morsels, Neo4j assigns
``(source, destination)-morsels'' to threads. Specifically, each thread is assigned a morsel of 1024 (source destination) pairs and independently computes
shortest paths between each pair
using an IFE-based computation, which extends
frontiers one step from the source node side and one step from the
destination node side. However, Neo4j does not dispatch work at the frontier level. As such, this approach behaves similarly
to the \OneTOneS\ policy. As we will explain in detail below, our queries 
limit the number of source nodes but not the destination nodes.
Since Neo4j assigns morsels based on (source, destination) pairs,
Neo4j becomes very inefficient on these queries. Therefore,
we test Neo4j on modified queries that limit the number of destination nodes to 1024.

\item Ligra: We use Ligra, which implements frontier-level parallelism, as an external baseline for the \nTOneS\ policy. 

\item \revision{\DuckPGQnTkS:  We tried using the original version of DuckPGQ as a baseline for comparison, however we encountered several problems. Similar to Neo4j, DuckPGQ also issues (source, destination) morsels. Even
when we limited the destination nodes in our queries, we 
found DuckPGQ's MS-BFS implementation to be very slow
and also not parallelize well. 
We reached out to the authors of DuckPGQ~\cite{DuckPGQ} and 
verified that this was expected behavior.
Instead, we modified the DuckPGQ code
and implemented our \nTkS\ policy considering only the source nodes and without the
MS-BFS optimization. We refer to this modified DuckPGQ as \DuckPGQnTkS. Our implementation can be found here~\cite{duckpgq-fork-github}. Unlike Kuzu, which has a disk-based CSR index that gets created once when the database is constructed, DuckPGQ creates an on-the-fly in-memory CSR for each query. We observed that the creation of these in-memory CSRs take a significant time,
so instead of end-to-end times, we only report the time taken by \DuckPGQnTkS\ 
to perform the IFE computation. 
}


\end{squishedlist}

\noindent {\bf Hardware:} 
We use a single machine with Intel Xeon E5-2670 CPU @2.60GHz 
processors and 512GB of memory. The machine has 16 physical cores (32 virtual cores) across 2
NUMA nodes with 32 KB of L1 data cache, 32KB L1 instruction
cache, 256 KB L2 cache and 20 MB L3 cache sizes. We use the systems in their default settings, e.g., \Kuzu\ sets its
buffer manager size by default to 80\% of the host memory (410 GB).


\noindent {\bf Datasets:} Table~\ref{tab:datasets} lists the datasets used in our experiments, 
which include the following: (i) LDBC100 is a synthetic graph generated with LDBC social network benchmark~\cite{ldbc} at scale 100; (ii) LiverJournal is the LiveJournal social network graph
from the SNAP graph dataset repository~\cite{livejournal};
(iii) Spotify is a graph released by the music streaming platform Spotify representing songs and pairs of songs listened together in listening sessions~\cite{spotify-dataset}; and (iv) Graph500-28 is one of the a synthetic graphs released by the LDBC benchmark at scale 28.
We choose these datasets as they vary in their sizes from 20 million to 4.2 billion edges,
but each one is large enough that even recursive queries from a few sources
have room for benefiting from parallelization.

\begin{table}[ht]
	\centering
	\bgroup
	\setlength{\tabcolsep}{4pt}
	\def\arraystretch{1.2}
	\begin{tabular}{|c|c|c|c|}
		\hline
		\textbf{Name}   & \textbf{$|V|$} & \textbf{$|E|$}  & \textbf{Avg Degree}\\
		\hline
		\textbf{LDBC100}      &  448,626  & 19,941,198 & 44  \\
		\hline
		\textbf{LiveJournal}   &  4,847,571    &  68,993,773 & 14 \\
		\hline
		\textbf{Spotify}       &  3,604,454    &   1,927,482,013 & 535 \\
		\hline
            \textbf{Graph500-28}   &  121,242,388  &  4,236,163,958 & 35 \\
            \hline
	\end{tabular}
	\egroup
	\captionsetup{justification=centering}
	\caption{Datasets.}
	\label{tab:datasets}
\end{table}

\noindent {\bf Query workloads:} We used query workloads that find shortest paths starting from a set of source nodes to the rest of the nodes 
in the graph, and returning the lengths of the paths or actual paths.
In Cypher, these queries have the below structure:
\begin{lstlisting}[language=SQL,style=mystyleSQL,label=microbenchmark-query]
MATCH p = (a:Node)-[r:Rel* SHORTEST]->(b:Node)
WHERE a.id IN [s1, ... sn]
RETURN len(p)
\end{lstlisting}
$si, ..., sn$ values above identify the IDs of sources. \revision{We picked each source node in our workloads randomly and ensured that we can perform at least 3 levels of IFE computation.} We used 3 values for $n$ that limit the number of source nodes as follows:
\begin{squishedlist}

\item {\bf 1-source workloads} in which the queries contain 1 source.

\item {\bf 8-source workloads} in which the queries contain 8 sources.

\item {\bf 64-source workloads} in which the queries contain 64 sources. 
    
\end{squishedlist}

\noindent In Section~\ref{subsection:ms-bfs-morsel-optimization}, when we evaluate \Kuzu-\nTkMS\ policy, we use workloads with 128 and 256 sources as well.

Ligra does not have a query language
and its existing shortest path algorithm finds shortest path lengths from a single source. We modified the code slightly to run a separate IFE subroutine from multiple source one at a time.
Ligra's existing shortest paths algorithm only computes path lengths and not the actual paths, so we did not use queries that return paths for Ligra.

As discussed above, for Neo4j, we also modified the queries in 1/8/64-source workloads by putting a separate predicate to limit the number of destinations to 1024.  
Limiting the queries to contain
1024 destinations ensures that these queries have the same number of
(source, destination) morsels in Neo4j as there are source morsels in
\Kuzu\ configurations. With this workload change,
we can no longer compare the raw performance
of Neo4j with our other baselines. 
However our primary goal  is to compare the parallelism behaviors of these systems under representative workloads instead of comparing their raw performances.


Our experiments 
measure the systems we study with 1, 2, 4, 8, 16, and 32 threads. Tables ~\ref{tab:shortest-path-length-workloads} and ~\ref{tab:shortest-path-workloads} report numbers only for the experiments that contain 1, 8, and 32 threads, since these are the representative parallelism levels to explain the behaviors we want to highlight.
\iflong
Figure \ref{fig:1-source-exp}, \ref{fig:8-source-exp}, \ref{fig:64-source-exp}
report performance numbers for each thread level.
\else
\revision{The longer version of our paper contains additional figures that show the performances of these systems with 
2, 4, and 16 threads.}
\fi
\revision{
All reported numbers are based on execution on a warmed up database. We first run each query once to warm up each system's buffer manager cache, discard the first execution number, then report the average of an additional 3 more executions. The average runtime deviation was  
8\% across our experiments. The median deviations were 12.5\%, 11\%, 5\%, and 3\%
on any experiment, respectively, on
LDBC, LJ, Spotify and Graph500-28.}

\begin{table*}[t] 
    \centering
    \footnotesize 
    \setlength{\tabcolsep}{2pt} 
    \begin{minipage}[t]{0.32\linewidth} 
    \centering
        \begin{tabular}{|c|c|c|c|c|c|c|}
      \hline
      & \begin{tabular}[c]{@{}c@{}}Threads\\(→)\end{tabular}
        & 1      & 8             & \begin{tabular}[c]{@{}c@{}}CPU\\(\%)\end{tabular}
        & 32            & \begin{tabular}[c]{@{}c@{}}CPU\\(\%)\end{tabular} \\ \hline
     
      & nTkS     & 331     & 95   (3.5×) & 18 & 73   (4.5×) & 67 \\ \cline{2-7}
      & nT1S     & 310     & 91   (3.4×) & 18 & 62   (5.0×) & 70 \\ \cline{2-7}
      \multicolumn{1}{|c|}{LDBC}
      & 1T1S     & 312     & 304  (1.0×) & 3  & 310  (1.0×) & 3 \\ \cline{2-7}
      & Ligra    & 130     & 17   (7.6×) & 21 & 14   (9.3×) & 82 \\ \cline{2-7}
      & Neo4j    & 121     & 122  (1.0×) & 3  & 121  (1.0×) & 3 \\ \cline{2-7}
      & D-nTkS   & 69      & 29   (2.4×) & 12 & 27   (2.6×) & 12 \\ 
      \noalign{\global\arrayrulewidth=0.85pt}\cline{1-7}\noalign{\global\arrayrulewidth=.4pt}
    
      & nTkS     & 1975    & 440  (4.5×) & 22
                & 296  (6.7×) & 85 \\ \cline{2-7}
      & nT1S     & 1890    & 354  (5.3×) & 23
                & 274  (6.9×) & 85 \\ \cline{2-7}
      \multicolumn{1}{|c|}{LJ}
        & 1T1S     & 1814    & 1789 (1.0×) & 3
                  & 1811 (1.0×) & 3 \\ \cline{2-7}
      & Ligra    & 1411    & 190  (7.4×) & 24
                & 140  (10.1×)& 92 \\ \cline{2-7}
      & Neo4j    & 324     & 343  (1.0×) & 3
                & 323  (1.0×) & 3 \\ \cline{2-7}
      & D-nTkS   & 518     & 123  (4.2×) & 23
                & 71   (7.3×) & 89 \\ 
      \noalign{\global\arrayrulewidth=0.85pt}\cline{1-7}\noalign{\global\arrayrulewidth=.4pt}
    
      & nTkS     & 15012   & 2007 (7.5×) & 24
                & 1184 (12.7×)& 95 \\ \cline{2-7}
      & nT1S     & 14389   & 2106 (6.8×) & 24
                & 1121 (12.8×)& 95 \\ \cline{2-7}
      \multicolumn{1}{|c|}{Sp}
        & 1T1S     & 14372   & 14289(1.0×) & 3
                  & 14197 (1.0×)& 3 \\ \cline{2-7}
      & Ligra    & 20428   & 3394 (6.1×) & 24
                & 1243 (16.4×)& 99 \\ \cline{2-7}
      & Neo4j    & 3481    & 3562 (1.0×) & 2
                & 3561 (1.0×) & 2 \\ \cline{2-7}
      & D-nTkS   & 7976    & 1203 (6.6×) & 24
                & 630  (12.7×)& 95 \\ 
      \noalign{\global\arrayrulewidth=0.85pt}\cline{1-7}\noalign{\global\arrayrulewidth=.4pt}
    
      & nTkS     & 131963  & 18662 (7.1×)& 22
                & 14854 (8.9×) & 91 \\ \cline{2-7}
      & nT1S     & 128090  & 17980 (7.1×)& 23
                & 14032 (9.1×) & 92 \\ \cline{2-7}
      \multicolumn{1}{|c|}{G-28}
        & 1T1S     & 126929  & 125116(1.0×)& 3
                  & 120198(1.0×)& 3 \\ \cline{2-7}
      & Ligra    & 113276  & 17116 (6.6×)& 22
                & 12582 (9.0×) & 92 \\ \cline{2-7}
      & Neo4j    & 136929  & 135116(1.0×)& 3
                & 119092(1.0×)& 3 \\ \cline{2-7}
      & D-nTkS   & 61591   & 10557 (5.8×)& 22
                & 6288  (9.8×)& 91 \\ \hline
    \end{tabular}

    \vspace{0.1cm}
    \begin{center}
    \par \textbf{(a)} \footnotesize 1-Source Workload
    \end{center}
    \label{tab:single-source-pathlen-workloads}
  \end{minipage}%
  \hfill \hspace{0.5cm}
  \begin{minipage}[t]{0.32\linewidth} 
    \raggedleft
    \begin{tabular}{|c|c|c|c|c|}
  \hline
  1 & 8 & \begin{tabular}[c]{@{}c@{}}CPU\\(\%)\end{tabular}
           & 32 & \begin{tabular}[c]{@{}c@{}}CPU\\(\%)\end{tabular} \\ \hline

  2177    & 320   (6.8×) & 24 & 190  (11.5×) & 97 \\ \cline{1-5}
  2089    & 485   (4.3×) & 17 & 401  (5.2×)  & 66 \\ \cline{1-5}
  2077    & 358   (5.8×) & 22 & 346  (5.8×)  & 23 \\ \cline{1-5}
  635     & 118   (5.4×) & 21 & 86   (7.4×)  & 79 \\ \cline{1-5}
  808     & 157   (5.1×) & 22 & 159  (5.1×)  & 23 \\ \cline{1-5}
  602     & 174   (3.5×) & 13 & 173  (3.5×)  & 13 \\ 
  \noalign{\global\arrayrulewidth=0.85pt}\cline{1-5}\noalign{\global\arrayrulewidth=.4pt}

  15800   & 2311  (6.8×) & 24 & 1258 (12.6×) & 99 \\ \cline{1-5}
  15708   & 3141  (5.0×) & 18 & 2244 (7.0×)  & 70 \\ \cline{1-5}
  15749   & 2581  (6.1×) & 23 & 2316 (6.8×)  & 24 \\ \cline{1-5}
  8803    & 1543  (5.7×) & 20 & 871  (10.0×) & 78 \\ \cline{1-5}
  14993   & 2541  (5.9×) & 23 & 2458 (6.1×)  & 24 \\ \cline{1-5}
  4062    & 745   (5.5×) & 23 & 404  (10.1×) & 98 \\ 
  \noalign{\global\arrayrulewidth=0.85pt}\cline{1-5}\noalign{\global\arrayrulewidth=.4pt}

  121401  & 16845 (7.2×) & 24 & 8723 (14.0×) & 98 \\ \cline{1-5}
  121307  & 16617 (7.3×) & 24 & 8985 (13.5×) & 93 \\ \cline{1-5}
  121703  & 20627 (5.9×) & 23 & 19503 (6.2×) & 24 \\ \cline{1-5}
  131963  & 23992 (5.5×) & 22 & 10828(12.2×) & 95 \\ \cline{1-5}
  63138   & 12428 (5.1×) & 22 & 12380 (5.2×) & 23 \\ \cline{1-5}
  67712   & 10213 (6.6×) & 23 & 5514 (12.3×) & 98 \\ 
  \noalign{\global\arrayrulewidth=0.85pt}\cline{1-5}\noalign{\global\arrayrulewidth=.4pt}

  716313  & 100548 (7.1×) & 23 & 64943 (11.0×) & 96 \\ \cline{1-5}
  713313  & 127377 (5.6×) & 17 & 90292 (7.9×)  & 69 \\ \cline{1-5}
  710989  & 106117 (6.7×) & 22 & 103557(7.0×)  & 23 \\ \cline{1-5}
  647141  & 156599 (4.1×) & 17 & 69137 (9.4×)  & 74 \\ \cline{1-5}
  1819017 & 271495 (6.7×) & 23 & 267502(6.9×)  & 23 \\ \cline{1-5}
  348438  & 58744  (6.0×) & 23 & 33725 (10.3×) & 95 \\ \hline
\end{tabular}
    
    \vspace{0.1cm}
    \begin{center}
    \par \textbf{(b)} \footnotesize 8-Source Workload
    \end{center}
    \label{tab:few-source-pathlen-workloads}
\end{minipage}%
    \hfill
\begin{minipage}[t]{0.32\linewidth} 
    \centering
    \begin{tabular}{|c|c|c|c|c|}
  \hline
  1 & 8 & \begin{tabular}[c]{@{}c@{}}CPU\\(\%)\end{tabular}
           & 32 & \begin{tabular}[c]{@{}c@{}}CPU\\(\%)\end{tabular} \\ \hline

  15683    & 2140  (7.3×) & 24  & 1073 (14.6×) &   98 \\ \cline{1-5}
  15598    & 3391  (4.6×) & 18  & 3058  (5.1×) &   67 \\ \cline{1-5}
  15609    & 2230  (7.0×) & 21  & 1334 (11.7×) &   80 \\ \cline{1-5}
  3463     & 815   (4.2×) & 19  & 498   (7.0×) &   75 \\ \cline{1-5}
  7581     & 1025  (7.4×) & 21  & 681  (11.1×) &   81 \\ \cline{1-5}
  4329     & 1178  (3.6×) & 12  & 1165 (3.7×)  &   12 \\ 
  \noalign{\global\arrayrulewidth=0.85pt}\cline{1-5}\noalign{\global\arrayrulewidth=.4pt}

  124197   & 18601 (6.7×) & 24  & 8952 (13.9×)  & 98 \\ \cline{1-5}
  123957   & 25824 (4.8×) & 19  & 16310 (7.6×)  & 72 \\ \cline{1-5}
  124087   & 20681 (6.0×) & 21  & 12285 (10.1×) & 82 \\ \cline{1-5}
  66141    & 11617 (5.7×) & 20  & 7622  (8.7×)  & 78 \\ \cline{1-5}
  113919   & 18374 (6.2×) & 21  & 10689 (10.7×) & 80 \\ \cline{1-5}
  34222    & 5692  (6.0×) & 24  & 3074 (11.1×)  & 98 \\ 
  \noalign{\global\arrayrulewidth=0.85pt}\cline{1-5}\noalign{\global\arrayrulewidth=.4pt}

  955169   & 132885 (7.2×) & 24  & 82740 (11.5×) & 91 \\ \cline{1-5}
  955109   & 136444 (7.0×) & 24  & 72109 (13.3×) & 94 \\ \cline{1-5}
  948231   & 155448 (6.1×) & 21  & 93884 (10.1×) & 80 \\ \cline{1-5}
  1065394  & 184212 (5.8×) & 20  & 91849 (11.6×) & 92 \\ \cline{1-5}
  2381885  & 590591 (4.0×) & 19  & 246495 (9.7×) & 80 \\ \cline{1-5}
  818891   & 133735 (6.1×) & 23  & 58913 (13.9×) & 93 \\ 
  \noalign{\global\arrayrulewidth=0.85pt}\cline{1-5}\noalign{\global\arrayrulewidth=.4pt}

  4959436  & 795891 (6.2×)  & 24  & 416996 (11.9×) & 95 \\ \cline{1-5}
  4929436  & 1095430 (4.5×) & 18  & 631978 (7.8×)  & 70 \\ \cline{1-5}
  4939436  & 809743 (6.1×)  & 21  & 519940 (9.5×)  & 73 \\ \cline{1-5}
  4407690  & 1275313 (3.5×) & 17  & 523052 (8.4×)  & 74 \\ \cline{1-5}
  5898289  & 966932 (6.1×)  & 22 & 620872 (9.5×)  & 78 \\ \cline{1-5}
  2748338  & 407708 (6.7×)  & 24  & 231375 (11.8×) & 96 \\ \hline
\end{tabular}     
    \vspace{0.01cm}
    \begin{center}
    \par \textbf{(c)} \footnotesize 64-Source Workload
    \end{center}
    \label{tab:many-source-pathlen-workloads}
\end{minipage}
  \caption{\revision{\large Runtime (ms) and CPU utilizations (for 8 and 32 threads) for path length queries. D-nTkS is \DuckPGQnTkS.}}
\label{tab:shortest-path-length-workloads}
\end{table*}

\begin{table*}[t] 
    \centering
    \footnotesize 
    \setlength{\tabcolsep}{1.5pt} 
    \begin{minipage}[t]{0.32\linewidth}
        \centering
        \begin{tabular}{|p{0.1cm}|c|c|c|c|c|c|}
  \hline
  & \begin{tabular}[c]{@{}c@{}}Threads\\($\rightarrow$)\end{tabular}
    & 1      & 8             & \begin{tabular}[c]{@{}c@{}}CPU\\(\%)\end{tabular}
    & 32            & \begin{tabular}[c]{@{}c@{}}CPU\\(\%)\end{tabular} \\ \hline

  & nTkS     & 971     & 152  (6.4×)  & 18 & 89   (10.9×) & 87 \\ \cline{2-7}
  & nT1S     & 951     & 158  (6.1×)  & 18 & 87   (11.0×) & 89 \\ \cline{2-7}
  \multicolumn{1}{|c|}{LDBC}
    & 1T1S     & 959     & 961  (1.0×)  & 3  & 965  (1.0×) & 3  \\ \cline{2-7}
  & Neo4j    & 172     & 175  (1.0×)  & 3  & 178  (1.0×) & 3  \\ \cline{2-7}
  & D-nTkS   & 202     & 67   (3.0×)  & 13 & 67   (3.0×) & 13 \\ 
  \noalign{\global\arrayrulewidth=0.85pt}\cline{1-7}\noalign{\global\arrayrulewidth=.4pt}

  & nTkS     & 5129    & 954  (5.4×)  & 22 & 540  (9.5×)  & 87 \\ \cline{2-7}
  & nT1S     & 5117    & 979  (5.2×)  & 23 & 531  (9.6×)  & 88 \\ \cline{2-7}
  \multicolumn{1}{|c|}{LJ}
    & 1T1S     & 5095    & 5079 (1.0×)  & 3  & 5158 (1.0×) & 3  \\ \cline{2-7}
  & Neo4j    & 1110    & 1069 (1.0×)  & 3  & 1057 (1.0×) & 3  \\ \cline{2-7}
  & D-nTkS   & 1345    & 250  (5.4×)  & 23 & 141  (9.6×)  & 88 \\ 
  \noalign{\global\arrayrulewidth=0.85pt}\cline{1-7}\noalign{\global\arrayrulewidth=.4pt}

  & nTkS     & 24471   & 3437 (7.1×)  & 24 & 1847 (13.3×) & 93 \\ \cline{2-7}
  & nT1S     & 24171   & 3597 (6.7×)  & 24 & 1778 (13.6×) & 94 \\ \cline{2-7}
  \multicolumn{1}{|c|}{Sp}
    & 1T1S     & 24729   & 24829(1.0×)  & 3  & 24419(1.0×) & 3  \\ \cline{2-7}
  & Neo4j    & 5671    & 5579 (1.0×)  & 2  & 5694 (1.0×)  & 2  \\ \cline{2-7}
  & D-nTkS   & 13001   & 1826 (7.3×)  & 24 & 981  (13.2×) & 94 \\ 
  \noalign{\global\arrayrulewidth=0.85pt}\cline{1-7}\noalign{\global\arrayrulewidth=.4pt}

  & nTkS     & 212099  & 28984 (7.3×) & 22 & 19325 (11.0×)& 91 \\ \cline{2-7}
  & nT1S     & 215005  & 29019 (7.4×) & 23 & 19201 (11.2×)& 92 \\ \cline{2-7}
  \multicolumn{1}{|c|}{G-28}
    & 1T1S     & 210019  & 208298(1.0×) & 3  & 205902(1.0×) & 3  \\ \cline{2-7}
  & Neo4j    & 30192   & 31777 (1.0×) & 3  & 30395 (1.0×) & 3  \\ \cline{2-7}
  & D-nTkS   & 98992   & 13527 (7.4×) & 22 & 9019  (11.2×)& 91 \\ \hline
\end{tabular}
    \vspace{0.1cm}
    \begin{center}
    \par \textbf{(a)} \footnotesize 1-Source Workload
    \end{center}
    \label{tab:single-source-path-workloads}
  \end{minipage}
  \hfill \hspace{0.5cm}
  \begin{minipage}[t]{0.32\linewidth}
    \raggedleft
    \begin{tabular}{|c|c|c|c|c|}
  \hline
  1 & 8 & \begin{tabular}[c]{@{}c@{}}CPU\\(\%)\end{tabular}
           & 32 & \begin{tabular}[c]{@{}c@{}}CPU\\(\%)\end{tabular} \\ \hline

  5175    & 727   (7.1×)  & 24 & 391   (13.3×) & 99 \\ \cline{1-5}
  4961    & 844   (5.9×)  & 17 & 586   (8.5×)  & 68 \\ \cline{1-5}
  4991    & 745   (6.7×)  & 23 & 674   (7.4×)  & 24 \\ \cline{1-5}
  3485    & 601   (5.8×)  & 22 & 580   (6.1×)  & 23 \\ \cline{1-5}
  1431    & 485   (3.0×)  & 13 & 477   (3.0×)  & 13 \\ 
  \noalign{\global\arrayrulewidth=0.85pt}\cline{1-5}\noalign{\global\arrayrulewidth=.4pt}

  39981   & 5701  (7.1×)  & 24 & 3236  (12.4×) & 99 \\ \cline{1-5}
  40343   & 6165  (6.5×)  & 18 & 4101  (9.8×)  & 71 \\ \cline{1-5}
  39091   & 6023  (6.5×)  & 23 & 5505  (7.1×)  & 25 \\ \cline{1-5}
  50195   & 9842  (5.1×)  & 23 & 9126  (5.5×)  & 24 \\ \cline{1-5}
  10278   & 1465  (7.0×)  & 24 & 830   (12.5×) & 99 \\ 
  \noalign{\global\arrayrulewidth=0.85pt}\cline{1-5}\noalign{\global\arrayrulewidth=.4pt}

  194690  & 27462 (7.1×)  & 24 & 13235 (14.7×) & 97 \\ \cline{1-5}
  193905  & 25392 (7.6×)  & 24 & 13063 (15.0×) & 98 \\ \cline{1-5}
  192998  & 25733 (7.5×)  & 23 & 24775 (7.8×)  & 24 \\ \cline{1-5}
  34044   & 5158  (6.6×)  & 22 & 5007  (6.8×)  & 24 \\ \cline{1-5}
  108589  & 15317 (7.2×)  & 23 & 7081  (14.9×) & 98 \\ 
  \noalign{\global\arrayrulewidth=0.85pt}\cline{1-5}\noalign{\global\arrayrulewidth=.4pt}

  1236007 & 167390 (7.4×)  & 23 & 99185  (12.5×) & 95 \\ \cline{1-5}
  1229780 & 168026 (7.3×)  & 17 & 119275 (10.3×) & 65 \\ \cline{1-5}
  1231109 & 163930 (7.5×)  & 22 & 167270 (7.4×)  & 24 \\ \cline{1-5}
  247201  & 41898  (5.9×)  & 23 & 40524  (6.1×)  & 23 \\ \cline{1-5}
  601234  & 81423  (7.8×)  & 23 & 48246  (12.6×) & 95 \\ \hline
\end{tabular}
    \vspace{0.1cm} 
    \begin{center}
    \par \textbf{(b)} \footnotesize 8-Source Workload
    \end{center}
    \label{tab:few-source-path-workloads}
    \end{minipage}
    \hfill
  \begin{minipage}[t]{0.32\linewidth}
    \centering
    \begin{tabular}{|c|c|c|c|c|}
  \hline
  1 & 8 & \begin{tabular}[c]{@{}c@{}}CPU\\(\%)\end{tabular}
           & 32 & \begin{tabular}[c]{@{}c@{}}CPU\\(\%)\end{tabular} \\ \hline

  35827    & 4808  (7.4×)  & 24 & 2315  (15.5×) & 98 \\ \cline{1-5}
  35992    & 6209  (5.8×)  & 18 & 4530   (8.0×) & 68 \\ \cline{1-5}
  35920    & 5434  (6.6×)  & 21 & 3292  (11.0×) & 79 \\ \cline{1-5}
  29474    & 4151  (7.1×)  & 21 & 2631  (11.2×) & 80 \\ \cline{1-5}
  9889     & 3015  (3.3×)  & 12 & 2996  (3.3×)  & 13 \\ 
  \noalign{\global\arrayrulewidth=0.85pt}\cline{1-5}\noalign{\global\arrayrulewidth=.4pt}

  318855   & 44892 (7.1×)  & 24 & 22089 (14.5×) & 97 \\ \cline{1-5}
  321421   & 48396 (6.6×)  & 19 & 32281 (9.8×)  & 70 \\ \cline{1-5}
  321879   & 49519 (6.5×)  & 21 & 27025 (11.9×) & 80 \\ \cline{1-5}
  478855   & 61391 (7.8×)  & 21 & 36140 (13.2×) & 80 \\ \cline{1-5}
  87859    & 12369 (7.8×)  & 24 &  6086 (14.9×) & 97 \\ 
  \noalign{\global\arrayrulewidth=0.85pt}\cline{1-5}\noalign{\global\arrayrulewidth=.4pt}

  1578916  & 217542 (7.3×) & 24 & 126666 (12.5×) & 90 \\ \cline{1-5}
  1550691  & 203740 (7.6×) & 24 & 110323 (14.1×) & 95 \\ \cline{1-5}
  1561234  & 207060 (7.5×) & 21 & 148688 (10.5×) & 81 \\ \cline{1-5}
  2023172  & 293213 (6.9×) & 19 & 175928 (11.5×) & 82 \\ \cline{1-5}
  1188343  & 163729 (7.2×) & 23 &  98210 (12.2×) & 90 \\ 
  \noalign{\global\arrayrulewidth=0.85pt}\cline{1-5}\noalign{\global\arrayrulewidth=.4pt}

  8315386  & 1285103 (6.5×)  & 24 & 615954 (13.5×) & 94 \\ \cline{1-5}
  8419386  & 1326979 (6.3×)  & 18 & 889690 (9.5×)  & 69 \\ \cline{1-5}
  8419193  & 1360128 (6.2×)  & 21 & 764686 (11.0×) & 75 \\ \cline{1-5}
  10759211 & 1605852 (6.7×)  & 22 & 847182 (12.7×) & 77 \\ \cline{1-5}
  4608082  & 712157  (6.6×)  & 24 & 301182 (15.3×) & 94 \\ \hline
\end{tabular}    
    \vspace{0.1cm}
    \begin{center}
    \par \textbf{(c)} \footnotesize 64-Source Workload
    \end{center}
    \label{tab:many-source-path-workloads}
\end{minipage}
  \caption{\revision{\large Runtime (ms) and CPU utilizations (for 8 and 32 threads) for path queries. D-nTkS is \DuckPGQnTkS.}}
\label{tab:shortest-path-workloads}
\end{table*}

\iflong
\begin{figure*}[t!]
    \centering    
    \begin{minipage}{\dimexpr 0.25\linewidth - 0.5\tabcolsep\relax}
        \centering
        \includegraphics[width=\linewidth]{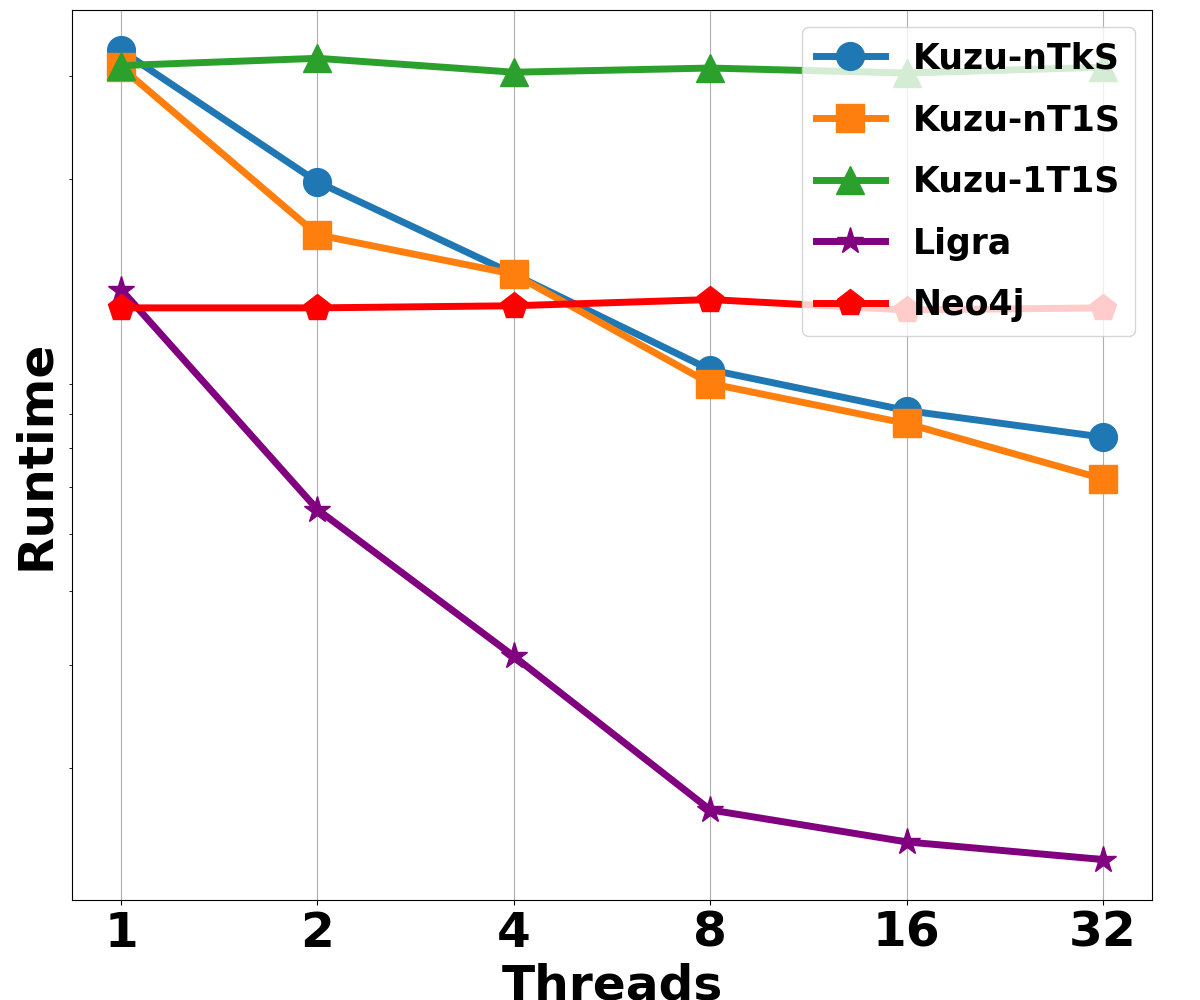}
        \textbf{a) LDBC100, Path Length}
    \end{minipage}\hfill
    \begin{minipage}{\dimexpr 0.25\linewidth - 0.5\tabcolsep\relax}
        \centering
        \includegraphics[width=\linewidth]{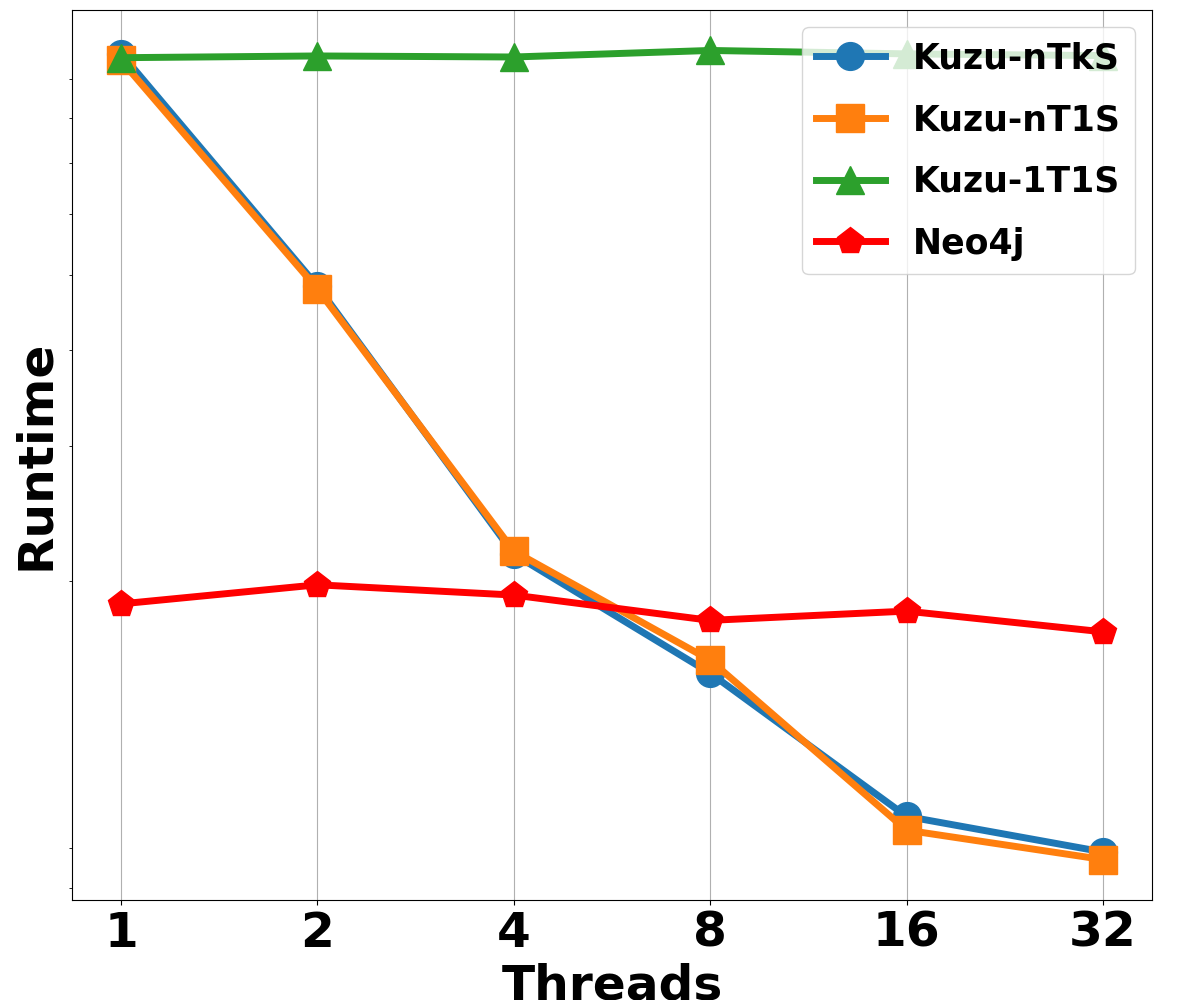}
        \textbf{b) LDBC100, Path}
    \end{minipage}\hfill
        \begin{minipage}{\dimexpr 0.25\linewidth - 0.5\tabcolsep\relax}
        \centering
        \includegraphics[width=\linewidth]{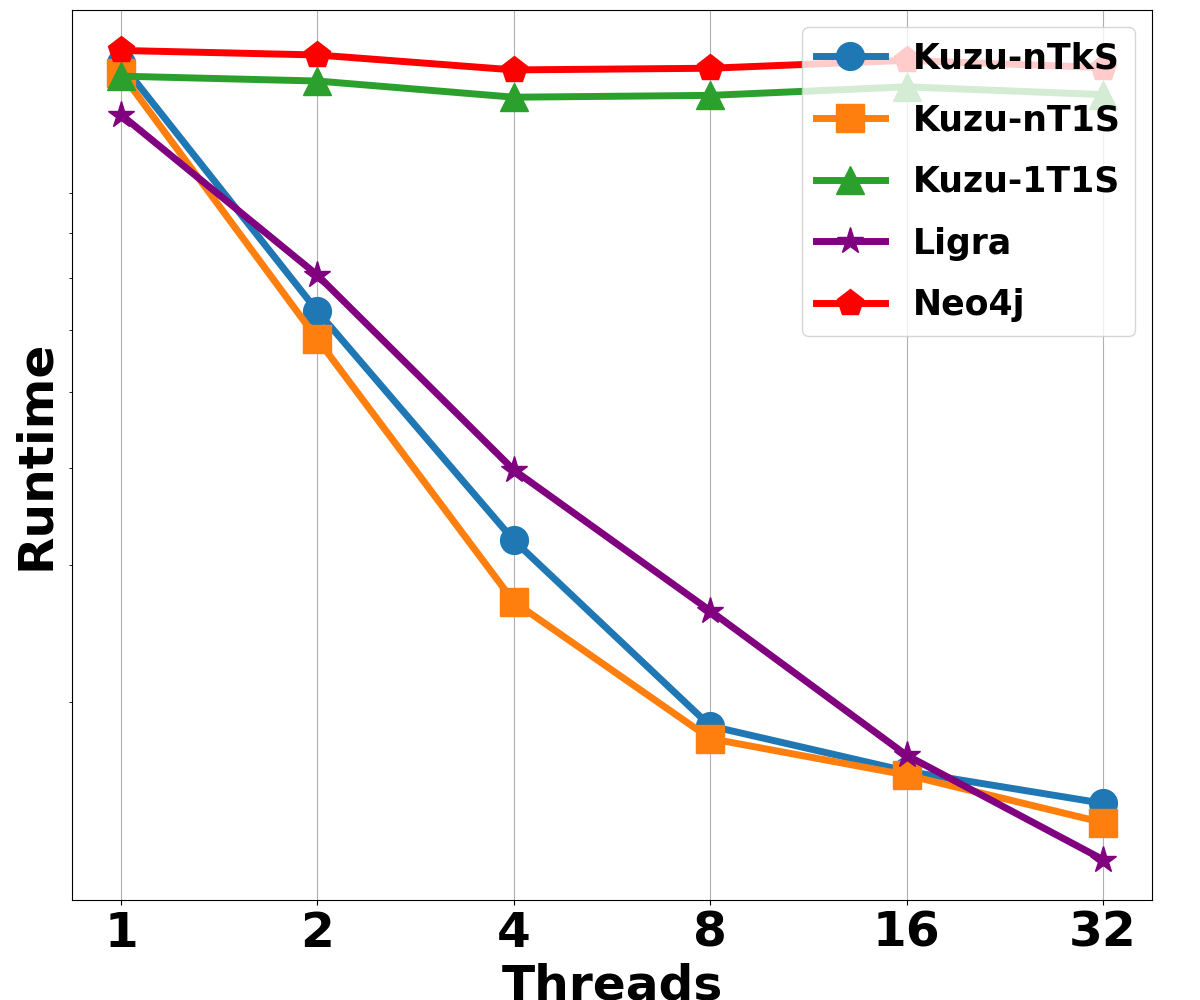}
        \textbf{c) Graph500-28, Path Length}
    \end{minipage}\hfill
    \begin{minipage}{\dimexpr 0.25\linewidth - 0.5\tabcolsep\relax}
        \centering
        \includegraphics[width=\linewidth]{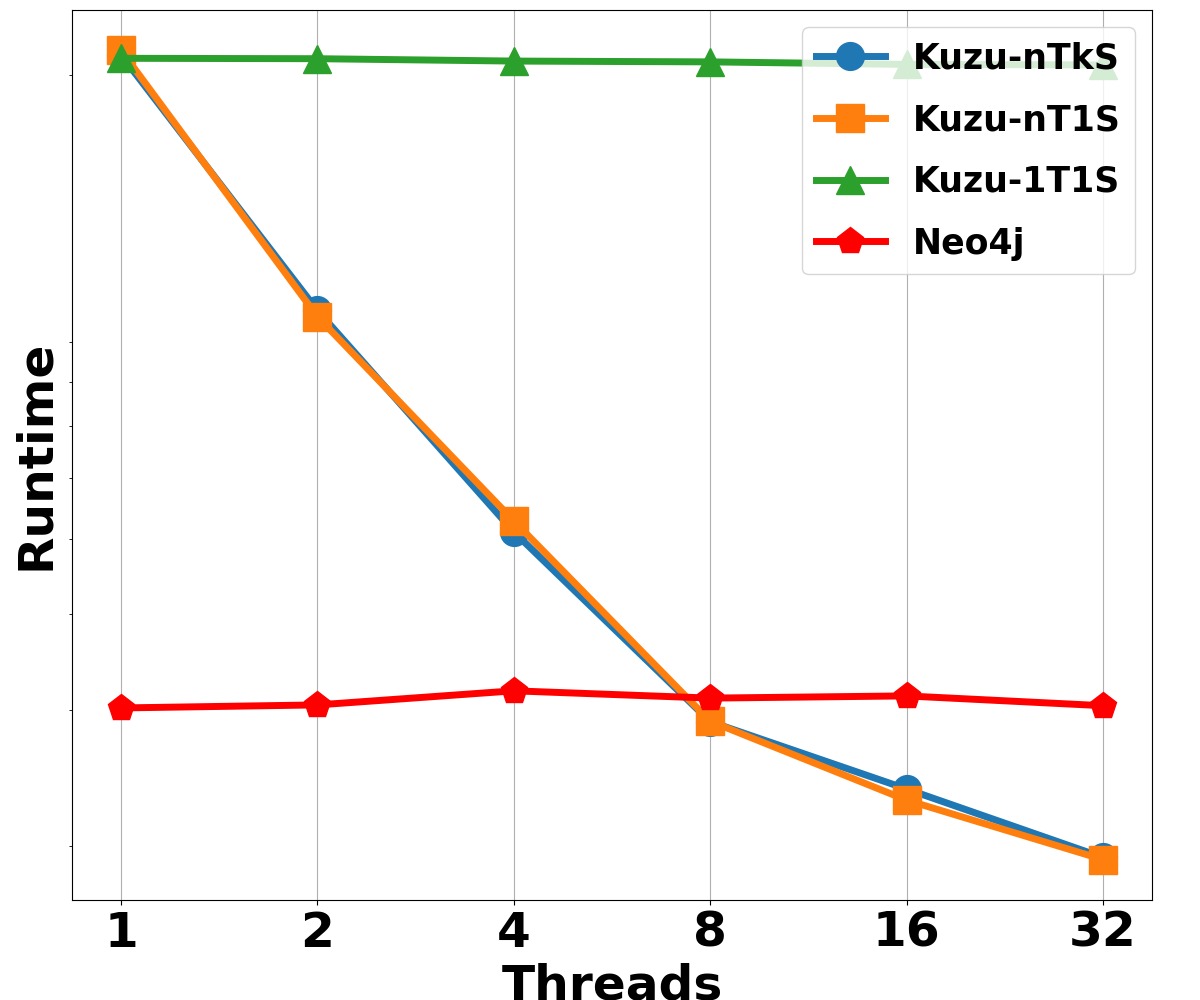}
         \textbf{d) Graph500-28, Path}
    \end{minipage}
    \captionsetup{justification=centering}
    \caption{Runtime vs. threads for 1-source query (runtimes in milliseconds).}
    \label{fig:1-source-exp}
\end{figure*}

\begin{figure*}[t!]
    \centering    
    \begin{minipage}{\dimexpr 0.25\linewidth - 0.5\tabcolsep\relax}
        \centering
        \includegraphics[width=\linewidth]{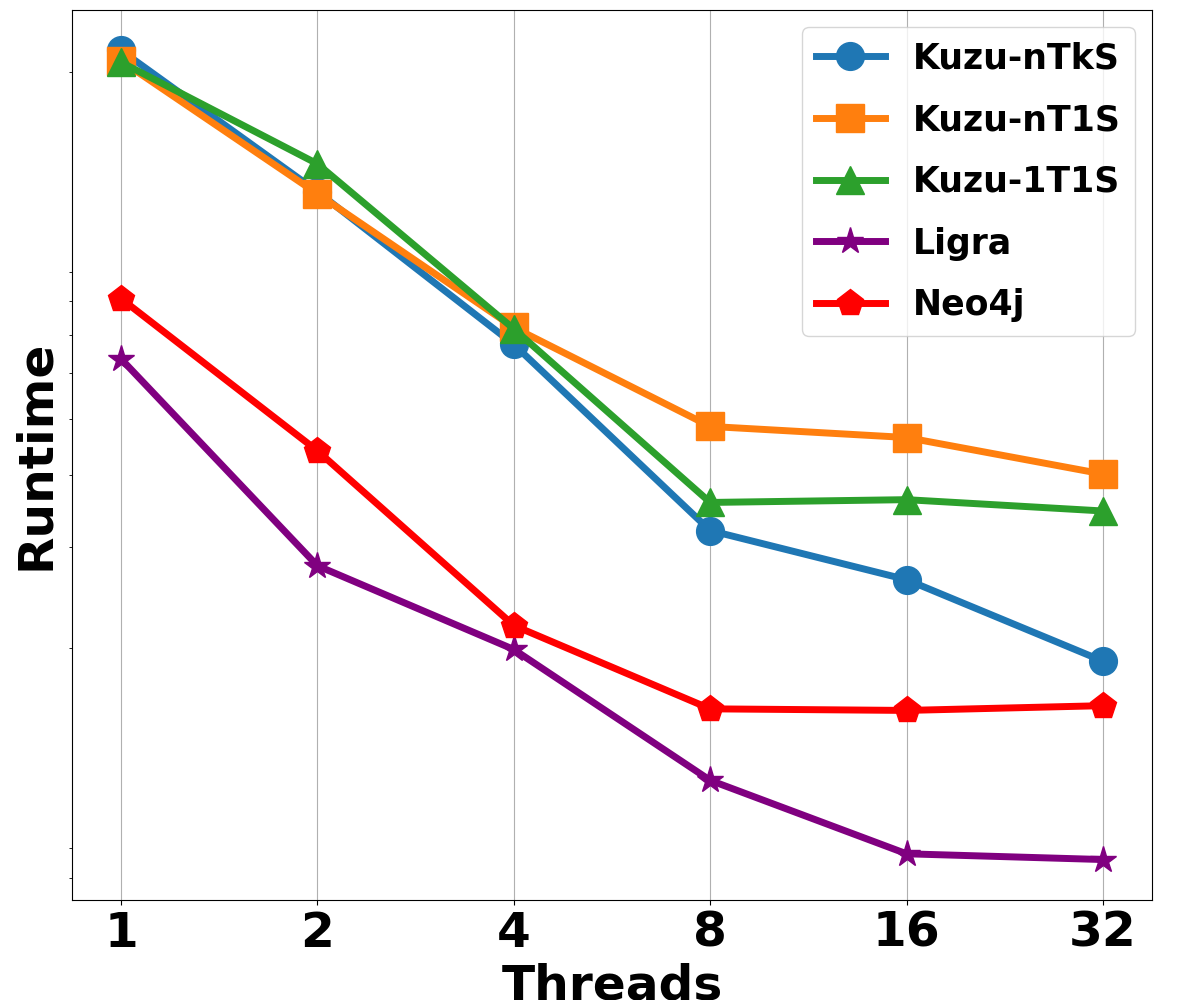}
         \textbf{a) LDBC100, Path Length}
    \end{minipage}\hfill
    \begin{minipage}{\dimexpr 0.25\linewidth - 0.5\tabcolsep\relax}
        \centering
        \includegraphics[width=\linewidth]{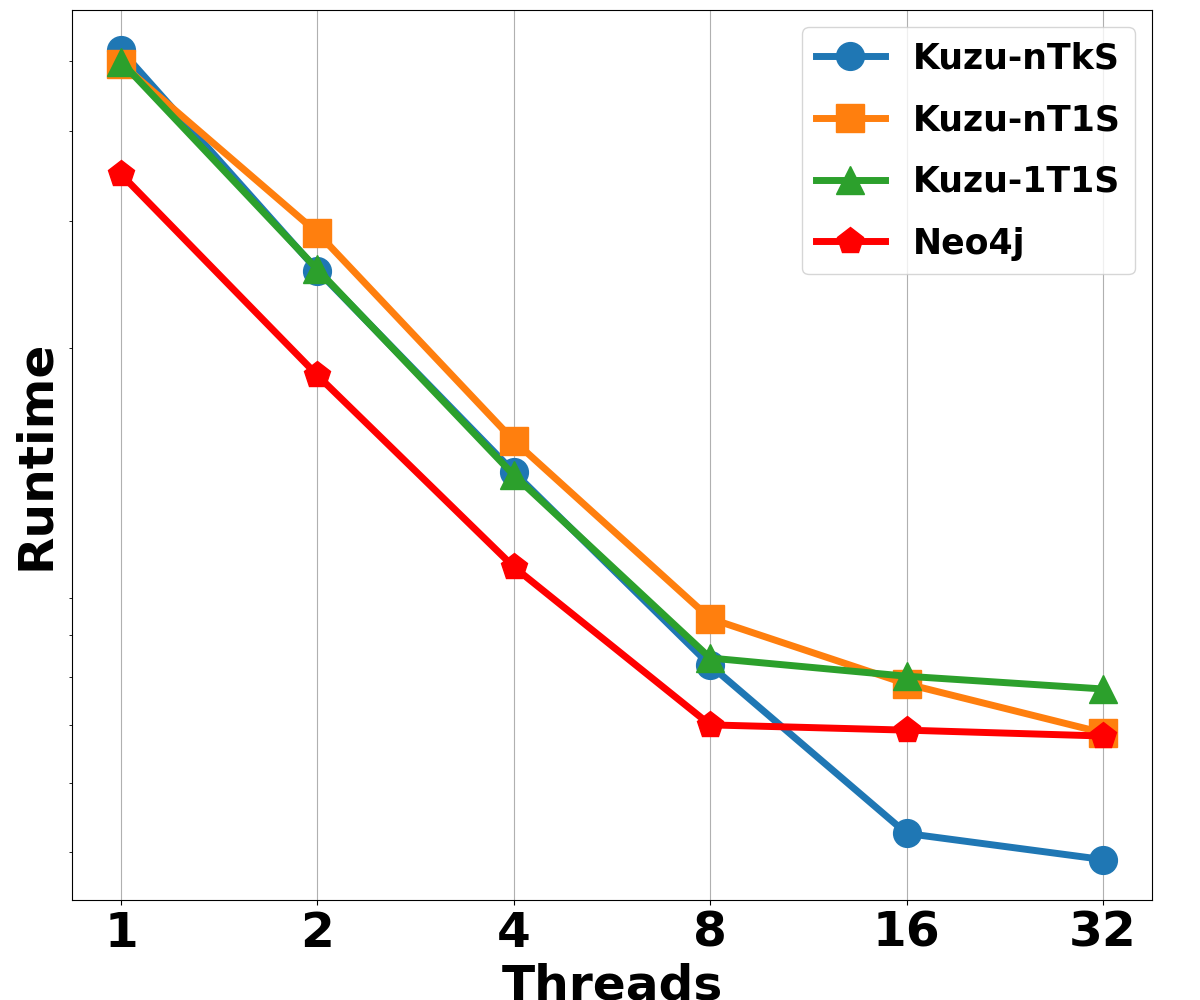}
        \textbf{b) LDBC100, Path}
    \end{minipage}\hfill
        \begin{minipage}{\dimexpr 0.25\linewidth - 0.5\tabcolsep\relax}
        \centering
        \includegraphics[width=\linewidth]{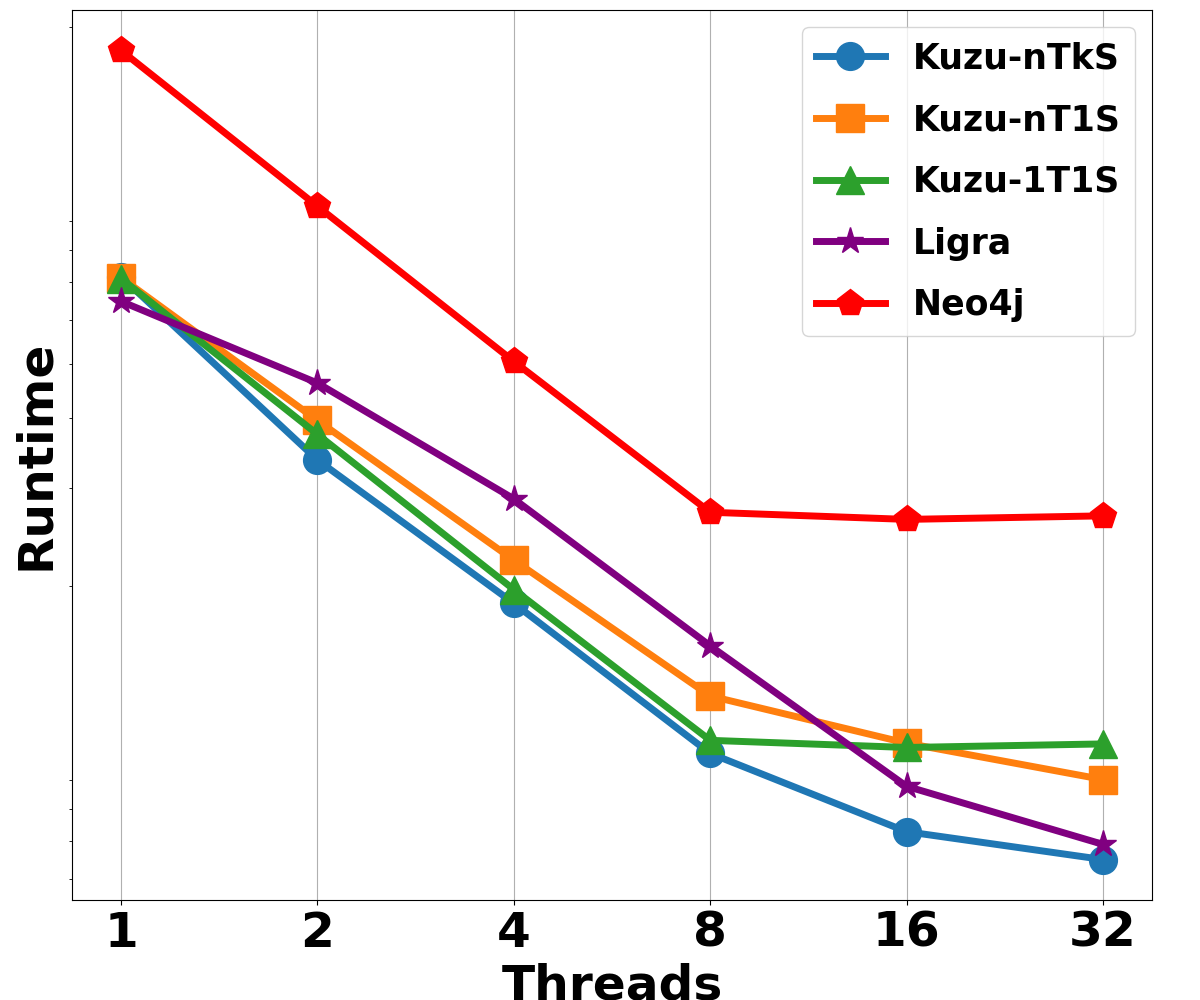}
        \textbf{c) Graph500-28, Path Length}
    \end{minipage}\hfill
    \begin{minipage}{\dimexpr 0.25\linewidth - 0.5\tabcolsep\relax}
        \centering
        \includegraphics[width=\linewidth]{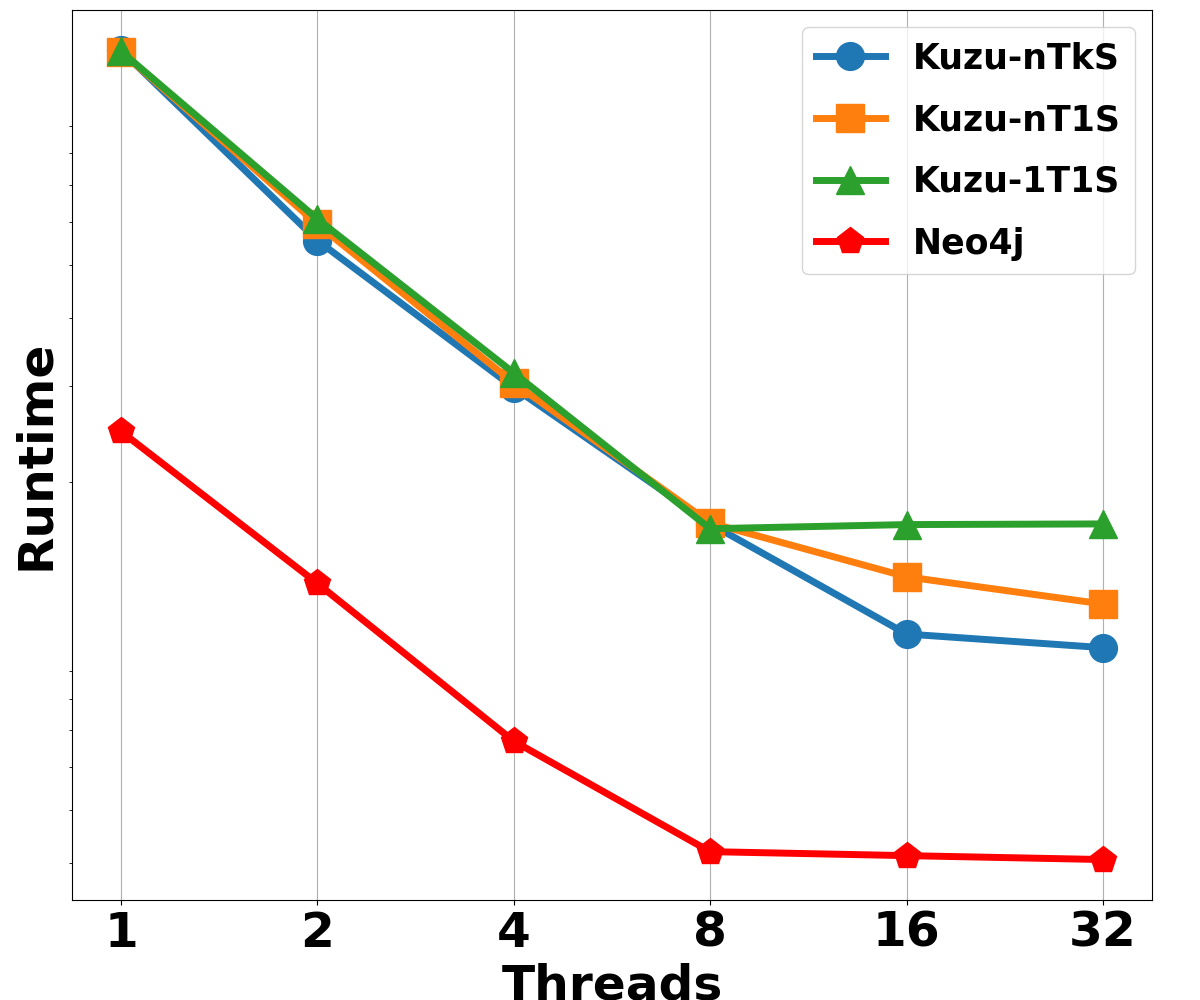}
        \textbf{d) Graph500-28, Path}
    \end{minipage}
    \captionsetup{justification=centering}
    \caption{Runtime vs. threads for 8-Source Workloads (runtime in milliseconds).}
    \label{fig:8-source-exp}
\end{figure*}

\begin{figure*}[t!]
    \centering    
    \begin{minipage}{\dimexpr 0.25\linewidth - 0.5\tabcolsep\relax}
        \centering
        \includegraphics[width=\linewidth]{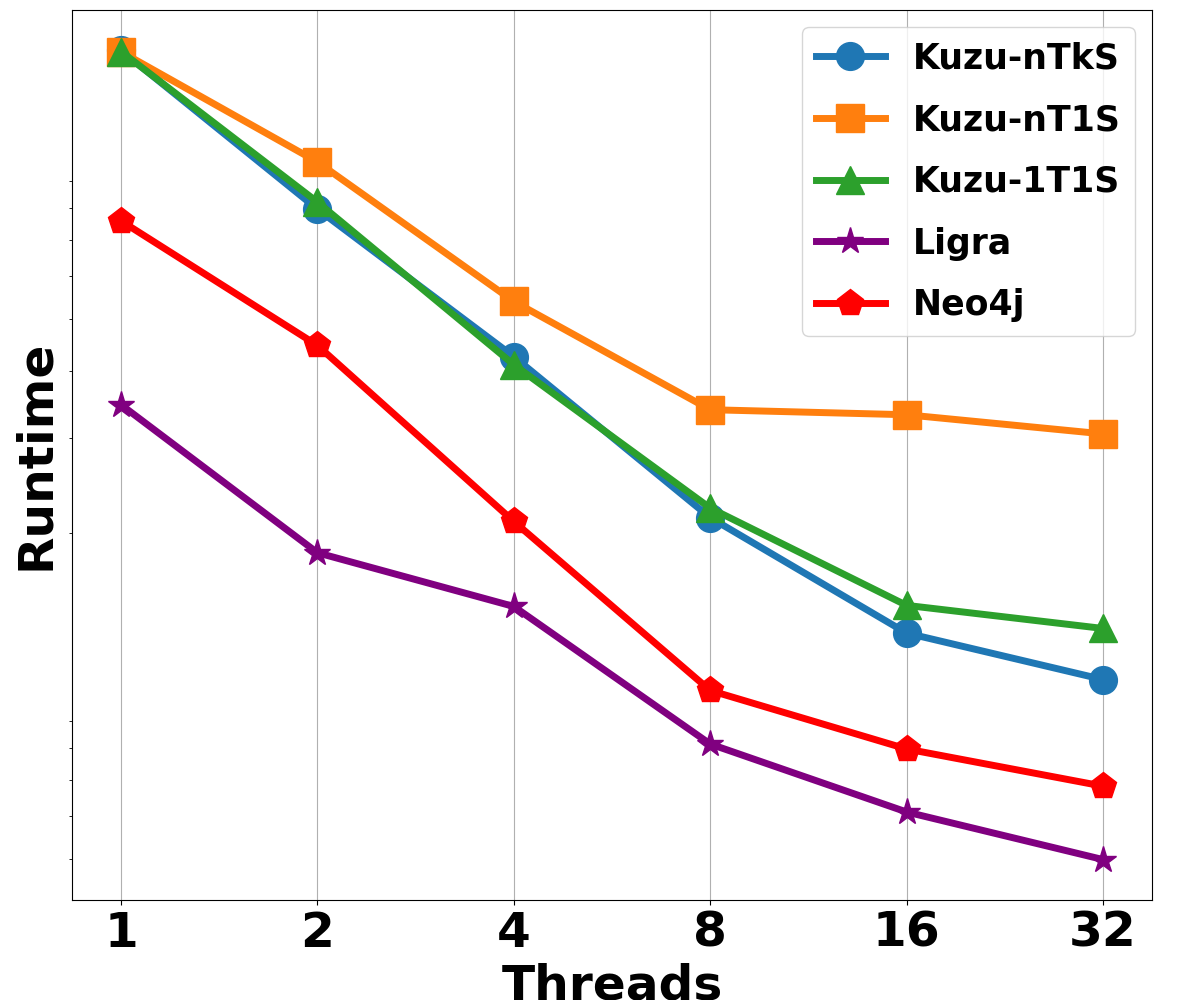}
         \textbf{a) LDBC100, Path Length}
    \end{minipage}\hfill
    \begin{minipage}{\dimexpr 0.25\linewidth - 0.5\tabcolsep\relax}
        \centering
        \includegraphics[width=\linewidth]{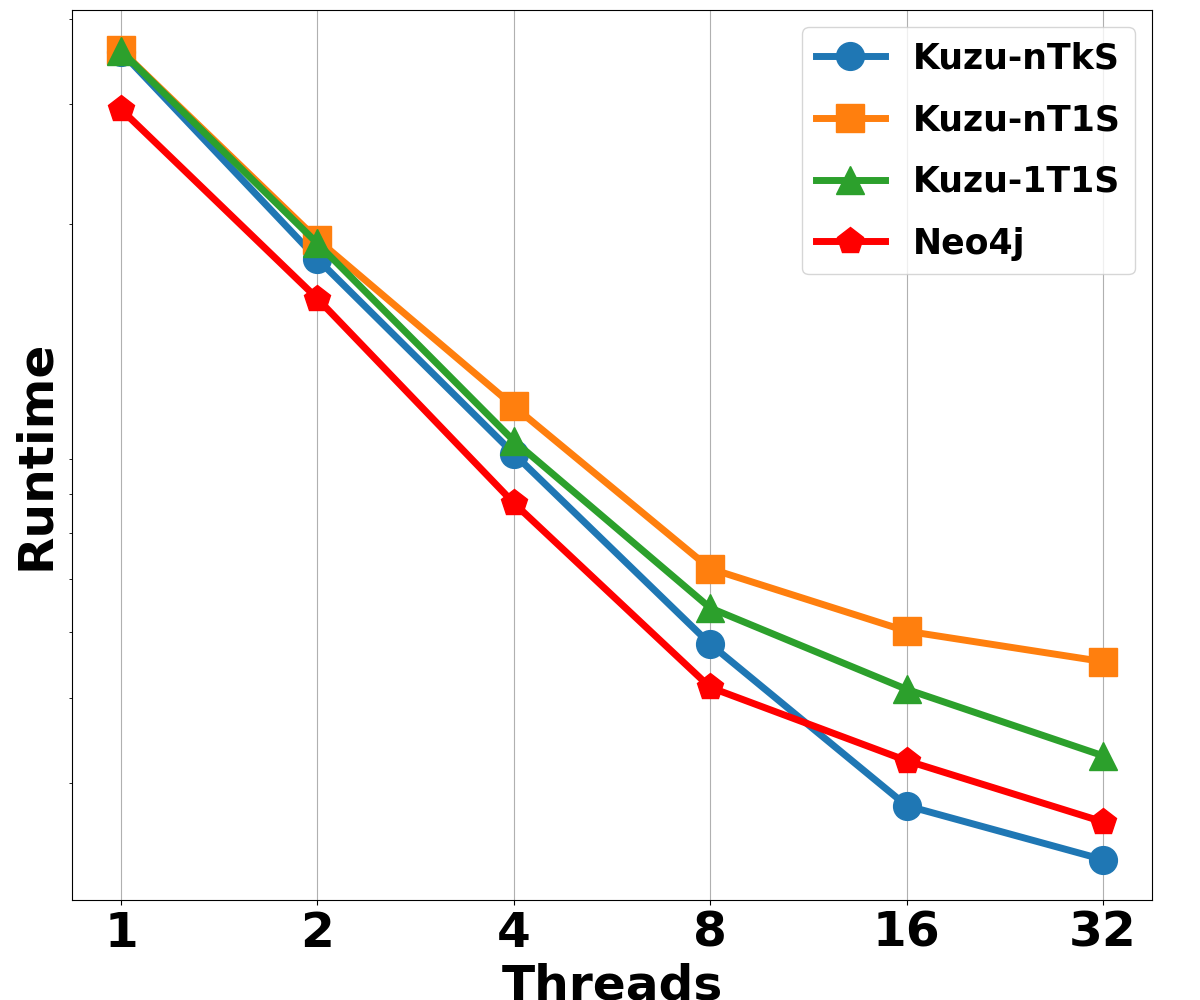}
        \textbf{b) LDBC100, Path}
    \end{minipage}\hfill
        \begin{minipage}{\dimexpr 0.25\linewidth - 0.5\tabcolsep\relax}
        \centering
        \includegraphics[width=\linewidth]{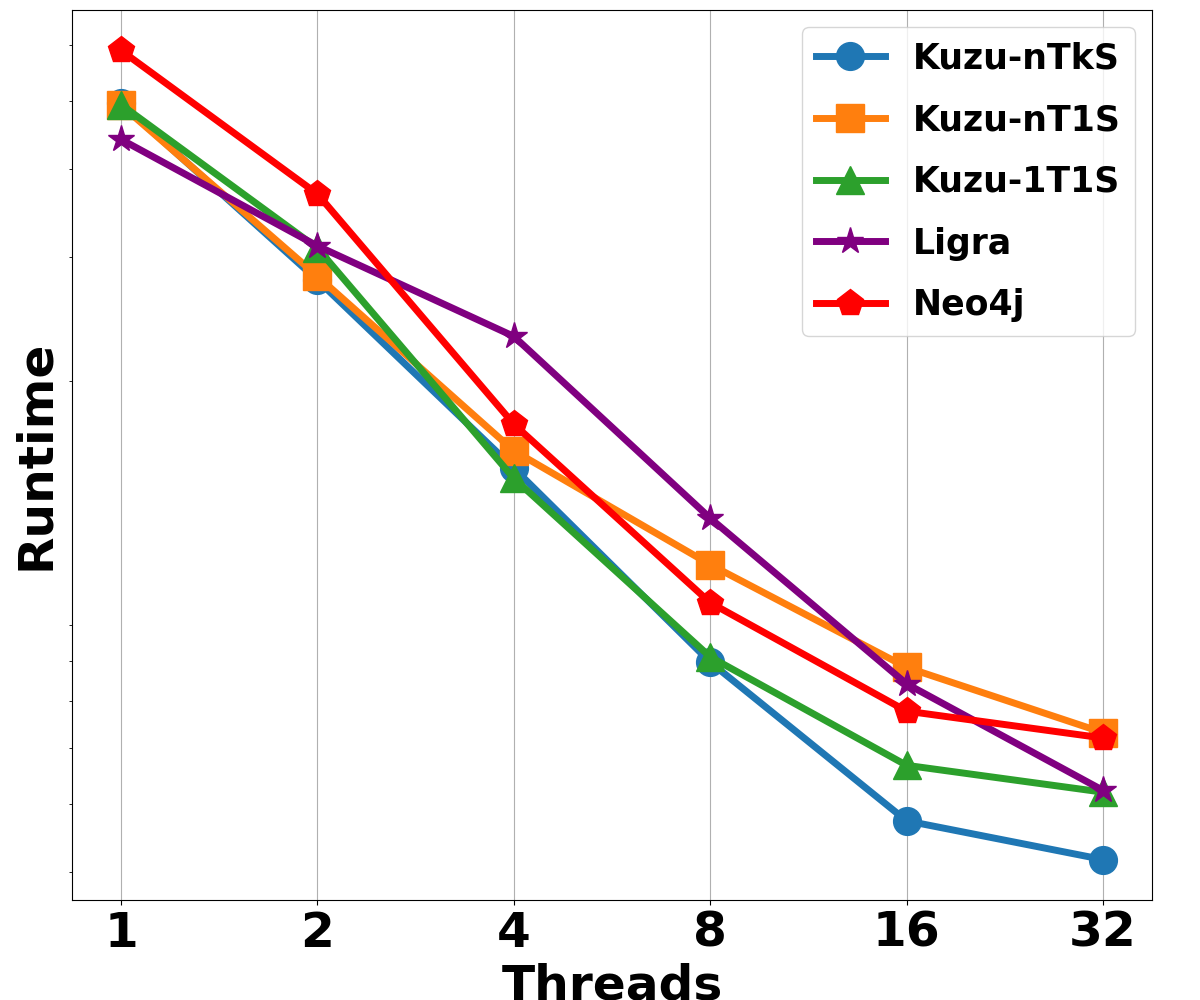}
        \textbf{c) Graph500-28, Path Length}
    \end{minipage}\hfill
    \begin{minipage}{\dimexpr 0.25\linewidth - 0.5\tabcolsep\relax}
        \centering
        \includegraphics[width=\linewidth]{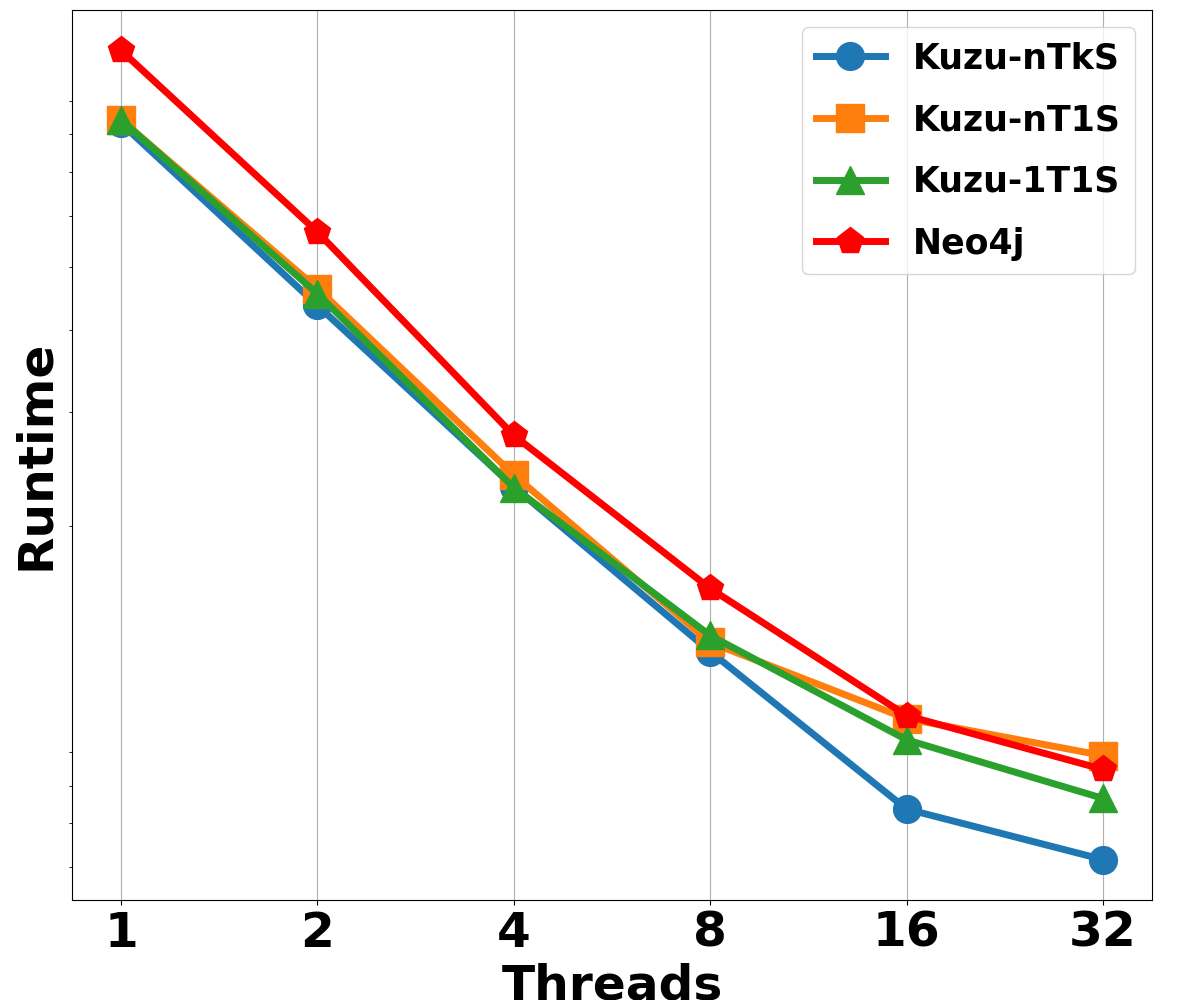}
        \textbf{d) Graph500-28, Path}
    \end{minipage}
    \captionsetup{justification=centering}
    \caption{Runtime vs. threads for 64-source workloads (runtime in milliseconds).}
    \label{fig:64-source-exp}
\end{figure*}
\fi

\subsection{1-Source Workloads}
\label{subsection:single-source-workloads}
In our first set of experiments we measure the runtime of our baselines
on our 1-source workloads as we increase the parallelism from 1 thread to 32 threads.
We expect that systems that implement \OneTOneS\ to not parallelize and those that implement frontier parallelism (i.e., \nTOneS\ policy) to parallelize with more threads. We also expect systems that implement the \nTkS\ policy to parallelize.
In particular,
we expect \Kuzu-\nTkS\ to mimic the behavior of \Kuzu-\nTOneS. 

Tables ~\ref{tab:shortest-path-length-workloads}a and~\ref{tab:shortest-path-workloads}a show the runtimes under our workloads that return path lengths and paths, respectively. \revision{The tables also report the  CPU utilizations of the systems.
Recall that the machine we use supports 32 virtual cores/threads. As an example, 
a 50\% utilization indicates that 16 of the threads were busy doing useful work.} 
\iflong
Figure~\ref{fig:1-source-exp} shows the same results for all parallelism
levels for our smallest LDBC100 and largest Graph500-28 graphs.
\fi

Observe that as we expect, both  \Kuzu-\OneTOneS\
and Neo4j, which implements a \OneTOneS\ policy, cannot benefit from additional threads. Instead, \Kuzu-\nTOneS\ and Ligra benefit from additional threads. On the queries that return path lengths, Ligra achieves between 9.0x to 16.4x improvement, while \Kuzu-\nTOneS\ achieves between 5.0x to 12.8x improvement. Importantly, \Kuzu-\nTkS\ mimics the behavior of \Kuzu-\nTOneS, producing almost identical runtime numbers. Similarly \DuckPGQnTkS\ demonstrates comparable scalability factors to \Kuzu-\nTOneS\ on  all datasets except for LDBC100. This is because the underlying DuckDB system allocates
as many threads to any pipelines as there are 
number of {\em row groups}~\cite{duckdb-rowgroups} in the underlying scanned table
of a pipeline. A row group in DuckDB
contains 122880 rows.
The pipeline for the IFE computation on LDBC scans from the node
table in DuckDB, which contains 448626 rows.
So the IFE pipeline is assigned only 4 threads
on LDBC and the CPU utilization of \DuckPGQnTkS\ is around 4/32=12.5\%.

We note that \Kuzu-\nTOneS\ and Ligra are two independent systems at different performance levels. Ligra is generally faster because it is a purely in-memory 
system and accesses adjacency lists directly, while \Kuzu-\nTOneS\ accesses them through the buffer manager. 
Note that the \DuckPGQnTkS\ numbers we report are also faster than
\Kuzu-\nTkS\ for two reasons. First, DuckPGQ also uses
an in-memory CSR and does not use a buffer manager. Second,
recall that we only report the IFE computation
pipeline of \DuckPGQnTkS, as \DuckPGQnTkS's end-to-end runtime
is dominated by its on-the-fly in-memory CSR construction pipeline.

\subsection{8-source Workloads}
\label{subsection:few-sources-workloads}
We next evaluate our baselines on workloads that contain multiple sources but smaller than
the number of available threads, using our 8-source workloads.
Here we expect that systems that implement \OneTOneS\ policies to benefit from parallelism up to 8 threads, which are the number of sources in these workloads. We expect \nTOneS\ policies to behave similar to the single-source experiments, since these policies
should repeat their previous behaviors on each source. \Kuzu-\nTkS\ should on the other hand
outperform both policies. This is because unlike \OneTOneS, it can use all threads when more than 8 threads are available and 
unlike \nTOneS\, it is not limited by the amount of parallelism achievable on a single frontier. 

Tables ~\ref{tab:shortest-path-length-workloads}b and~\ref{tab:shortest-path-workloads}b show our results for all of our datasets. 
\iflong
Figure ~\ref{fig:8-source-exp}
shows the same results for all parallelism
levels for our smallest LDBC100 and largest Graph500-28 graphs. 
\fi
Observe that now both \Kuzu-\OneTOneS\ and Neo4j
parallelize up to 8 threads and then flatten as there are not enough source morsels to issue to threads. This is why the CPU utilization for both these cases reach at most 25\%, which corresponds to 8 threads out of 32, across all datasets.
\Kuzu-\OneTOneS\ achieves between 5.8x and 7.8x improvements, while Neo4j achieves between 5.1x and 6.9x improvements. 
Further,
 \Kuzu-\nTOneS\ and Ligra achieve similar scalability levels as before: on the queries that return path lengths, Ligra achieves between 7.4x to 12.2x improvement, while \Kuzu-\nTOneS\ achieves between 5.2x to 13.5x improvement. 


\Kuzu-\nTkS\ achieves more robust parallelism on the same workloads, parallelizing between 11.0x-14.0x. 
\revision{\DuckPGQnTkS\ behaves similar to \Kuzu-\nTkS\ except on LDBC, where it again is assigned only 4 threads due to row group based processing.} The improvements of \Kuzu-\nTkS\ over \Kuzu-\nTOneS\ is especially visible on the LDBC graph, where \Kuzu-\nTOneS\ achieves a very limited parallelism of 5.2x. Instead, \Kuzu-\nTkS\ can achieve 11.5x 
improvement here, improving absolute runtime by 2.1x. This indicates that when the frontier of a specific IFE subroutine gets sparse and \Kuzu-\nTOneS\ 
starts keeping threads idle, \Kuzu-\nTkS\ can keep those threads active on other concurrently running IFE subroutines. \revision{This behavior is also reflected by the higher CPU utilization numbers for \Kuzu-\nTkS\ than other policies.} We observe similar patterns on the 
queries that return paths, where \Kuzu-\nTOneS\ achieves between 8.5x-14.1x improvement factors. In contrast, \Kuzu-\nTkS\ achieves more robust improvements factors of between 12.4x to 14.7x. Spotify is an outlier here, where the runtime and CPU utilization gap between \nTkS\ and \nTOneS\ policy is not significant. 
This is because Spotify has a very high average degree,
which leads to denser frontiers. Therefore, even if a single source's IFE computation is active at any point, there is enough work to keep all threads busy and we do not need to launch multiple concurrent IFE computations. As a result, \nTOneS\ scales better on Spotify than other datasets.

\subsection{64-source Workloads}
\label{subsection:many-sources-workloads}
Our next set of experiments evaluates the behavior of \nTkS\ and our baselines when the query has more sources than total threads in the system.
We now expect \nTkS\ to mimic the behavior of systems that implement the \OneTOneS\ policy, but we can also expect it to beat its performance.
The reason for this is that in the last phases of the computation for \OneTOneS, once number of available
sources go below 32, \OneTOneS\ policies start to keep
some threads idle. Instead in \nTkS, those threads start helping other IFE subroutines that are continuing to execute.
Finally, we expect the \nTOneS\ policies to behave similar to their previous behaviors.

Tables~\ref{tab:shortest-path-length-workloads}c and~\ref{tab:shortest-path-workloads}c show our results for all of our datasets. 
\iflong
Figure~\ref{fig:64-source-exp}
shows the same results for all parallelism
levels for our smallest LDBC100 and largest Graph500-28 graphs.
\fi
As expected, Ligra and \Kuzu-\nTOneS\ continue behaving similar to the previous experiments.
Observe that we no longer see Neo4j and \Kuzu-\OneTOneS\ policies flattening as there are enough source morsels to keep threads busy. Neo4j now achieves between
9.5x to 13.2x improvements while \Kuzu-\OneTOneS\ achieves between 7.8x to 11.9x improvements. 
Importantly, except on the Spotify dataset,
 \Kuzu-\OneTOneS\ now consistently outperforms \Kuzu-\nTOneS\ and achieves
 better scalability numbers. 

\revision{\Kuzu-\nTkS\ achieves
between 11.5x and 15.5x improvements.
\DuckPGQnTkS\ similarly achieves between 
11.1x and 15.3x improvements (again except on LDBC).}
Observe that \Kuzu-\nTkS\ is consistently competitive with or outperform \Kuzu-\OneTOneS. This is because once the number of active IFE subroutines reduces below the number of threads, \Kuzu-\OneTOneS\ starts keeping threads idle. \revision{This is why it consistently achieves
lower CPU utilization numbers than \Kuzu-\nTkS}.
Instead, \Kuzu-\nTkS\ keeps those threads busy by dispatching them work from IFE subroutines that have not yet finished. 

\revision{{\em In summary, our experiments so far demonstrate that the \nTkS\ policy is a robust approach to obtaining good scalability across query workloads that contain few source nodes to those with more source nodes than the number of threads.} This is because although \nTkS\ implements a mechanism
to dispatch work from the frontiers of each IFE subroutine,
whenever frontiers get sparse and there is contention, it moves threads to other frontiers from which more work can be given. Therefore, we recommend it to system developers as a robust morsel dispatching policy.}

\begin{figure}[t!]
    \centering
    \hspace*{-0.25cm}
    \begin{subfigure}{0.5\textwidth}
        \includegraphics[width=\linewidth]{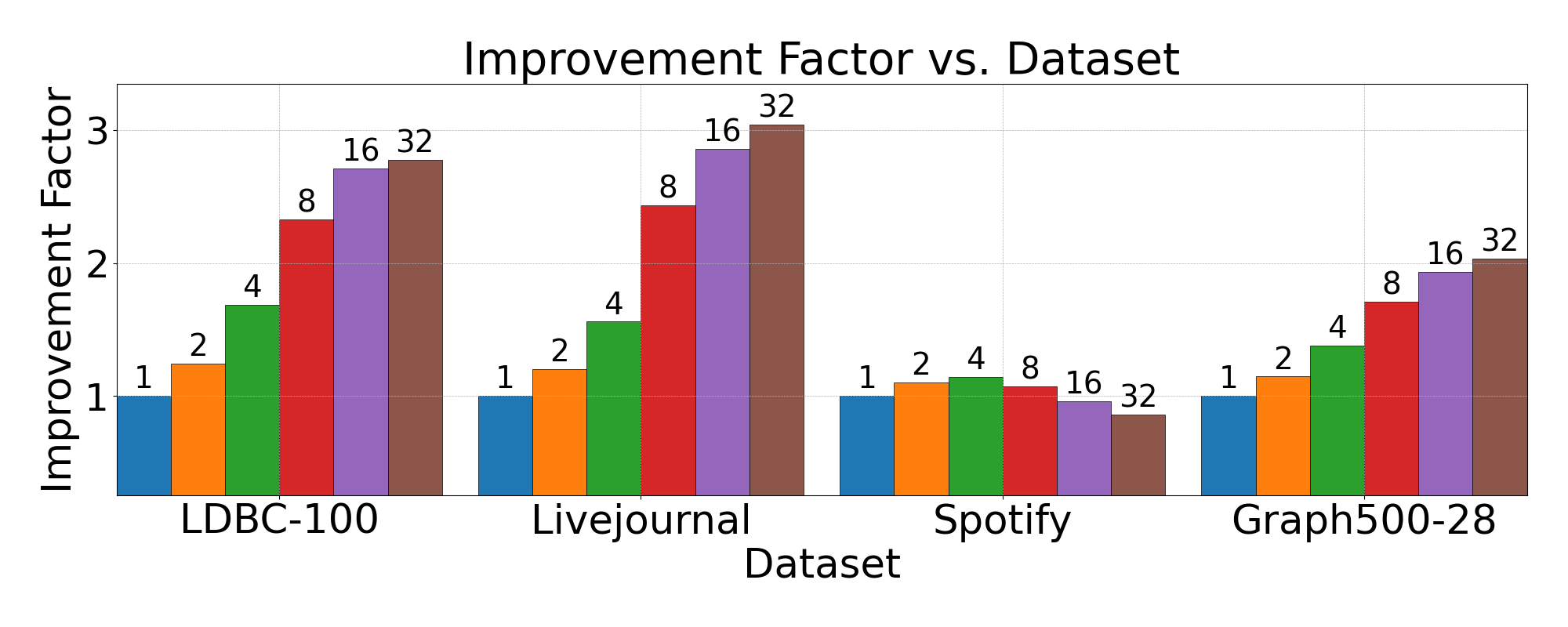}
        \caption{\nTkS\ experiments.}
        \label{fig:choice-of-k-single-source}
    \end{subfigure}
    \hfill
    \hspace*{-0.25cm}
    \begin{subfigure}{0.5\textwidth}
        \includegraphics[width=\linewidth]{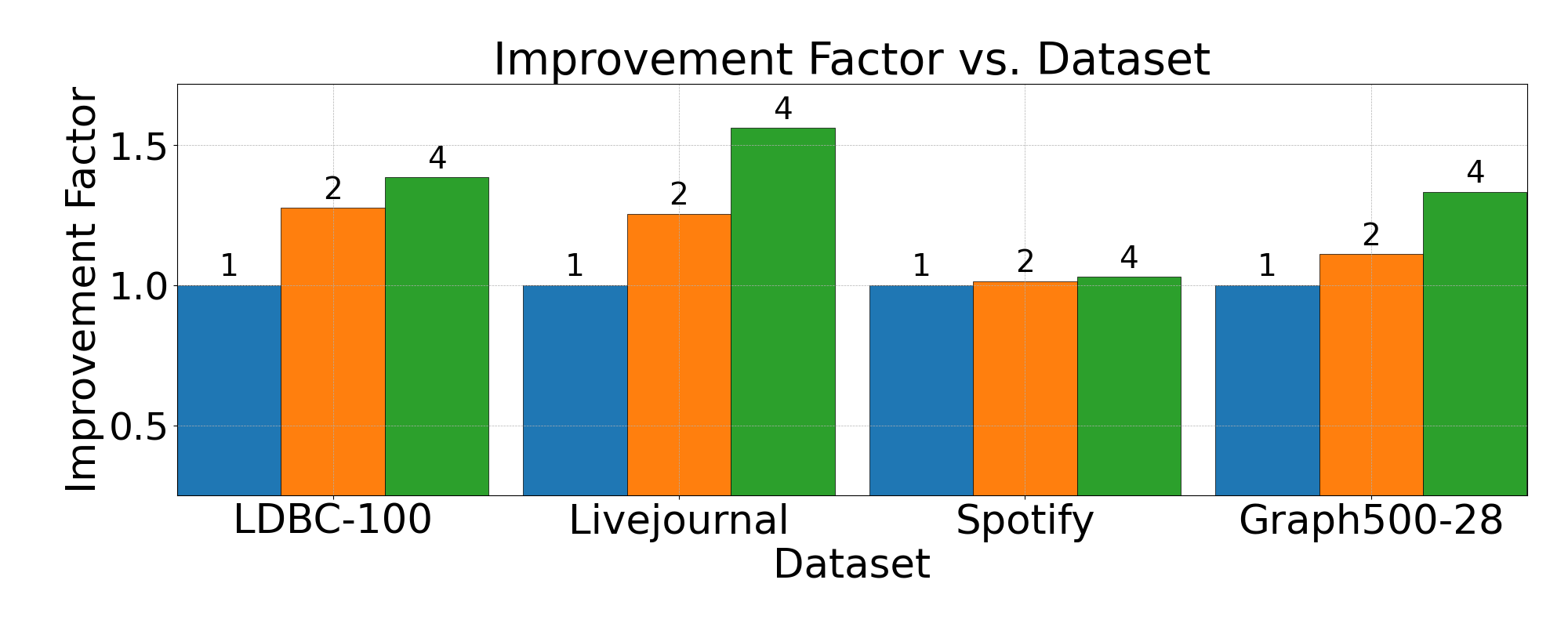}
        \caption{\nTkMS\ experiments.}
        \label{fig:choice-of-k-multi-source}
    \end{subfigure}
    \caption{(a) \Kuzu-\nTkS\ and (b) \Kuzu-\nTkMS\ performances under varying $k$. The values on top of the bars are $k$ values. Improvement factors are over using $k$=1.}
    \label{fig:choice-of-k}
\end{figure}

\subsection{Effects of the Choice of $k$}
\label{subsection:choice-of-k}

\begin{figure}[ht]
    \centering
    \includegraphics[width=\linewidth]{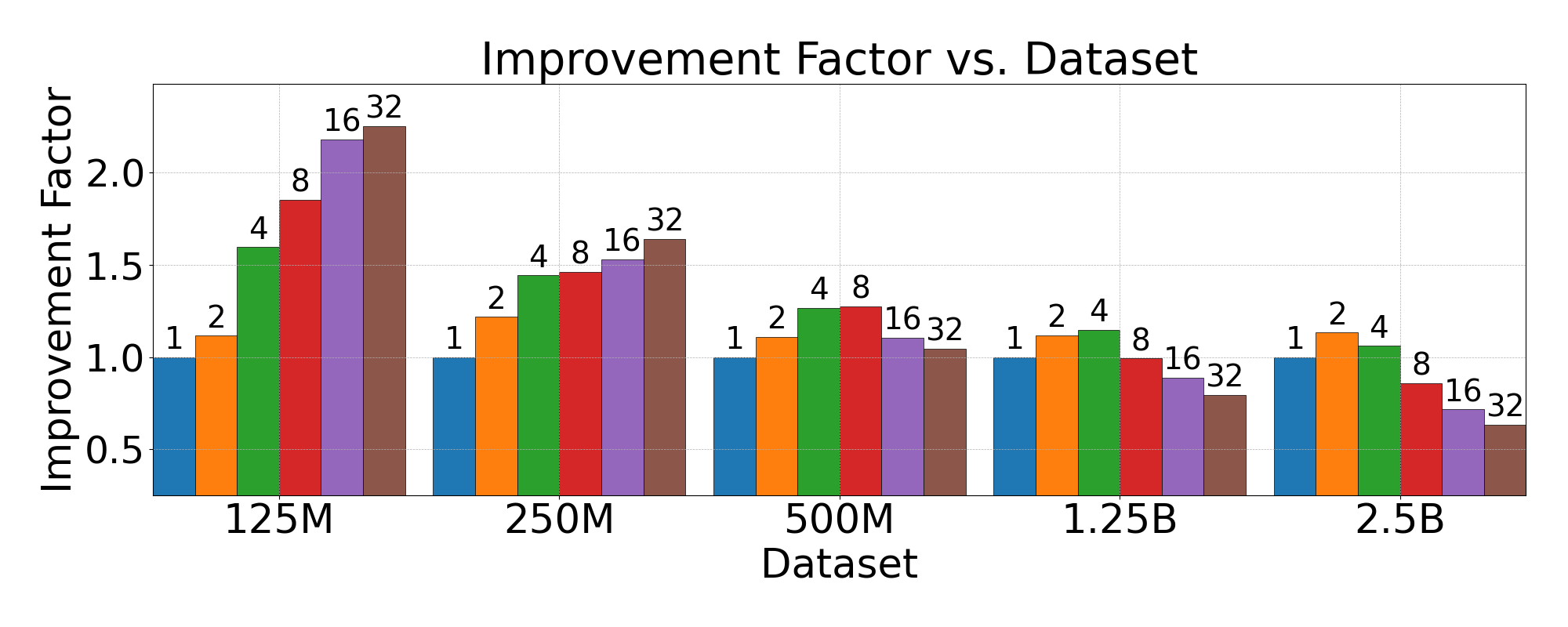}
    \caption{Varying $k$ for random Erd\H{o}s-Renyi graphs.}
    \label{fig:value-of-k-rmat}
\end{figure}

Recall that the parameter $k$ in the \nTkS\ policy determines the number of concurrent sources that will be dispatched to threads. So far in our experiments, 
we set this value to 32, which is the maximum number of threads we use in our experiments. This value ensures that threads work on separate sources when 
possible and only work together on the same source when the total active sources is constrained through $k$ or if there are no new sources left to launch. 
Our next set of experiments study the effects of the choice of $k$ on the performance of \nTkS. We use 32 threads and use our 64-sources workload that return 
path lengths and vary $k$ from 1 to 32. Figure~\ref{fig:choice-of-k-single-source} shows the improvement factors we get as we increase $k$ over
using $k$=1. 
\revision{We see that  increasing $k$ generally improves performance
by up to 3x on our datasets, except in Spotify, where we see a 1.15x performance loss.

To better understand this differing behavior, we analyzed the CPU metrics using Linux perf tool on the Spotify experiments. First,
no matter what the value of $k$ is, the CPU utilization is high across all Spotify experiments. CPU utilization is a good metric for how well
the computation is parallelizing, e.g., is
there contention on locks, but not necessarily 
whether the CPUs are doing actual work or stalled for memory accesses. Instead, the runtime behavior \nTkS\ policy 
on Spotify is related to the CPU cache hit rate. Table~\ref{tab:runtime-vs-cache2} shows the LLC-Load throughputs for each experiment from Figure~\ref{fig:choice-of-k}. 
LLC-Load is the number of memory requests that was served from
the L3 cache.
Observe that except for Spotify, LLC-Load increases in each dataset as $k$ increases, which correlates with the runtime pattern in Figure~\ref{fig:choice-of-k}. In contrast for Spotify, we see a decrease after $k=4$ (from 50.1M to 38.2M).

We hypothesized that this behavior must be driven by the average degree in these graphs. Spotify has an average degree of 535. This is much larger than other datasets, which have an average degree of at most 44. First, this explains why the LLC throughput is significantly larger in Spotify (between 38-50M loads/second) compared to other datasets (at most 24M loads/second). That is,
the IFE computation performs a lot more memory accesses in general.
More importantly, a high average degree implies that on average, nodes will have many common neighbors. Therefore, threads that are extending the frontiers of the same source morsel access the common locations in the same visited array structure to check if a neighbor is already visited or not. 
\iflong
To verify this, we measured the average number of times each node is visited when the densest frontier is extended during an IFE computation. This frontier is the most expensive frontier to extend since it has the most outgoing edges. 
The higher this is, the more L3 cache locality a computation obtains. Table ~\ref{tab:total-visits-factor} shows this metric for different datasets for a single source's IFE execution. As shown, in Spotify, the each visited array location is accessed on average 498.8 times, while this number is much lower in other datasets (at most 29.1 times).

\begin{table}[ht]
\centering
\bgroup
\setlength{\tabcolsep}{2pt}
\def\arraystretch{1.1}
\small 
\begin{tabular}{|c|c|c|}
\hline
     & Densest Frontier Total Visits & Factor (Visits / Total Nodes) \\ \hline
LDBC & 13,068,901                    & 29.1                          \\ \hline
LJ   & 34,848,468                    & 7.2                           \\ \hline
Sp   & 1,798,171,033                 & 498.8                         \\ \hline
G-28 & 2,442,428,724                 & 20.1                          \\ \hline
\end{tabular}
\egroup
\captionsetup{justification=centering}
\caption{Total Visits Factor}
\label{tab:total-visits-factor}
\end{table}

\else
To verify this, we provide additional metrics in the longer version of this paper~\cite{chakraborty:recjoins-tr} that directly measures the average number of times each node is accessed in the visited array across experiments on different datasets.
\fi
This leads to a higher L3 cache hits. 
However, as k increases, this cache locality decreases, since threads start working on different auxiliary data structures and access different visited arrays.

To verify that increasing $k$ can lead to lower cache locality as graphs get denser, we performed a further controlled experiment. We generated a set of random Erd\H{o}s-Renyi graphs, each with 5 million nodes, and increasing average degrees from 25 to 500.
Then we repeated our experiment studying the effect of $k$. Figure \ref{fig:value-of-k-rmat} shows our results. 
Starting with an average degree of 100 (500M edges), increasing $k$ can start degrading performance. 
Further, the denser the graph, the lower the value of $k$ at which further increases in $k$ start degrading the performance. Specifically at average degrees 100 (500M edges), 250 (1.25B edges), and 500 (2.5B edges), performance degrades at $k=16$, $k=8$, and $k=4$, respectively.
{\em In summary, the optimal choice of $k$ depends on how dense the input graph is. On the one hand, increasing $k$ increases the total parallelism we can extract from the \nTkS\ policy and improves performance.
At the same time, as the average degrees of input graphs increase, this gain can be offset by a loss of CPU cache locality and in balance degrade performance. Systems 
that use the \nTkS\ policy can use the average degrees as a proxy to select an appropriate $k$ value.}
}

\begin{table}[ht]
\centering
\bgroup
\setlength{\tabcolsep}{2pt}
\def\arraystretch{1.1}
\small 
\begin{tabular}{|c|c|c|c|c|c|c|c|}
\hline
          & \textbf{k $\rightarrow$} & 1 & 2 & 4 & 8 & 16 & 32 \\ \hline
\multirow{3}{*}{LDBC}    
          & Time  & 4.1  & 3.3  & 2.3  & 1.5  & 1.3  & 1.2  \\ \cline{2-8}
          & LLC Tp & 10.9  & 11.4  & 13.9  & 19.4  & 23.6  & 23.9  \\ \cline{2-8}
          & CPU \% & 66       & 78      & 85       & 88       & 95       & 98       \\ \hline
\multirow{3}{*}{LJ}      
          & Time  & 37.5 & 31.2 & 22.6 & 13.5 & 10.3 & 9.7  \\ \cline{2-8}
          & LLC Tp & 6.2   & 6.5   & 7.2   & 9.5   & 10.7  & 10.9  \\ \cline{2-8}
          & CPU \% & 87       & 90       & 92       &  93      &  96      &  98      \\ \hline
\multirow{3}{*}{Sp} 
          & Time  & 82.8 & 71.8 & 68.7 & 73.0 & 82.8 & 95.6 \\ \cline{2-8}
          & LLC Tp & 40.4  & 48.5  & 50.1  & 48.6  & 43.1  & 38.2  \\ \cline{2-8}
          & CPU \% & 94      & 96      & 98        & 97       & 95       &  91      \\ \hline
\multirow{3}{*}{G-28}    
          & Time  & 938.9 & 766.0 & 640.0 & 492.9 & 449.9 & 432.0 \\ \cline{2-8}
          & LLC Tp & 12.7  & 15.1  & 17.2  & 21.2  & 23.0  & 24.0  \\ \cline{2-8}
          & CPU \% & 70       &  80      & 87       & 92       & 95       &  96      \\ \hline
\end{tabular}
\egroup
\captionsetup{justification=centering}
\caption{Runtime (s) vs LLC Throughput (Tp, Million/s).  64-src workload}
\label{tab:runtime-vs-cache2}
\end{table}

\subsection{Multi Source Morsel Optimization}
\label{subsection:ms-bfs-morsel-optimization}
Our final set of experiments evaluate the performance benefits of \nTkMS\ policy. 
Recall that MS-BFS has the benefit of reducing the amount of scans by sharing scans across multiple IFE subroutines.
 At the same time, it has some overheads. Specifically, 
when a node 
$u$ is visited in an IFE, i.e., put into a frontier, its 64-bit visited status changes. We need an additional loop to find out all the frontiers that $u$ is part of to 
update the other data structures.
We first evaluate this optimization on query workloads that contain  1 to 256 sources using \Kuzu-\nTkMS\ configuration with $k=4$. 
As a baseline, we use the standard \Kuzu-\nTkS\ configuration
with $k=32$.

\begin{figure}[t!]
\includegraphics[keepaspectratio,height=5.5cm]{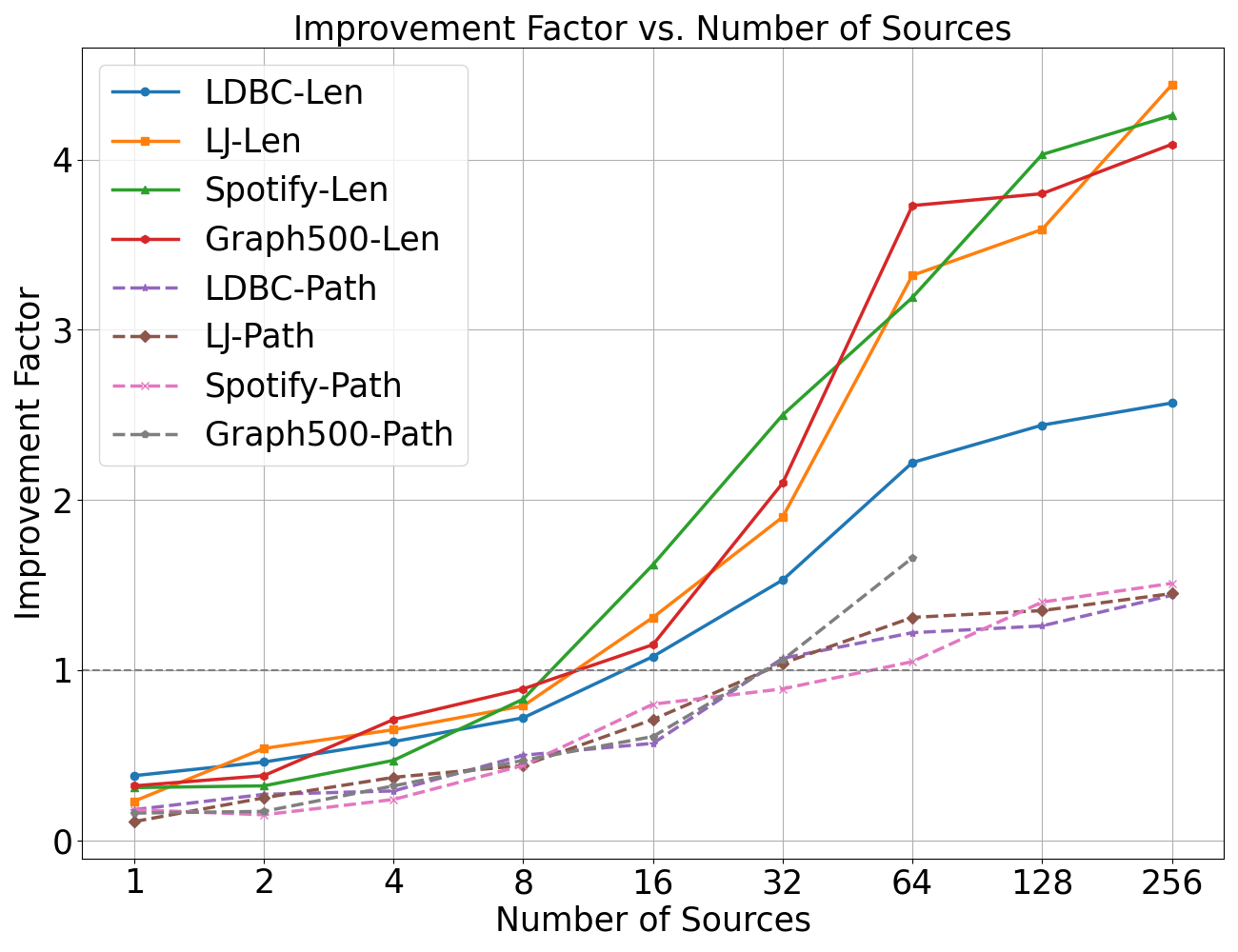}
\centering
\caption{\nTkMS\ ($k=4$) improvement over \nTkS\ ($k=32$).}
\label{fig:ntks-vs-msbfs}
\end{figure}

Figure~\ref{fig:ntks-vs-msbfs} shows the performance improvements of \Kuzu-\nTkMS\ over \Kuzu-\nTkS\ as we increase the number of sources in the queries.
The solid and dashed lines represent the results when using query workloads that return lengths and paths, respectively. First, observe that \Kuzu-\nTkMS\ is often 
slower than \Kuzu-\nTkS\ when using fewer than 32 sources. Unless enough sources are available for a single multi-source morsel, the overhead of \Kuzu-\nTkMS\ is 
higher than its benefits. When we saturate the ``lanes'', using the optimization improves performance, 1.4x-4.4x across the different datasets and query workloads. 

\revision{Observe that starting with 128 sources, experiments
that return paths run out of memory on Graph500-28. 
Recall from Section~\ref{subsec:data-structures}
that we pre-allocate 536 bytes per node, per multi-source
morsels. Therefore, the pre-allocated memory requirements for 2 multi-source morsels
for Graph500-28, which has $\sim$120 million nodes, is 128 GB. With the space needed to store the actual paths, this exceeds the memory capacity of our machine. 
Returning path lengths requires only 21 GB allocation for 2 multi-source morsels and the size of the memory footprint does not increase during computation.}

In our next and final set of experiments, we repeat our experiment
from Section~\ref{subsection:choice-of-k} and study the effect
of $k$ on \Kuzu-\nTkMS. We use the 256-source workload returning path
lengths and vary $k$. Figure~\ref{fig:choice-of-k-multi-source} shows that
as long as the system is not running out of resources, a larger $k$ improves performance. 
For \Kuzu-\nTkMS\ the improvement factors are less compared to \Kuzu-\nTkS\ policy because as long as there are more that 64 sources in the queries, even when $k=1$, a 
single multi-source morsel keeps threads busy. An advantage of packing multiple sources together is that frontiers generally remain denser because they represent the union of 64 IFE computations running at the same time. \revision{Finally, similar to our previous experiments we 
see that \Kuzu-\nTkMS\ benefits least on our Spotify graph, which has the highest average degree, on which the CPU utilization is consistently already very high for $k=1$.

Prior work ~\cite{DuckPGQ} has provided experiments for MS-BFS only on queries
that return lengths and only on workloads
that saturate at least one multi-source morsel. Our conclusions for MS-BFS are
more nuanced. {\em Specifically, we observe 
important runtime benefits only when the number of sources
are large enough to saturate many of the lanes in a multi-source morsel. Further, on queries
that return paths, the memory footprint of computing
the paths for a large number of sources
can be prohibitively large.} Systems implementing this
optimizations should decide when to trigger this
optimization based on the return type and the number of 
source nodes in the queries.
}


\section{Related Work}
\label{sec:rw}
Morsel-driven parallelism is related to the seminal Volcano system's~\cite{graefe:volcano, graefe:volcano-parallelism} parallelism
model.
Graefe has introduced several parallelism techniques in the context of the Volcano system.
Importantly, he introduces {\em horizontal parallelism}, which refers to two separate approaches. First is {\em bushy parallelism}, which runs
different subplans that do not depend on each other independently on separate threads at the same time. 
\revision{This is a form of {\em task-level parallelism}}.
Second is {\em intra-operator parallelism}, which splits the input of an operator into multiple partitions.
\revision{This is a form of {\em data-level parallelism}.} 
Morsel-driven parallelism a form of data parallelism that extends Volcano's intra-operator parallelism, where the inputs are partitioned
in a much more fine-grained manner than in the Volcano system (always at the leaves of the subplans/tasks).
\revision{While many modern systems adopt data-level parallelism, such as DBMSs adopting morsel-driven parallelism,
Spark~\cite{zaharia:spark}, and Timely Dataflow~\cite{murray:td},
there are also systems that adopt hybrid task- and data-level parallelism approaches~\cite{boehm:systemml}.} 

Our design space includes the vertex-centric parallelism approach of parallel or distributed graph analytics systems.
Systems such as Pregel~\cite{Pregel}, Ligra~\cite{Ligra}, Graph\-Chi~\cite{Graphchi}, GraphLab~\cite{Graphlab} 
have been based on this model of parallelism. 
These systems have vertex- or edge-centric APIs, such as Ligra's \texttt{edgeCompute()} function, 
that assume an IFE-based computation. 
Programmers implement these functions to describe the computation
that should be done per vertex or edge. Then, the system automatically parallelizes the executions of these functions
using vertex-centric parallelism that parallelize work at each frontier.
Our implementation uses Ligra's API to implement parallel versions of recursive algorithms,
including recursive path finding algorithms, such as transitive closure, shortest paths ~\cite{Pregel, Ligra, Graphchi, nole:giraph}. 
Implementing recursive algorithms in a DBMS with a vertex-centric abstraction however, deviates from
the tuple-based abstraction that is used in other physical operators in a DBMS. Another parallelism approach is to use OpenMP \cite{openmp08} or a similar
multi-threading library to automatically parallelize loops using their own thread pools. These threading libraries are adopted in several in-memory graph
analytics libraries, such as graph-tool~\cite{graphtool}, which support a suite of batch graph algorithms. \revision{IFE-based graph algorithms that can be
modeled as sparse matrix multiplication-based computations are also studied extensively in the high-performance computing literature. There is extensive work in this literature on communication-avoiding algorithms that aim to minimize the communication between a set of distributed/parallel compute nodes by advanced data partitioning approaches~\cite{solomonik:ca, demmel:ca}.
For example, in 2D partitioning approaches,
a matrix $M$, each row of which represents an adjacency lists, is partitioned into 
submatrices whose dimensions have size roughly the square root of $M$'s dimensions. 
Our work differs from this literature in that we assume an IFE algorithm executing inside a DBMS that has a standard data layout in which the threads can scan the entire adjacency lists of each node.}

\section{Conclusions}
\label{sec:conclusions}

In this paper we studied several approaches in prior literature to parallelize recursive path-finding queries
in GDBMSs: (i) vanilla morsel-driven parallelism of GDBMSs; (ii) frontier-level parallelism of graph analytics libraries; and (iii) the multi-source morsel optimization. \revision{We first showed that these approaches can all be integrated
to into DBMSs that adopt morsel-driven parallelism as different
morsel dispatching policies that assign source (or multi-source) morsels and frontier-morsels to threads.} We then extended them with hybrid policies, which we call \nTkS\ and \nTkMS\ policies that dispatch both source and frontier morsels. We then experimentally evaluated and analyzed their pros and cons on a variety of query workloads, especially recommending  the \nTkS\ policy as a robust policy to parallelize recursive queries.
Our work focused on queries using walk semantics of paths, which allows multiple nodes and edges to be repeated. 
One important venue for future work is to study queries under the two other path semantics in modern graph query languages, 
which are trail and acyclic. These semantics respectively limit edges and nodes from being repeated in the paths.
It is important to study the optimizations that can be applied to directly compute these semantics, especially
under parallel execution. 

\balance
\bibliographystyle{ACM-Reference-Format}
\bibliography{references}

\end{document}